\documentclass[twocolumn,traditabstract]{aa}
\usepackage{graphicx,amsmath}
\usepackage{epsf}
\usepackage{txfonts}
\usepackage[hyphens]{url}
\usepackage{longtable}
\usepackage{lscape}
\usepackage[hyperindex,breaklinks=true, colorlinks, citecolor=blue]{hyperref}  
\usepackage{breakurl}
\usepackage{comment}
\usepackage{natbib}
\usepackage{upgreek}
\usepackage{float}
\usepackage[switch,pagewise]{lineno}
\usepackage{multirow}
\usepackage{caption}
\usepackage{lipsum}
\usepackage{threeparttable}
\usepackage{multirow}

\newcommand{\juan}[1]{{#1}}

\newcommand{\commentproof}[1]{{\bf \color{green}#1}}
\renewcommand{\commentproof}[1]{} 

\def\setsymbol#1#2{\expandafter\def\csname #1\endcsname{#2}}
\def\getsymbol#1{\csname #1\endcsname}

\def\Planck{\textit{Planck}}

\def\all2013resultspapers{\nocite{planck2013-p01, planck2013-p02, planck2013-p02a, planck2013-p02d, planck2013-p02b, planck2013-p03, planck2013-p03c, planck2013-p03f, planck2013-p03d, planck2013-p03e, planck2013-p01a, planck2013-p06, planck2013-p03a, planck2013-pip88, planck2013-p08, planck2013-p11, planck2013-p12, planck2013-p13, planck2013-p14, planck2013-p15, planck2013-p05b, planck2013-p17, planck2013-p09, planck2013-p09a, planck2013-p20, planck2013-p19, planck2013-pipaberration, planck2013-p05, planck2013-p05a, planck2013-pip56, planck2013-p06b, planck2013-p01a}}

\newbox\tablebox    \newdimen\tablewidth
\def\leaderfil{\leaders\hbox to 5pt{\hss.\hss}\hfil}
\def\tablenote#1 #2\par{\begingroup \parindent=0.8em
    \abovedisplayshortskip=0pt\belowdisplayshortskip=0pt
    \noindent
    $$\hss\vbox{\hsize\tablewidth \hangindent=\parindent \hangafter=1 \noindent
    \hbox to \parindent{$^#1$\hss}\strut#2\strut\par}\hss$$
    \endgroup}

\def\L2{\ifmmode L_2\else $L_2$\fi}

\def\DeltaT{\ifmmode \Delta T\else $\Delta T$\fi}
\def\deltat{\ifmmode \Delta t\else $\Delta t$\fi}
\def\fknee{\ifmmode f_{\rm knee}\else $f_{\rm knee}$\fi}
\def\Fmax{\ifmmode F_{\rm max}\else $F_{\rm max}$\fi}
\def\solar{\ifmmode{\rm M}_{\mathord\odot}\else${\rm M}_{\mathord\odot}$\fi}
\def\Msolar{\ifmmode{\rm M}_{\mathord\odot}\else${\rm M}_{\mathord\odot}$\fi}
\def\Lsolar{\ifmmode{\rm L}_{\mathord\odot}\else${\rm L}_{\mathord\odot}$\fi}
\def\inv{\ifmmode^{-1}\else$^{-1}$\fi}
\def\mo{\ifmmode^{-1}\else$^{-1}$\fi}
\def\sup#1{\ifmmode ^{\rm #1}\else $^{\rm #1}$\fi}
\def\expo#1{\ifmmode \times 10^{#1}\else $\times 10^{#1}$\fi}
\def\,{\thinspace}
\def\lsim{\mathrel{\raise .4ex\hbox{\rlap{$<$}\lower 1.2ex\hbox{$\sim$}}}}
\def\gsim{\mathrel{\raise .4ex\hbox{\rlap{$>$}\lower 1.2ex\hbox{$\sim$}}}}

\def\simprop{\mathrel{\raise .4ex\hbox{\rlap{$\propto$}\lower 1.2ex\hbox{$\sim$}}}}
\def\deg{\ifmmode^\circ\else$^\circ$\fi}
\def\pdeg{\ifmmode $\setbox0=\hbox{$^{\circ}$}\rlap{\hskip.11\wd0 .}$^{\circ}
          \else \setbox0=\hbox{$^{\circ}$}\rlap{\hskip.11\wd0 .}$^{\circ}$\fi}
\def\arcs{\ifmmode {^{\scriptstyle\prime\prime}}
          \else $^{\scriptstyle\prime\prime}$\fi}
\def\arcm{\ifmmode {^{\scriptstyle\prime}}
          \else $^{\scriptstyle\prime}$\fi}
\newdimen\sa  \newdimen\sb
\def\parcs{\sa=.07em \sb=.03em
     \ifmmode \hbox{\rlap{.}}^{\scriptstyle\prime\kern -\sb\prime}\hbox{\kern -\sa}
     \else \rlap{.}$^{\scriptstyle\prime\kern -\sb\prime}$\kern -\sa\fi}
\def\parcm{\sa=.08em \sb=.03em
     \ifmmode \hbox{\rlap{.}\kern\sa}^{\scriptstyle\prime}\hbox{\kern-\sb}
     \else \rlap{.}\kern\sa$^{\scriptstyle\prime}$\kern-\sb\fi}
\def\ra[#1 #2 #3.#4]{#1\sup{h}#2\sup{m}#3\sup{s}\llap.#4}
\def\dec[#1 #2 #3.#4]{#1\deg#2\arcm#3\arcs\llap.#4}
\def\deco[#1 #2 #3]{#1\deg#2\arcm#3\arcs}
\def\rra[#1 #2]{#1\sup{h}#2\sup{m}}

\def\dots{\relax\ifmmode \ldots\else $\ldots$\fi}
\def\WHzsr{\ifmmode $W\,Hz\mo\,sr\mo$\else W\,Hz\mo\,sr\mo\fi}
\def\mHz{\ifmmode $\,mHz$\else \,mHz\fi}
\def\GHz{\ifmmode $\,GHz$\else \,GHz\fi}
\def\mKs{\ifmmode $\,mK\,s$^{1/2}\else \,mK\,s$^{1/2}$\fi}
\def\muKs{\ifmmode \,\mu$K\,s$^{1/2}\else \,$\mu$K\,s$^{1/2}$\fi}
\def\muKRJs{\ifmmode \,\mu$K$_{\rm RJ}$\,s$^{1/2}\else \,$\mu$K$_{\rm RJ}$\,s$^{1/2}$\fi}
\def\muKHz{\ifmmode \,\mu$K\,Hz$^{-1/2}\else \,$\mu$K\,Hz$^{-1/2}$\fi}
\def\MJysr{\ifmmode \,$MJy\,sr\mo$\else \,MJy\,sr\mo\fi}
\def\MJysrmK{\ifmmode \,$MJy\,sr\mo$\,mK$_{\rm CMB}\mo\else \,MJy\,sr\mo\,mK$_{\rm CMB}\mo$\fi}
\def\microns{\ifmmode \,\mu$m$\else \,$\mu$m\fi}

\def\muK{\ifmmode \,\mu$K$\else \,$\mu$\hbox{K}\fi}
\def\microK{\ifmmode \,\mu$K$\else \,$\mu$\hbox{K}\fi}
\def\muW{\ifmmode \,\mu$W$\else \,$\mu$\hbox{W}\fi}
\def\kms{\ifmmode $\,km\,s$^{-1}\else \,km\,s$^{-1}$\fi}
\def\kmsMpc{\ifmmode $\,\kms\,Mpc\mo$\else \,\kms\,Mpc\mo\fi}
\providecommand{\sorthelp}[1]{}

\newcommand{\planck}{\Planck}  

\providecommand{\sorthelp}[1]{}

\newcommand{\prs}{$V$}
\newcommand{\mrv}{$r$}
\newcommand{\kps}{km\,s$^{-1}$}
\newcommand{\vhi}{$v_{\rm HI}$}
\newcommand{\vco}{$v_{\rm 13CO}$}
\newcommand{\vlsr}{$v_{\rm LSR}$}
\newcommand{\vlos}{$v_{\rm LOS}$}

\begin{document}

\title{Histogram of oriented gradients: a technique for the study of molecular cloud formation.}
\titlerunning{Spatial correlation of H{\sc i} and $^{13}$CO as revealed by the HOG method.}
    \author{
        J.\,D.~Soler$^{1}$\thanks{Corresponding author, \email{soler@mpia.de}},
        H.~Beuther$^{1}$,
        M.~Rugel$^{1}$,
        Y.~Wang$^{1}$,        
        P.\,C.~Clark$^{2}$,
        S.\,C.\,O.~Glover$^{3}$,
        P.\,F.~Goldsmith$^{4}$,
        M.~Heyer$^{5}$,
        L.\,D.~Anderson$^{6}$,
        A.~Goodman$^{7}$,
        Th.~Henning$^{1}$,
        J.~Kainulainen$^{8}$,
        R.\,S.~Klessen$^{3,9}$,
        S.\,N.~Longmore$^{10}$, 
        N.\,M.~McClure-Griffiths$^{11}$,
        K.\,M.~Menten$^{12}$,
        J.\,C.~Mottram$^{1}$,
        J.~Ott$^{13}$,
        S.\,E.~Ragan$^{2}$,
        R.\,J.~Smith$^{14}$,
        J.\,S.~Urquhart$^{15}$,        
        F.~Bigiel$^{16,3}$,
        P.~Hennebelle$^{17}$,   
        N.~Roy$^{18}$,
        P.~Schilke$^{19}$. 
} 
\institute{
1. Max Planck Institute for Astronomy, K\"{o}nigstuhl 17, 69117, Heidelberg, Germany.\\
2. School of Physics and Astronomy, Cardiff University, Queen's Buildings, The Parade, Cardiff, CF24 3AA, UK.\\
3. Universit\"{a}t Heidelberg, Zentrum f\"{u}r Astronomie, Institut f\"{u}r Theoretische Astrophysik, Albert-Ueberle-Str. 2, 69120, Heidelberg, Germany.\\
4. Jet Propulsion Laboratory, California Institute of Technology, 4800 Oak Grove Drive, Pasadena CA, 91109, USA.\\
5. Department of Astronomy, University of Massachusetts, Amherst, MA 01003-9305, USA.\\
6. Department of Physics and Astronomy, West Virginia University, Morgantown, WV 26506, USA.\\
7. Harvard-Smithsonian Center for Astrophysics, 60 Garden Street, MS 42, Cambridge, MA 02138, USA.\\
8. Dept. of Space, Earth and Environment, Chalmers University of Technology, Onsala Space Observatory, 439 92 Onsala, Sweden.\\
9. Universit\"at Heidelberg, Interdiszipli\"ares Zentrum f\"ur Wissenschaftliches Rechnen, Im Neuenheimer Feld 205, 69120 Heidelberg, Germany.\\
10. Astrophysics Research Institute, Liverpool John Moores University, 146 Brownlow Hill, Liverpool L3 5RF, UK.\\
11. Research School of Astronomy and Astrophysics, The Australian National University, Canberra, ACT, Australia.\\
12. Max Planck Institute for Radio Astronomy, Auf dem H\"{u}gel 69, 53121 Bonn, Germany.\\
13. National Radio Astronomy Observatory, PO Box O, 1003 Lopezville Road, Socorro, NM 87801, USA.\\
14. Jodrell Bank Centre for Astrophysics, School of Physics and Astronomy, University of Manchester, Oxford Road, Manchester M13 9PL, UK.\\
15. Centre for Astrophysics and Planetary Science, University of Kent, Canterbury CT2 7NH, UK.\\
16. Argelander-Institut f\"{u}r Astronomie, Universit\"at Bonn, Auf dem H\"{u}gel 71, 53121 Bonn, Germany.\\
17. Laboratoire AIM, Paris-Saclay, CEA/IRFU/SAp - CNRS - Universit\'{e} Paris Diderot, 91191, Gif-sur-Yvette Cedex, France.\\
18. Department of Physics, Indian Institute of Science, 560012 Bangalore, India.\\
19. Physikalisches Institut der Universit\"{a}t zu K\"{o}ln, Z\"{u}lpicher Str. 77, 50937 K\"{o}ln, Germany.\\
}
\authorrunning{Soler,\,J.\,D. and the THOR collaboration}

\date{Received 21 SEP 2018 / Accepted 28 DEC 2018}

\abstract{
We introduce the histogram of oriented gradients (HOG), a tool developed for machine vision that we propose as a new metric for the systematic characterization of spectral line observations of atomic and molecular gas and the study of molecular cloud formation models.
In essence, the HOG technique takes as input extended spectral-line observations from two tracers and provides an estimate of their spatial correlation across velocity channels.

We characterize HOG using synthetic observations of H{\sc i} and $^{13}$CO\,($J$\,$=$\,1\,$\rightarrow$\,0) emission from numerical simulations of magnetohydrodynamic (MHD) turbulence leading to the formation of molecular gas after the collision of two atomic clouds.
We find a significant spatial correlation between the two tracers in velocity channels where \vhi\,$\approx$\,\vco, almost \juan{independent} of the orientation of the collision with respect to the line of sight.

Subsequently, we use HOG to investigate the spatial correlation of the H{\sc i}, from The H{\sc i}/OH/Recombination line survey of the inner Milky Way (THOR), and the $^{13}$CO\,($J$\,$=$\,1\,$\rightarrow$\,0) emission from the Galactic Ring Survey (GRS), toward the portion of the Galactic plane 33\pdeg75\,$\leq$\,$l$\,$\leq$\,35\pdeg25 and $|b|$\,$\leq$\,1\pdeg25.
We find a significant spatial correlation between the two tracers in extended portions of the studied region.
Although some of the regions with high spatial correlation are associated with H{\sc i} self-absorption (HISA) features, suggesting that it is produced by the cold atomic gas, the correlation is not exclusive to this kind of region.

The HOG results derived for the observational data indicate significant differences between individual regions: some show spatial correlation in channels around \vhi\,$\approx$\,\vco\ while others present spatial correlations in velocity channels separated by a few kilometers per second.
We associate these velocity offsets to the effect of feedback and to the presence of physical conditions that are not included in the atomic-cloud-collision simulations, such as more general magnetic field configurations, shear, and global gas infall.
}

\maketitle
\clearpage

\section{Introduction}\label{section:introduction}

Molecular clouds (MCs) are the main reservoir of cold gas from which stars are formed in the Milky Way and similar spiral galaxies \citep[see for example,][]{bergin2007,dobbs2014,molinari2014}.
Hence the study of the formation, evolution, and destruction of MCs is crucial for any understanding of the star formation process.
 
Much of the interstellar medium (ISM) in disk galaxies is in the form of neutral atomic hydrogen (H{\sc i}), which is the matrix within which many MCs reside \citep{ferriere2001,dickey2003,kalberla2009}.
Much of the H{\sc i} is observed to be either warm neutral medium (WNM) with $T$\,$\approx$\,$10^{4}$\,K or cold neutral medium (CNM) with $T$\,$\approx$\,$10^{2}$\,K \citep{kulkarni1987,dickey1990,heilesANDtroland2003b}.
The transition between the H{\sc i} and the molecular gas is primarily driven by changes in the density and extinction \citep{reach1994,draineANDbertoldi1996,glover2011}.
Consequently, the first step for MC formation is the gathering of sufficient gas in one place to raise the column density above the value needed to provide effective shielding against the photodissociation produced by the interstellar radiation field \citep{krumholz2008,krumholz2009,sternberg2014}.
There are multiple processes that intervene in the accumulation of the parcels of gas out of the diffuse ISM to make dense MCs \citep[for reviews see][and references therein]{hennebelle2012,klessen2016}.
However, despite the increasing number of models and observations, it is still unclear what are the dominant processes that lead to MC formation and what are the observational signatures with which to identify them.

Some of the MC formation mechanisms that have been proposed are converging flows driven by feedback or turbulence, agglomeration of smaller
clouds, gravitational instability and magneto-gravitational instability, and instability involving differential buoyancy \citep[see][and references therein]{dobbs2014}.
Each one of these processes produces morphological and kinematic imprints over different spatial and time scales.
Some are related to the spatial distribution of the atomic and molecular emission \citep[e.g.,][]{dawson2013}, some are associated with the relative velocity \citep[e.g.,][]{motte2014} or the spatial correlation between these two components \citep[e.g.,][]{gibson2005a,goldsmith2005}.
However, most of these imprints remain to be discovered.

An idealized spherical cloud of diffuse gas and dust immersed in a bath of isotropic interstellar radiation begins to form an MC when the column density gets sufficiently high that the gas/dust can self-shield, the H{\sc i} converts to H$_{2}$, and the $^{13}$CO appears toward the center.
In this \juan{ideal} cloud, it is expected that the H{\sc i} and $^{13}$CO emission match at exactly the same velocities, but that is not necessarily the case for a real MC, where the density and velocity structures are much more complex, the spectra of both tracers are affected by optical depth and self-absorption, and the simple inspection of the emission lines may not be sufficient to assess the association between the atomic and the molecular gas.
Yet, there is important information about the dynamics of the MC formation process encoded in the relation between the extended emission from both tracers.

To systematically study the density and velocity information in extended spectral line observations and characterize the imprint of MC formation scenarios in numerical simulations, we introduce the histogram of oriented gradients (HOG), a technique developed for machine vision that we employ to study the spatial correlation between different tracers of the ISM.  
In a nutshell, HOG takes as input extended spectral line observations from two ISM tracers and provides an estimate of their spatial correlation across velocity channels.
We use HOG to study three aspects of the correlation between atomic and molecular gas.
First, we evaluate if there is a spatial correlation between the two tracers, which would indicate the relation between the MC and its associated atomic gas. 
Second, we evaluate the distribution of such a spatial correlation across velocity channels, which can reveal details about the kinematics of both gas phases.
Third, we compare the spatial correlation and its distribution across velocity channels in different regions and evaluate if they are similar to the synthetic observations of one of the multiple MC formation scenarios.

\begin{figure}[ht!]
\centerline{
\includegraphics[width=0.5\textwidth,angle=0,origin=c]{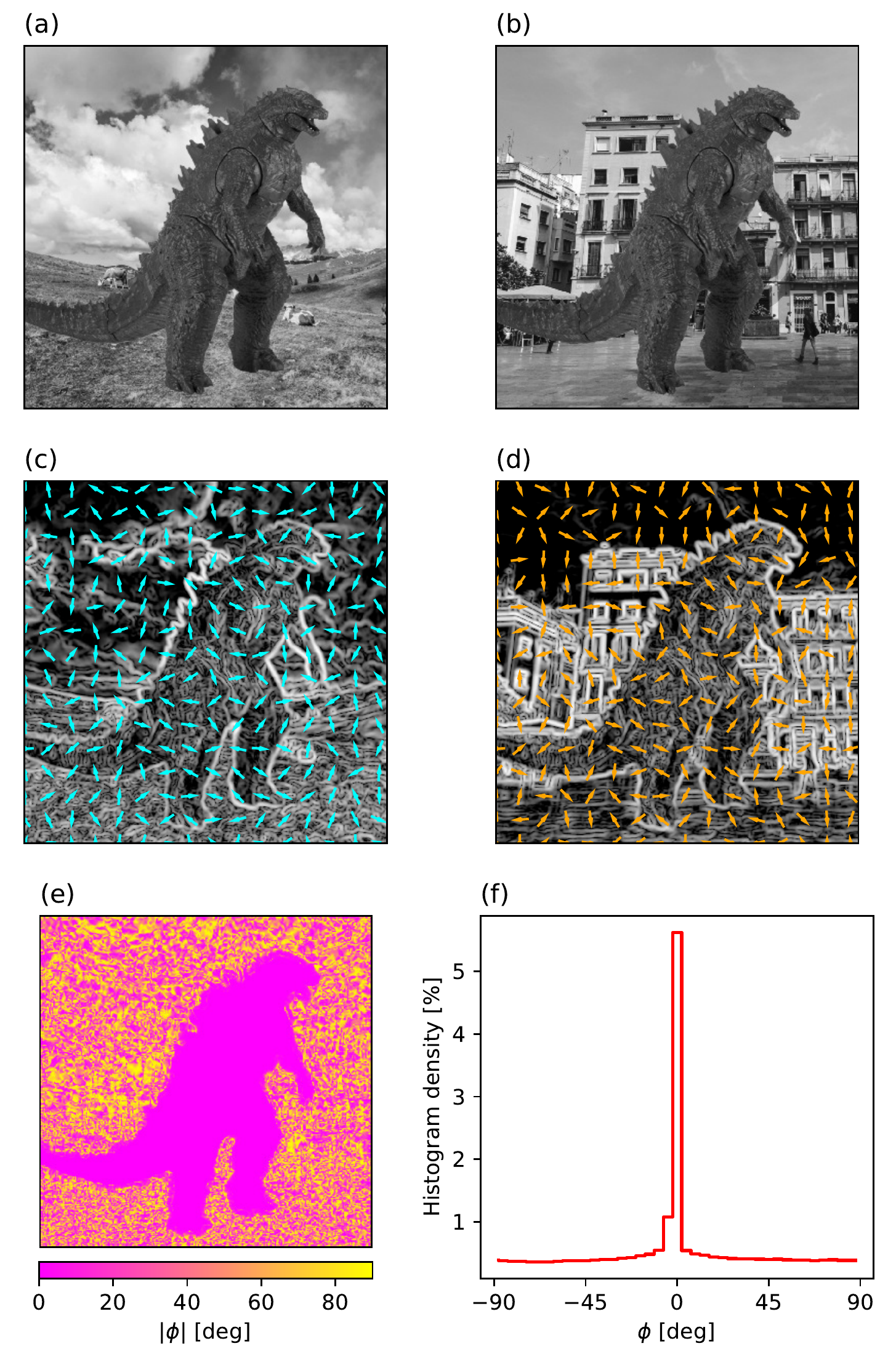}
}
\vspace{-0.2cm}
\caption{
Illustration of the histogram of oriented gradients (HOG) method.
A pair of images, (\emph{a}) and (\emph{b}), are characterized by the norm and the orientation of their gradients, (\emph{c}) and (\emph{d}).
The relative orientation angles between their gradients, (\emph{e}), are summarized in the histogram of oriented gradients, (\emph{f}).  
The number of histogram counts around 0\deg\ corresponds to the coincidence of gradient orientations in both images, which is quantified by using the mean resultant vector, Eq.~\eqref{eq:mymrv}, and the projected Rayleigh statistic, Eq.~\eqref{eq:myprs}.
Two completely uncorrelated images would produce a flat histogram while two identical images would produce a histogram having the form of a Dirac delta function centered at 0\deg.
}
\label{fig:HOGexample}
\end{figure}

In this work, we characterize HOG using a set of synthetic H{\sc i} and $^{13}$CO($J$\,$=$\,1\,$\rightarrow$\,0) emission observations obtained from the numerical simulation of magnetohydrodynamic (MHD) turbulence and MC formation in the collision of two atomic clouds presented in \cite{clarkInPrep}.
Then, we apply HOGs to the observations of the 21-cm H{\sc i} emission, from the H{\sc i}/OH/Recombination line survey of the inner Milky Way \citep[THOR,][]{beuther2016} and the $^{13}$CO($J$\,$=$\,1\,$\rightarrow$\,0) emission, from the Galactic Ring Survey \citep[GRS,][]{jackson2006}, toward a selected portion of the Galactic plane.
Finally, we detail the results of HOG toward some of the MC candidates identified in the GRS observations presented in \cite{rathborne2009}.
\juan{All of the routines used for the HOG analysis presented in this paper, including the example presented in Fig.~\ref{fig:HOGexample} and other illustrative cases, are publicly available at \url{http://github.com/solerjuan/astrohog}}.

This paper is organized as follows.
Section~\ref{section:method} describes our implementation of the HOG technique.
Section~\ref{section:mhd} presents the characterization of HOG using the colliding flow simulations.
Section~\ref{section:observations} introduces the H{\sc i} and $^{13}$CO($J$\,$=$\,1\,$\rightarrow$\,0) observations used for this study.
We report the results of the HOG analysis of the observations in Sec.~\ref{section:results}.
We discuss the origin of the spatial correlations and the MC characteristics revealed by HOG in Sec.~\ref{section:discussion}.
Finally, Sec.~\ref{section:conclusions} presents our main conclusions and the future prospects of this approach.
We reserve the technical details of the HOG technique for a set of appendices.
Appendix~\ref{app:HOG} describes details of the HOG method, such as the calculation of the gradient and the circular statistics used to evaluate the HOG results.
Appendix~\ref{app:HOGstats} presents a series of tests of the statistical significance of the HOG method.
Finally, Appendix~\ref{app:MHDsims} presents further analysis of the synthetic observations of MHD simulations.

\section{The histogram of oriented gradients}\label{section:method}

The histogram of oriented gradients (HOG) is a feature descriptor used in machine vision and image processing for object detection and image classification processes \citep{ml:mcconnell1986,ml:dalal2006}. 
A feature descriptor is a representation of an image or an image patch that simplifies the image by extracting one or more characteristics.
In the case of HOG, the method is based on the assumption that the local appearance and shape of an object in an image can be well characterized by the distribution of local intensity gradients or edge directions, which are by definition perpendicular to the direction of the gradient.
The HOG method is widely applied in the detection of objects in a variety of applications such as recognition of hand gestures \citep{ml:freeman1994}, detection of humans \citep{ml:zhu2006}, and use of sketches for searching and indexing digital image libraries \citep{ml:hu2010}.

One of the simplest applications of the HOG method is quantifying the spatial correlation between two images. 
The HOG is a representation of the occurrences of the relative orientations between local gradient orientations in the two images, thus it is a representation of how the edges in the images match each other.
Given that we are interested in evaluating the correlation between observations of astronomical objects through different tracers, we do not need to match the scales of the images or assume a prior on the shape of the objects that we are investigating.

Although the maps of extended atomic and molecular emission are not dominated by sharp edges, the HOG systematically characterizes and correlates the intensity contours that human vision recognizes as their main features, such as clumps or filaments.
We do not assume any physical interpretation for the origin of the velocity-channel map gradients, as it is the case in other gradient methods, such as those presented in the family of papers represented by \cite{lazarian2018}.
We use the velocity-channel map gradients to compare systematically the intensity contours that might be common to two ISM tracers.

An application of HOG has been previously introduced in astronomical research in the study of the correlation between the column density structures and the magnetic field orientation in both synthetic observations of simulations of MHD turbulence and \planck\ polarization observations \citep{soler2013,planck2015-XXXV}.
\juan{Other potential applications of the HOG technique in astronomy include, for example, characterizing the directionality of structures in an astronomical image, evaluating the morphological changes across velocity channels in a single PPV cube, and, in general, quantifying the spatial correlation between two or more ISM tracers.}

\subsection{Using HOG to quantify correlations between position-position-velocity cubes.}

We use the HOG method to quantify the spatial correlation between maps of H{\sc i} and $^{13}$CO emission across radial velocities, better known in astronomy as position-position-velocity (PPV) cubes.
Explicitly, we calculate the correlation between the two PPV cubes by following the steps described below.

\subsubsection{Computation of the HOG}

We align and re-project a pair of PPV cubes, $I^{\rm A}_{ij,l}$ and $I^{\rm B}_{ij,m}$, into a common spatial grid by using the {\tt reproject} routine included in the {\tt Astropy} package \citep{AstropyCollaboration2013}.
Throughout this paper, the indexes $i$ and $j$ correspond to the spatial coordinates, Galactic longitude and latitude, and the indexes $l$ and $m$ correspond to the velocity channels in the respective PPV cube.
Given that we are comparing the spatial gradients of each velocity channel map, the HOG technique does not require the same velocity resolution in the PPV cubes.
For a pair of velocity-channel maps $I^{\rm A}_{ij,l}$ and $I^{\rm B}_{ij,m}$, we calculate the relative orientation angle $\phi$ between intensity gradients by evaluating
\begin{equation}\label{eq:phi}
\phi_{ij,lm}=\arctan\left(\frac{(\nabla I^{\rm A}_{ij,l} \times \nabla I^{\rm B}_{ij,m})\cdot\hat{z}}{\nabla I^{\rm A}_{ij,l} \cdot \nabla I^{\rm B}_{ij,m}}\right),
\end{equation}
where the differential operator $\nabla$ corresponds to the gradient.
The term $(\nabla I^{\rm A}_{ij,l} \times \nabla I^{\rm B}_{ij,m})\cdot \hat{z} \equiv |\nabla I^{\rm A}_{ij,l}|\,|\nabla I^{\rm B}_{ij,m}|\sin\phi_{ij,lm}$ is the $z$-axis projection of the cross product.
The term $\nabla I^{\rm A}_{ij,l} \cdot \nabla I^{\rm B}_{ij,m}\equiv |\nabla I^{\rm A}_{ij,l}|\,|\nabla I^{\rm B}_{ij,m}|\cos\phi_{ij,lm}$ is the scalar product of vectors, or dot product.
We choose the representation in Eq.~\ref{eq:phi} because it is numerically better-behaved than the expression that would be obtained by using just the dot product and the $\arccos$ function.
Equation~\ref{eq:phi} implies that the relative orientation angles are in the range $[-\pi/2,\pi/2)$, thus accounting for the orientation of the gradients and not their direction.
The value of $\phi$ is only meaningful in regions when both $\nabla I^{\rm A}_{ij,l}$ and $\nabla I^{\rm B}_{ij,m}$ are significant, that is, their norm is greater than zero or above thresholds that are estimated according to the noise properties of the each PPV cube. 

We compute the gradients using Gaussian derivatives, explicitly, by applying the multidimensional Gaussian filter routines in the {\tt filters} package of {\tt Scipy}. 
The Gaussian derivatives are the result of the convolution of the image with the spatial derivative of a two-dimensional Gaussian function.
The width of the Gaussian determines the area of the vicinity over which the gradient is calculated.
Varying the width of the Gaussian kernel enables the sampling of different scales and reduces the effect of noise in the pixels \citep[see][and references therein]{soler2013}.

For the sake of clarity, we illustrate the aforementioned procedure in a pair of mock velocity-channel maps presented in Fig.~\ref{fig:HOGexample}.
We there present the two velocity-channel maps, panels (a) and (b); their corresponding gradients, panels (c) and (d); the relative orientation angles, $\phi$, panel (e); and the histograms of oriented gradients, panel (f), which we evaluate by using the tools of circular statistics presented in the next section.

\subsubsection{Evaluation of the correlation}

Once we calculate the relative orientation angles $\phi_{ij,lm}$ for a pair of channels centered on velocities $v^{\rm A}_{l}$ and $v^{\rm B}_{m}$, we summarize the spatial correlation contained in these angles by marginalizing over the spatial coordinates, indexes $i$ and $j$.
For that purpose, we use two tools from circular statistics: the mean resultant vector ($r$) and the projected Rayleigh statistic (\prs), both described in detail in Appendix~\ref{app:StatSignificance}.

In our application, we use the definition of the mean resultant vector 
\begin{equation}\label{eq:mymrv}
r_{lm}=\frac{\left(\left[\sum_{ij}w_{ij,lm}\cos(2\phi_{ij,lm})\right]^{2}+\left[\sum_{ij}w_{ij,lm}\sin(2\phi_{ij,lm})\right]^{2}\right)^{1/2}}{\sum_{ij}w_{ij,lm}},
\end{equation}
where the indexes $i$ and $j$ run over the pixel locations in the two spatial dimensions and $w_{ij,lm}$ is the statistical weight of each angle $\phi_{ij,lm}$.
We account for the spatial correlations introduced by the telescope beam by choosing $w_{ij,lm}$\,$=$\,$(\delta x/\Delta)^{2}$, where $\delta x$ is the pixels size and $\Delta$ is the diameter of the derivative kernel that we use to calculate the gradients.
For pixels where the norm of the gradient is negligible or can be confused with the signal produced by noise, we choose $w_{ij,lm}$\,$=$\,0 (see Appendix~\ref{app:HOG} for a description of the gradient selection).

The mean resultant vector, $r$, is a descriptive quantity that can be interpreted as the percentage of vectors pointing in a preferential direction.
However, it does not provide any information on the shape of the angle distribution.
The optimal statistic to test if the distribution of angles is non-uniform and peaked at 0\deg\ is the projected Rayleigh statistic
\begin{equation}\label{eq:myprs}
V_{lm} = \frac{\sum_{ij}w_{ij,lm}\cos(2\phi_{ij,lm})}{\sqrt{\sum_{ij}[(w_{ij,lm})^{2}/2]}},
\end{equation}
which follows the same conventions introduced in Eq.~\eqref{eq:mymrv}.
Each value $V_{lm}$ represents the likelihood test against a von Mises distribution, which is the circular normal distribution centered on 0\deg, or in other words, the likelihood that the gradients of the emission maps $I^{\rm A}_{ij,l}$ and $I^{\rm B}_{ij,m}$ are mostly parallel.
The ensemble of $V_{lm}$ values, which we denominate the correlation plane, represents the correlation between the emission maps centered on velocities $v^{\rm A}_{l}$ and $v^{\rm B}_{m}$.
For the sake of simplicity, we designate the HOG correlation between tracers A and B as $V(v^{\rm A},v^{\rm B})$, but this is just an approximation given that we can only estimate the discrete values of $V_{lm}$, which depend on the spectral resolution of the observations and the width of the velocity channels.

We present the results of our analysis in terms of both \prs\ and \mrv.
The values of the latter are only meaningful for our purposes when they are validated by \prs; because large values of the mean resultant vector only indicate a preferential orientation, not necessarily $\phi$\,$=$\,0\deg.
We note that the gradient vectors in each individual velocity-channel map are not statistically independent, that is, even if the observations were made with infinite angular resolution, the physical phenomena governing the ISM; that is, gravity, turbulence, and the magnetic fields; impose correlations across multiple spatial scales.
And so it is not possible to draw conclusions from the values of \prs\ alone, but its statistical significance should be assessed by comparing its value to the values obtained in maps with similar statistical properties.

Given the difficulties in reproducing the statistical properties of each velocity-channel map, we use the mean value,
\begin{equation}
\left<V\right> \equiv \frac{\sum^{[l_{\rm min},l_{\rm max}]}_{l} \sum^{[m_{\rm min},m_{\rm max}]}_{m} V_{lm}}{(l_{\rm max}-l_{\rm min})(m_{\rm max}-m_{\rm min})}, 
\end{equation}
and the population variance,
\begin{equation}\label{eq:sigmav}
\varsigma^{2}_{V} \equiv \frac{\sum^{[l_{\rm min},l_{\rm max}]}_{l} \sum^{[m_{\rm min},m_{\rm max}]}_{m}\left(V_{lm} - \left<V\right>\right)^{2}}{(l_{\rm max}-l_{\rm min})(m_{\rm max}-m_{\rm min})}, 
\end{equation}
in the velocity ranges defined by the indexes $[l_{\rm min},l_{\rm max}]$ and $[m_{\rm min},m_{\rm max}]$ to assess the statistical significance of \prs.
If we assume that most of the channel maps in a particular velocity range are uncorrelated, each of them would correspond to an independent realization of a scalar field with a spatial correlation given by the properties of the ISM and the angular resolution of the observations, $\varsigma_{V}$ would represent the chance correlation between those maps. 
Evidently that is not the case in reality, unless we consider channels separated by tens of \kps\ in a Galactic target, but still $\varsigma_{V}$ characterizes the $V$ population variance within the selected range of velocities.
There is of course a variance of $V$ for each particular pair of velocity-channel maps, $(\sigma_{V})_{lm}$, but it is in most cases smaller than $\varsigma_{V}$, as shown in App.~\ref{app:PRS}.

In this work, we report the values of $V_{lm}$ always in relation with the corresponding $\varsigma_{V}$, as inferred from Eq.~\eqref{eq:sigmav}, in a particular velocity range.
An alternative method for evaluating the statistical significance of $V$ is based on estimating the population variance using velocity-channel maps that are uncorrelated by construction, for example, two PPV cubes that are not coincident in the sky or one PPV cube flipped with respect to the other in one of the spatial coordinates.
This method is crucial for determining the validity of our method since the values of $V$ in the cases mentioned above should be exclusively dominated by chance correlation, as we show in App.~\ref{app:HOGstatsJacknives}.
But the direct estimation of $\varsigma_{V}$ using these null-tests is computationally demanding and does not lead to significant differences with respect to the values obtained with Eq.~\ref{eq:sigmav}. 

\section{HOG analysis of MHD simulations}\label{section:mhd}

We characterize HOG by analyzing a set of synthetic observations of H{\sc i} and $^{13}$CO emission from the numerical simulations of MC formation in a colliding flow presented in \cite{clarkInPrep}.
These simulations include a simplified treatment of the chemical and thermal evolution of the interstellar medium (ISM), which makes them well suited for obtaining synthetic observations of both tracers. 
Although the numerical setup and the chemistry treatment are not indisputable \citep[see for example,][]{levrier2012}, we use this simplified physical scenario to gain insight into the behavior of the HOG technique before we apply it to the observations. 

\begin{figure}[ht!]
\centerline{
\includegraphics[width=0.5\textwidth,angle=0,origin=c]{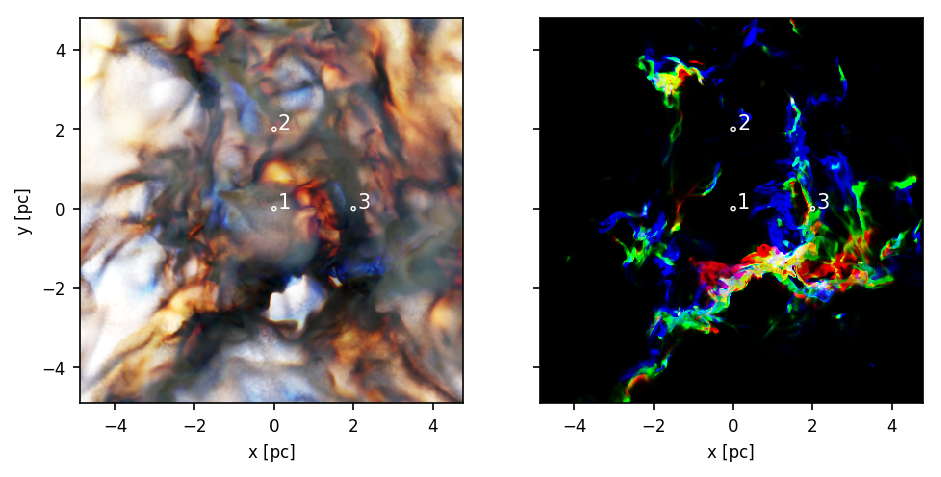}
}
\vspace{-0.1cm}
\centerline{
\includegraphics[width=0.5\textwidth,angle=0,origin=c]{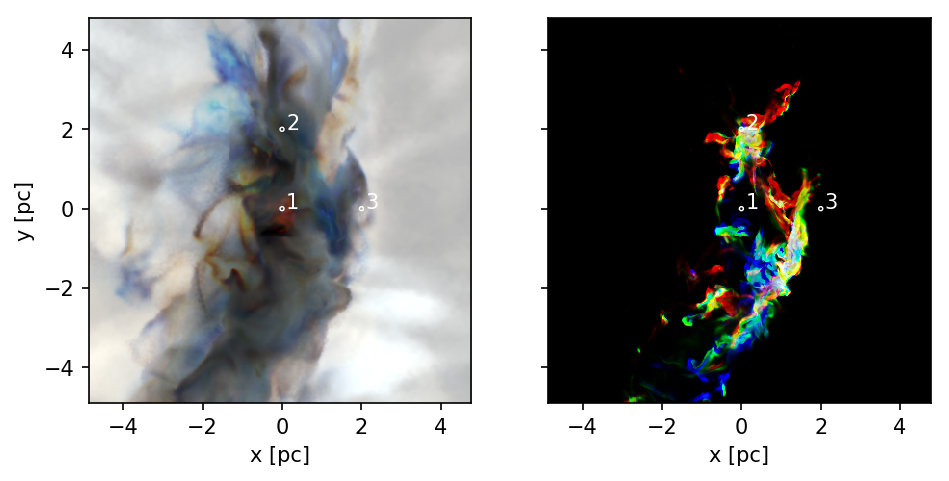}
}
\vspace{-0.1cm}
\caption{
Synthetic observations of H{\sc i} with a 100\,K background (left) and $^{13}$CO (right) emission from the MHD turbulence simulation of two colliding atomic clouds presented in \cite{clarkInPrep}.
In both maps the colors represent the integrated intensities in three groups of velocity channels: red for $-$0.6\,$\leq$\,\vlsr\,$\leq$\,$-$0.2\,\kps, green for $-$0.2\,$\leq$\,\vlsr\,$\leq$\,0.2\,\kps, and blue for 0.2\,$\leq$\,\vlsr\,$\leq$\,0.6\,\kps.
The numbers correspond to the positions of the spectra presented in Fig.~\ref{fig:spectraPCsims}.
\juan{The top and bottom panels correspond to the synthetic observations made with the line of sight parallel (face-on) and perpendicular to the collision axis (edge-on), respectively}.
}\label{fig:imagesPCsims}
\end{figure}

\subsection{Initial conditions}

The simulations considered were carried out using the {\tt AREPO} moving mesh code \citep{springel2010}.
They represent two 38-pc-diameter atomic clouds with an initial particle density $n^{\rm C}_{0}$\,$=$\,$10$\,cm$^{-3}$ that collide head-on along the $x$-axis of the simulation domain at 7.5\,\kps\ with respect to each other.
The clouds are given a turbulent velocity field with a 1 \kps\ amplitude and a $P(k)\propto k^{-4}$ scaling law.
The simulation includes a uniform initial magnetic field $B_{0}$\,$=$\,3\,$\mu$G oriented along the $x$-axis, that is, parallel to the collision axis.

The clouds are initially set one cloud radius apart (19\,pc) in a cubic computational domain of side 190\,pc and initial number density $n^{\rm S}_{0}$\,$=$\,$0.1$\,cm$^{-3}$.
The boundaries of the box are periodic, but self-gravity is not periodic.
The initial cell mass is approximately 5\,$\times$\,10$^{-3}$\,M$_{\odot}$, both in the clouds and in the low-density surrounding medium.
The cell refinement is set such that the thermal Jeans length is resolved by at least 16 {\tt AREPO} cells at all times.

The simulations follow the thermal evolution of the gas using a cooling function based on \citet{glover2010} and \citet{glover2012a}. 
The chemical evolution of the gas is modelled using a simplified H-C-O network based on \citet{glover2007} and \citet{nelson1999}, updated as described in \citet{glover2015}.
The effects of H$_{2}$ self-shielding and dust shielding are accounted for using the {\sc TreeCol} algorithm \citep{clark2012}.

The metallicity of the gas is taken to be solar with elemental abundances of oxygen and carbon set to $x_{\rm O}$\,$=$\,3.2\,$\times$\,$10^{-4}$ and $x_{\rm C}$\,$=$\,1.4\,$\times$\,$10^{-4}$ \citep{sembach2000}.
The three simulations presented in \cite{clarkInPrep} are designed to probe the effect of different interstellar radiation fields (ISRFs) and cosmic rate ionization rates (CRIRs).
For the characterization of HOG we have chosen the simulation with ISRF $G_{0}$\,=\,17 and CRIR\,$=$\,3\,$\times$\,10$^{-16}$\,s$^{-1}$.
This ISRF implies that the H$_{2}$ and the CO are found at higher column densities than in the other two simulations presented in \cite{clarkInPrep}, but it does not imply any loss of generality in our results.

\begin{figure}[ht!]
\centerline{
\includegraphics[width=0.5\textwidth,angle=0,origin=c]{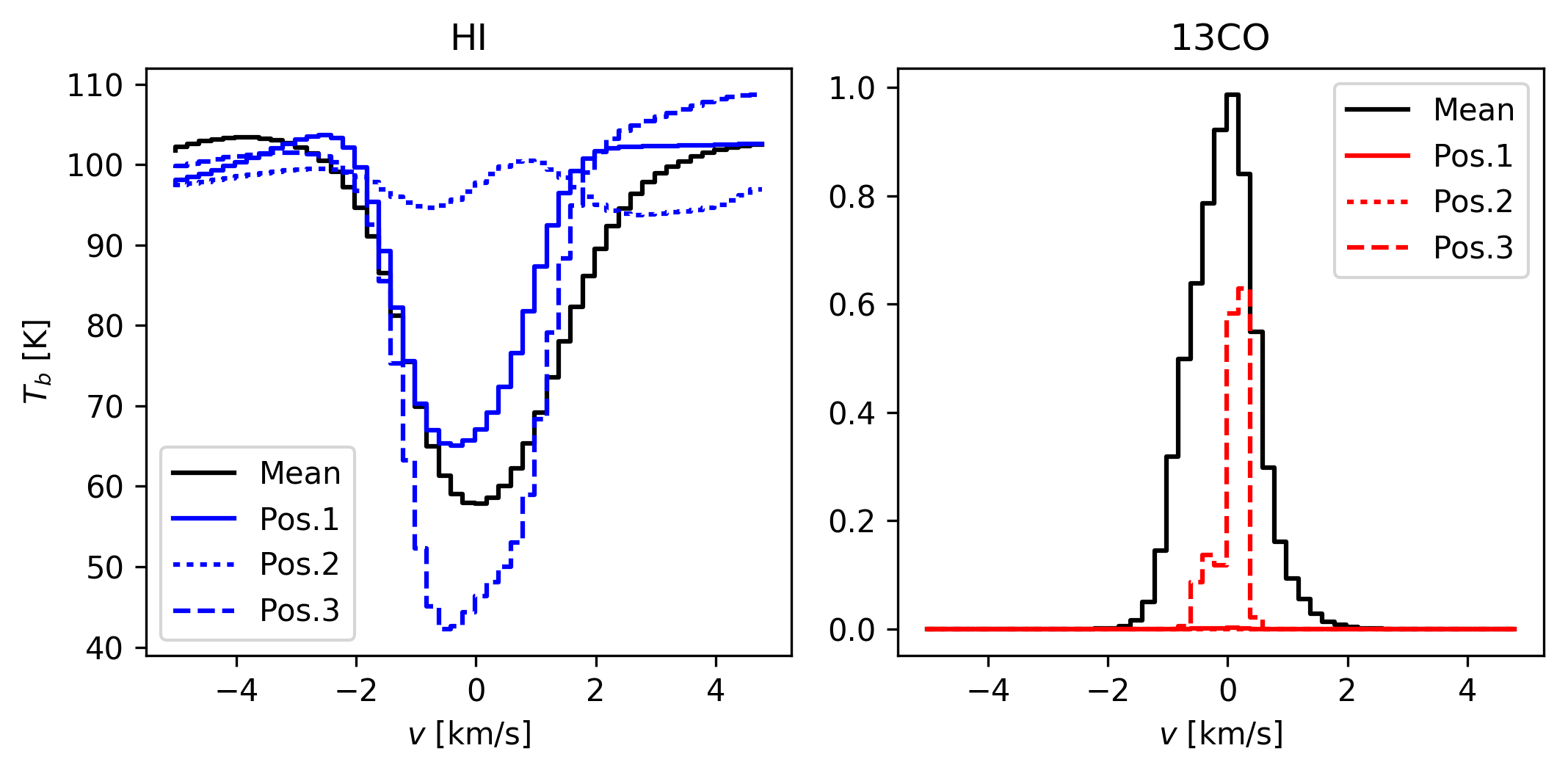}
}
\centerline{
\includegraphics[width=0.5\textwidth,angle=0,origin=c]{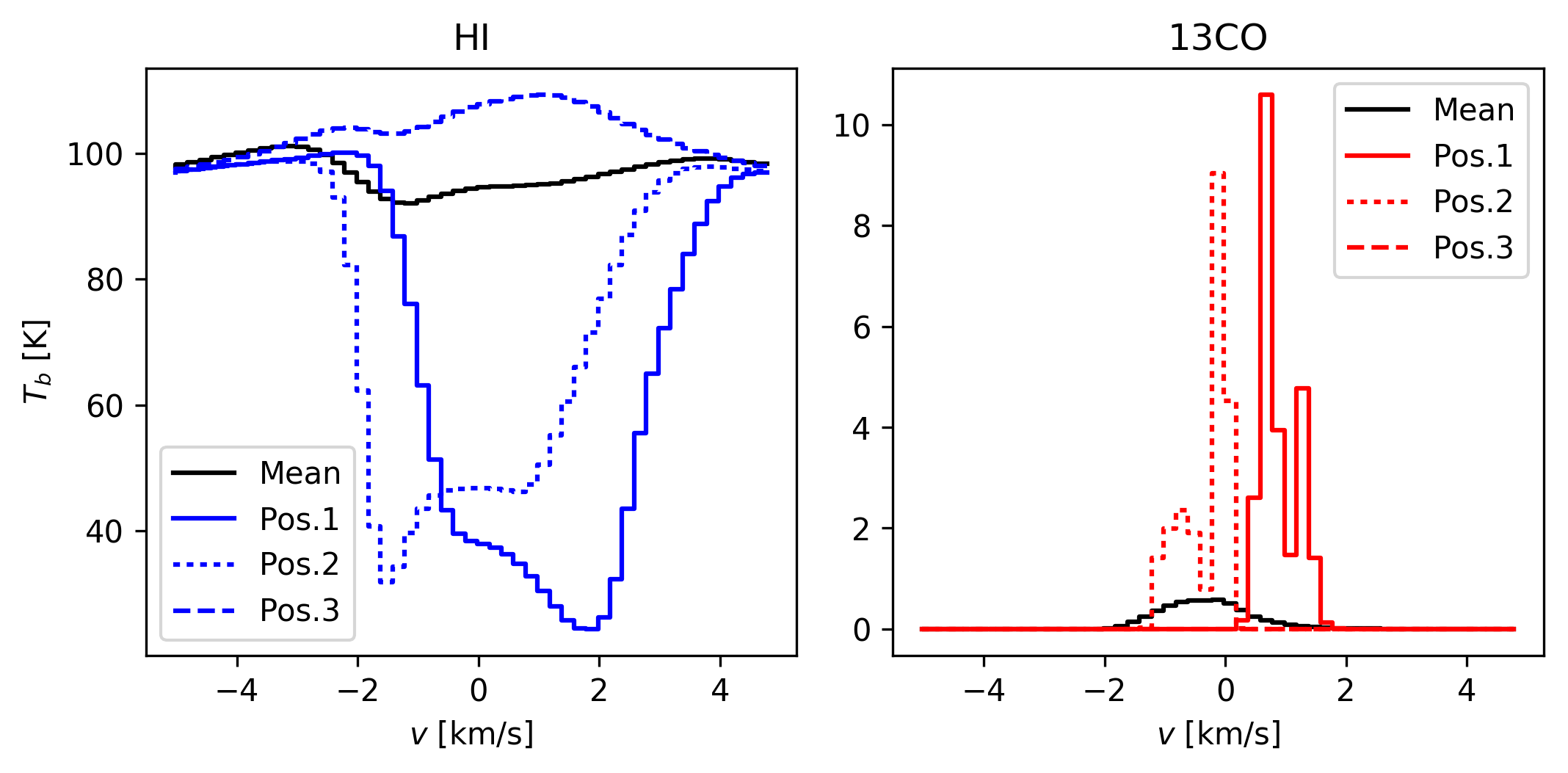}
}
\caption{
Spectra from the synthetic observations of H{\sc i} with a 100\,K background (left) and $^{13}$CO (right) emission presented in Fig.~\ref{fig:imagesPCsims}.
The black lines correspond to the average spectra over the whole map.
The solid, dashed and segmented colored lines correspond to the spectra toward the positions indicated in Fig.~\ref{fig:imagesPCsims}.  
The top and bottom panels correspond to the \juan{face-on and edge-on synthetic observations}, respectively.
}\label{fig:spectraPCsims}
\end{figure}

\begin{figure}[ht!]
\centerline{
\includegraphics[width=0.5\textwidth,angle=0,origin=c]{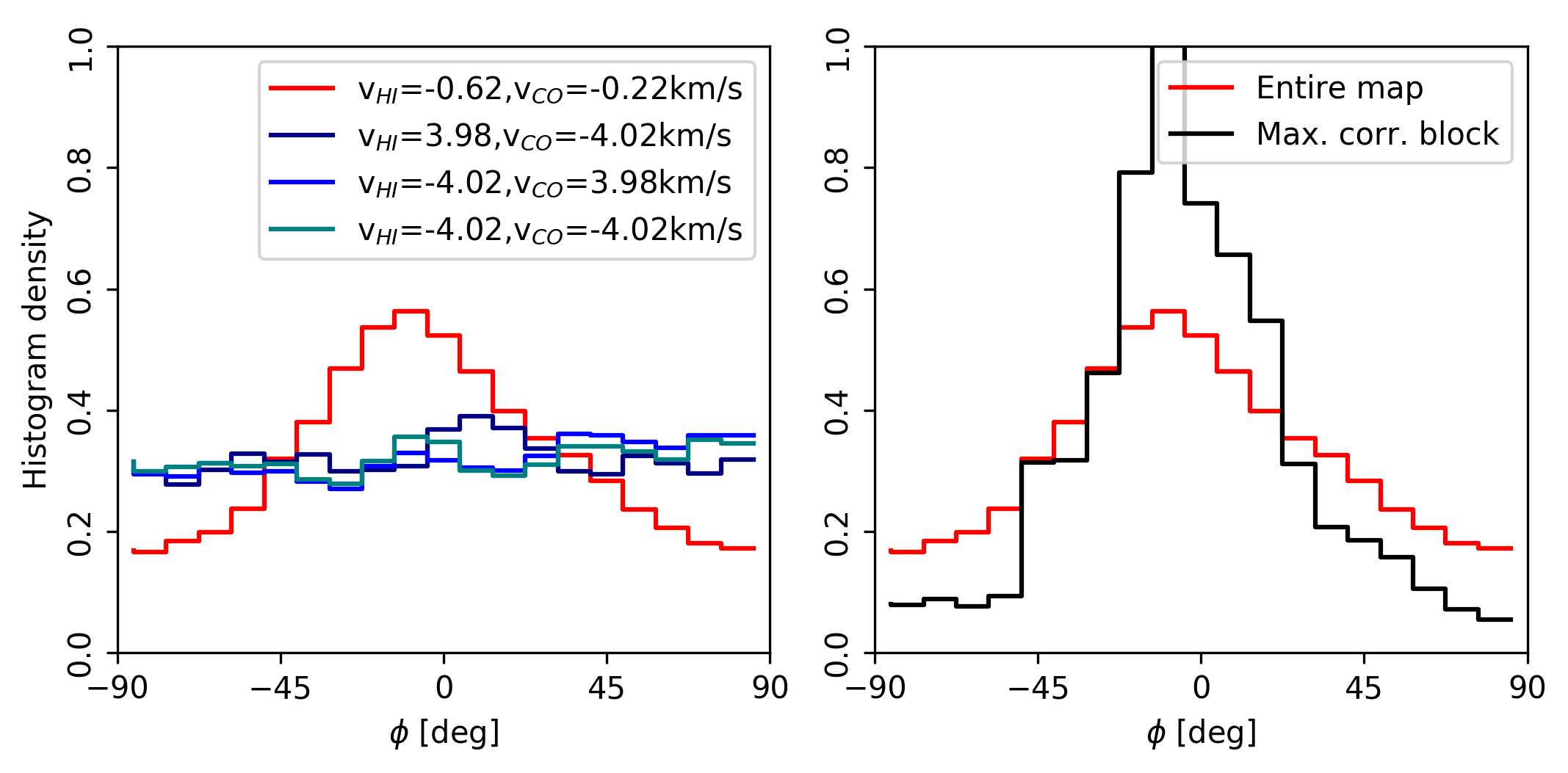}
}
\centerline{
\includegraphics[width=0.5\textwidth,angle=0,origin=c]{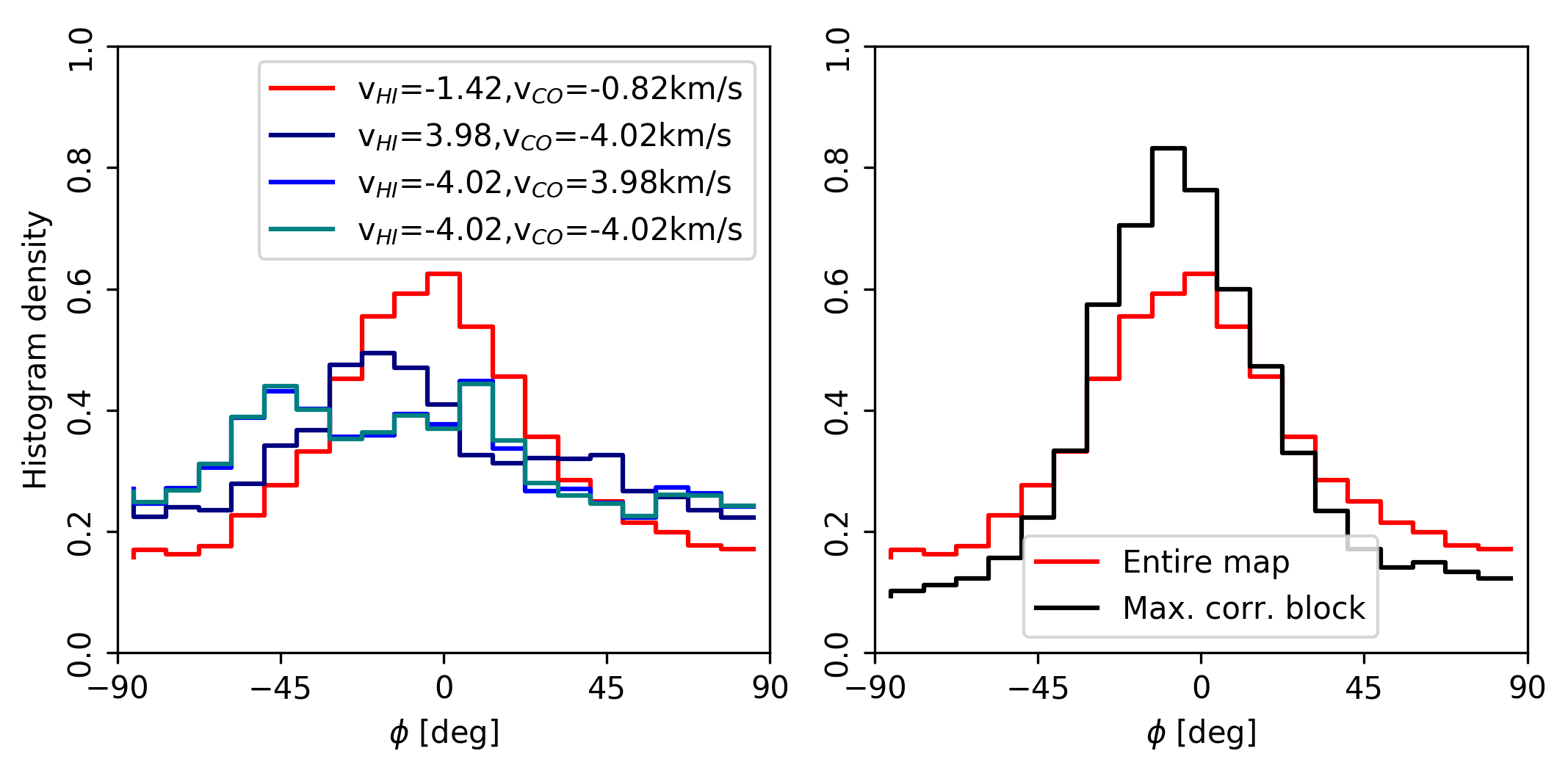}
}
\caption{
\emph{Left}. Histograms of oriented gradients (HOGs) corresponding to the pair of velocity-channel maps with the largest spatial correlation, as inferred from the \prs\ values shown in Fig.~\ref{fig:HOGcorrPCsims}, and three pairs of arbitrarily selected velocity channels in the synthetic observations presented in Fig.~\ref{fig:imagesPCsims}.
\emph{Right}. For the pair of velocity-channel maps with the largest spatial correlation, HOGs corresponding to the entire map and just the block with the largest \prs\ indicated in Fig.~\ref{fig:HOGpanelPCsims}.
The top and bottom panels correspond to the \juan{face-on and edge-on synthetic observations}, respectively.
}\label{fig:HOG_PCsims}
\end{figure}

\begin{figure*}[ht!]
\centerline{
\includegraphics[width=1.0\textwidth,angle=0,origin=c]{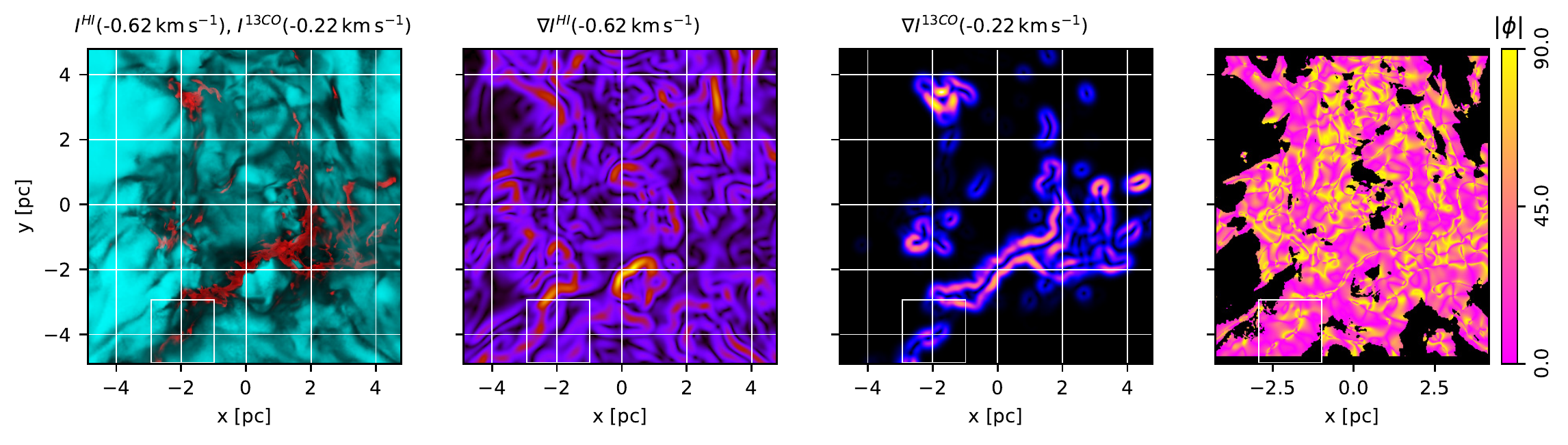}
}
\vspace{-0.1cm}
\centerline{
\includegraphics[width=1.0\textwidth,angle=0,origin=c]{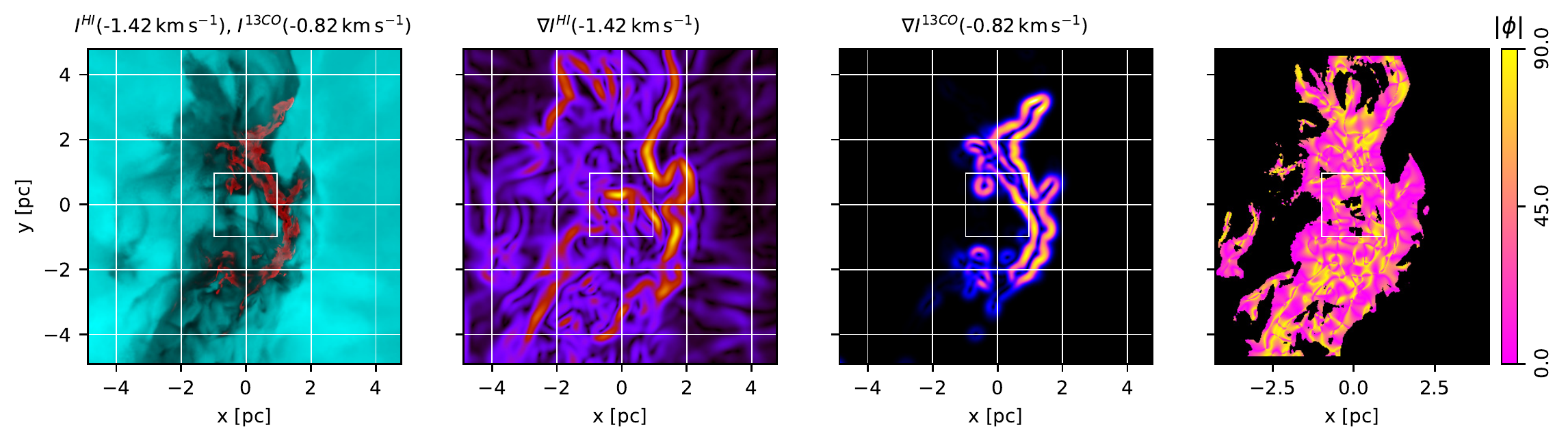}
}
\vspace{-0.2cm}
\caption{
Intensity, intensity gradients, and relative orientation angle maps from the synthetic observations presented in Fig.~\ref{fig:imagesPCsims}.
\emph{Left.} H{\sc i} (teal) and $^{13}$CO emission (red) in the velocity channels with the largest spatial correlation, as inferred from the \prs\ values shown in Fig.~\ref{fig:HOGcorrPCsims}.
\emph{Middle left.} Norm of the gradient of the H{\sc i} intensity map in the indicated velocity channel.
\emph{Middle right.} Norm of the gradient of the $^{13}$CO intensity map in the indicated velocity channel.
\emph{Right.} Relative orientation angle $\phi$, Eq.~\eqref{eq:phi}, between the gradients of the H{\sc i} and $^{13}$CO intensity maps in the indicated velocity channels.
The white color in the $\phi$ map corresponds to areas with no significant gradient in either tracer. 
The square indicates the block, selected from a 7\,$\times$\,7 spatial grid, with the largest values of \prs.
The top and bottom panels correspond to the \juan{face-on and edge-on synthetic observations}, respectively.
}\label{fig:HOGpanelPCsims}
\end{figure*}

\begin{figure}[ht!]
\centerline{
\includegraphics[width=0.25\textwidth,angle=0,origin=c]{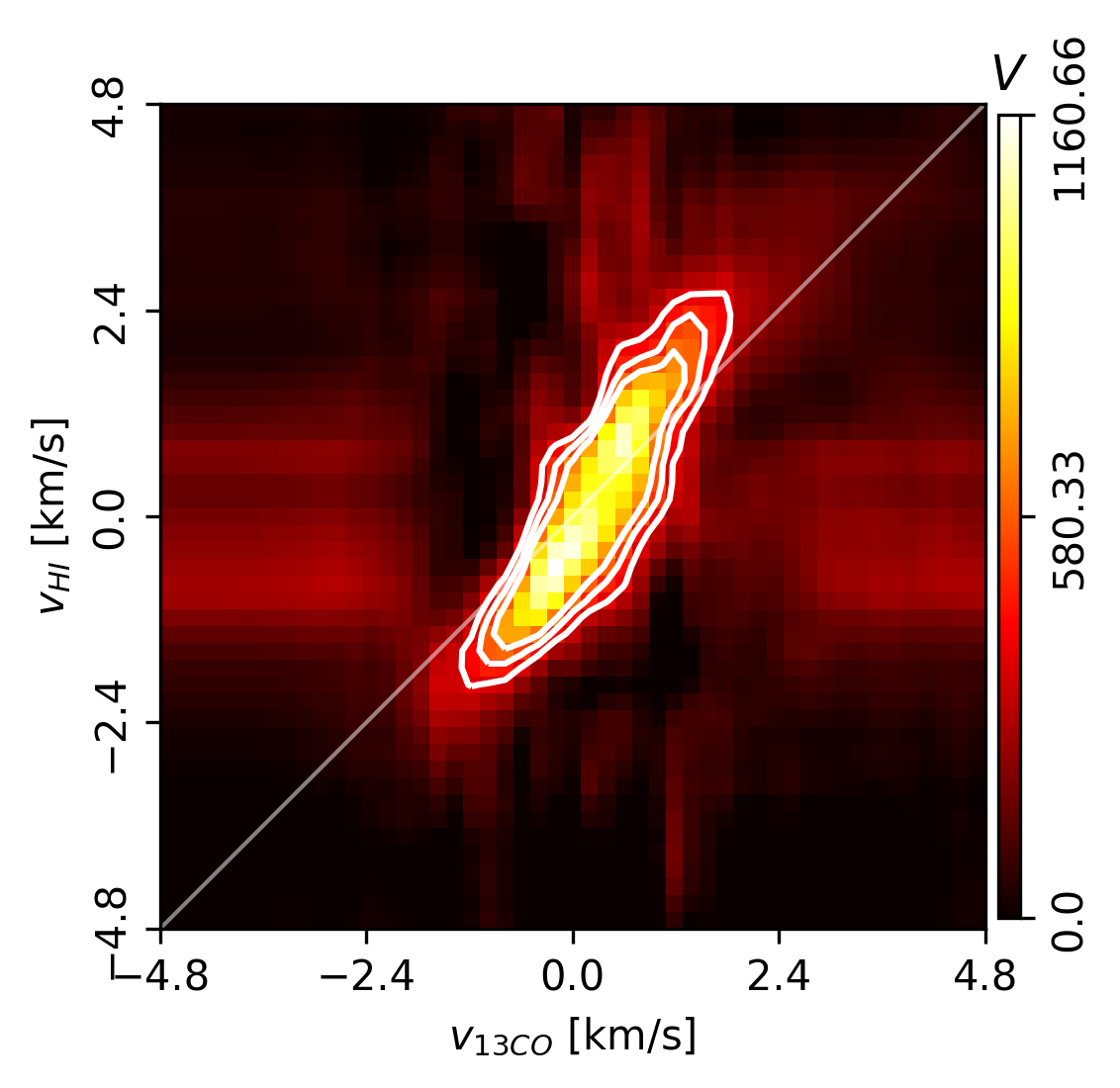}
\includegraphics[width=0.25\textwidth,angle=0,origin=c]{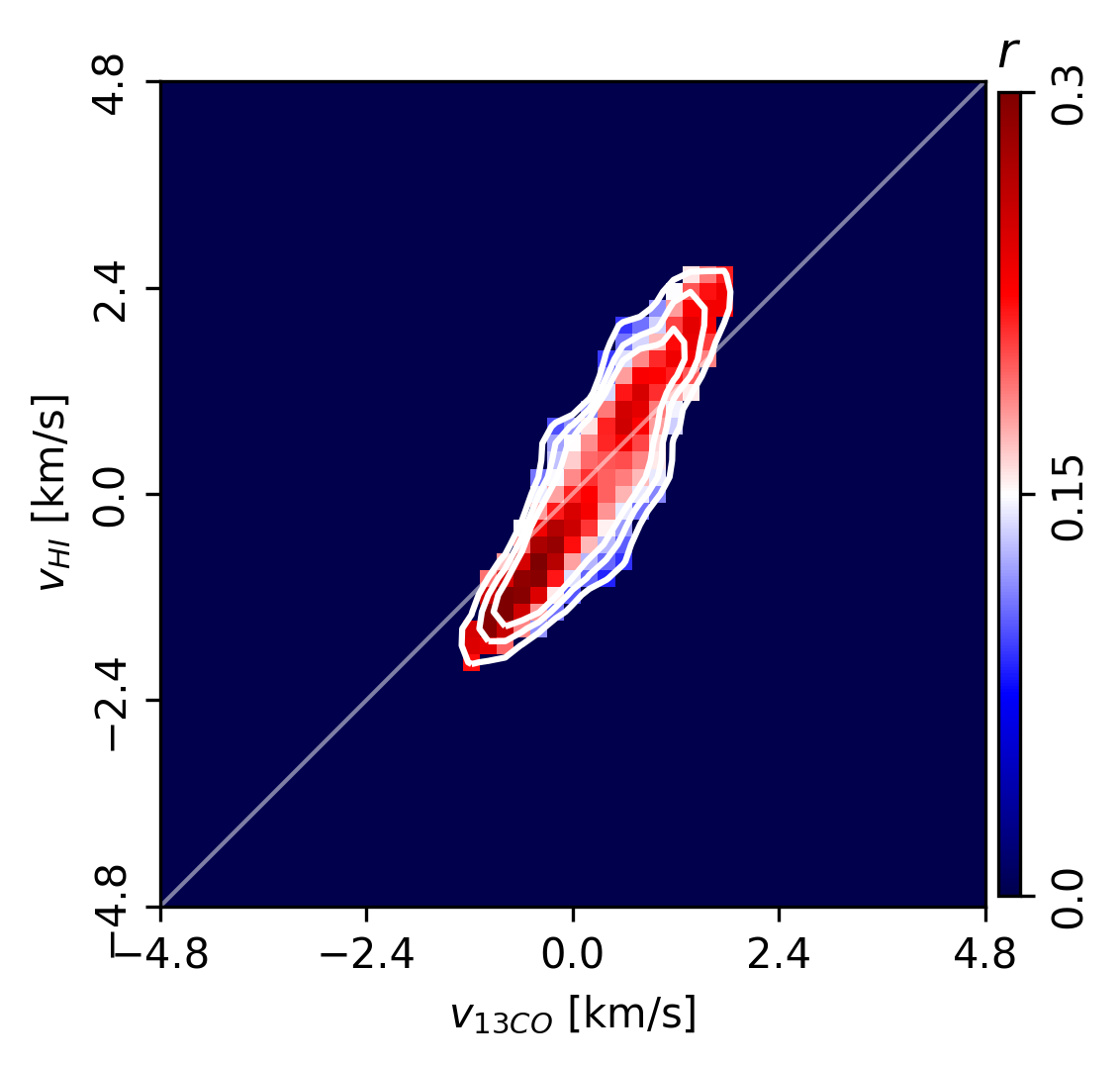}
}
\centerline{
\includegraphics[width=0.25\textwidth,angle=0,origin=c]{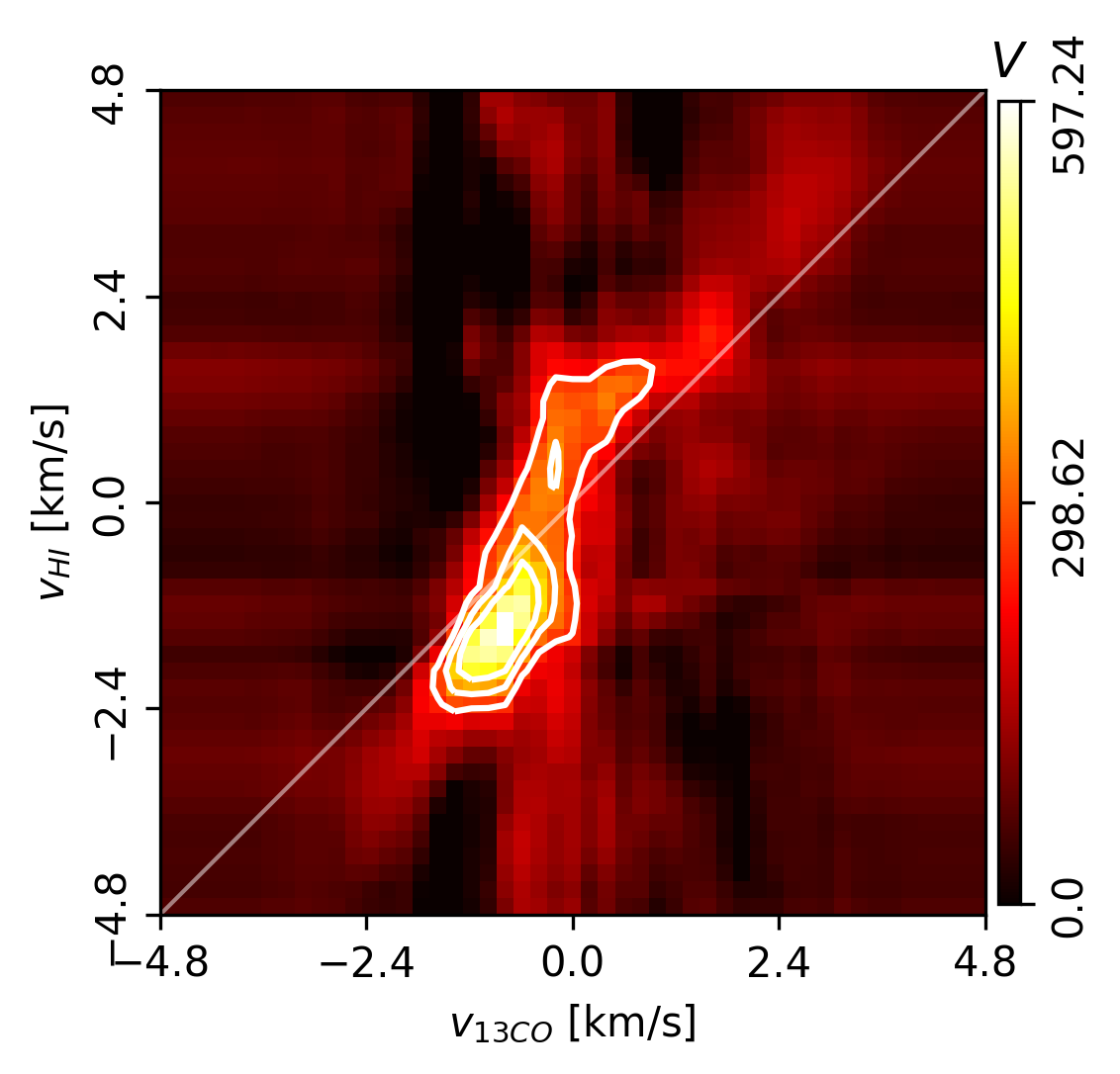}
\includegraphics[width=0.25\textwidth,angle=0,origin=c]{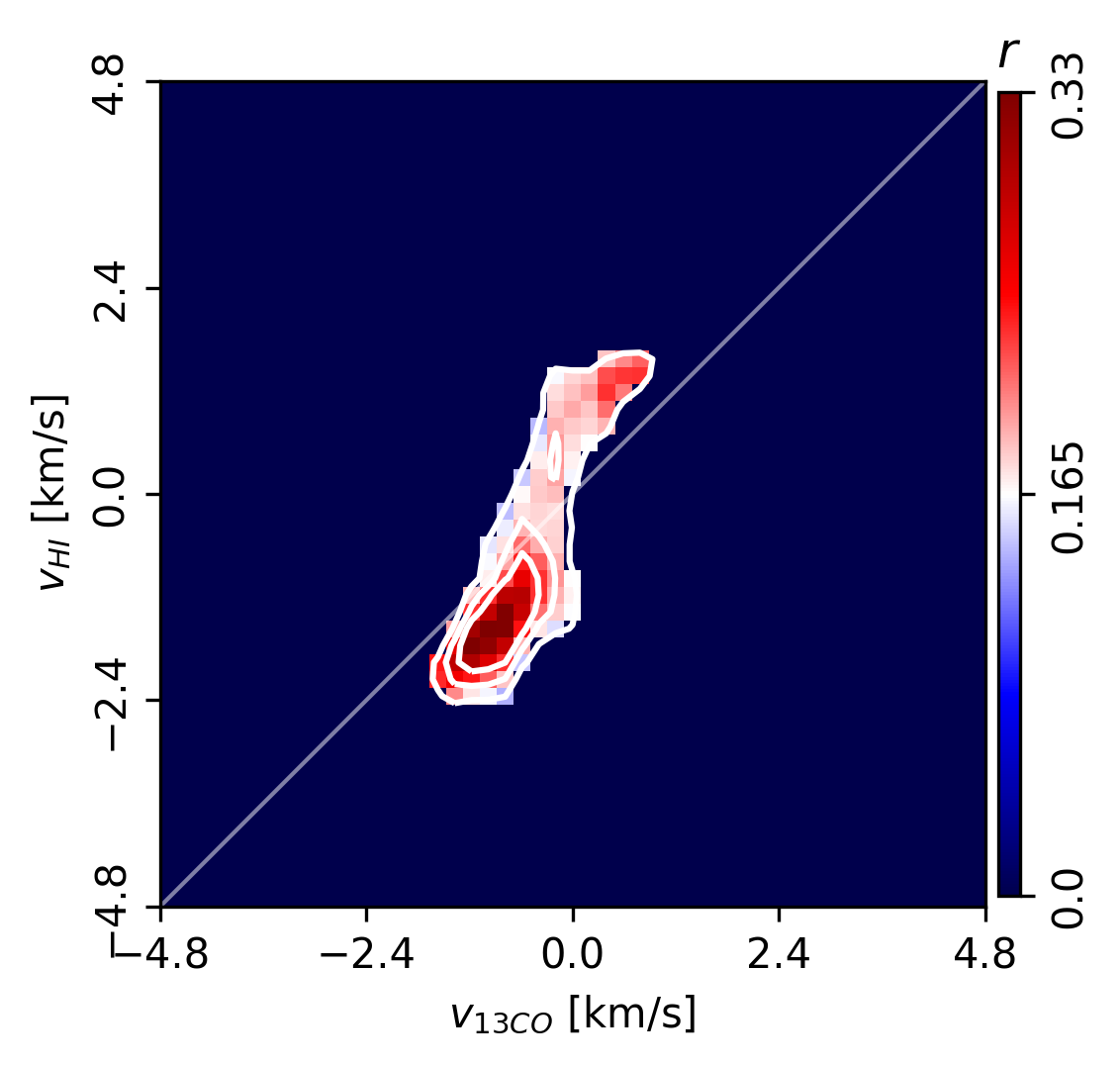}
}
\caption{
Results of the HOG analysis of the H{\sc i} and $^{13}$CO synthetic observations presented in Fig.~\ref{fig:imagesPCsims}.
\emph{Left.} Projected Rayleigh statistic, $V(v_{\rm 13CO},v_{\rm HI})$, the HOG statistical test of spatial correlation between H{\sc i} and $^{13}$CO velocity-channel maps, defined in Eq.~\eqref{eq:myprs}.
\juan{The contours indicate the 3$\varsigma_{V}$, 4$\varsigma_{V}$, and 5$\varsigma_{V}$ levels in the corresponding velocity range.}
\emph{Right.} Mean resultant vector length, $r(v_{\rm 13CO},v_{\rm HI})$, within the 3$\varsigma_{V}$ confidence interval, a HOG metric that is roughly equivalent to the percentage of gradient pairs that imply the spatial correlation between the velocity-channel maps, defined in Eq.~\eqref{eq:mymrv}.
The top and bottom panels correspond to the \juan{face-on and edge-on synthetic observations}, respectively.
}\label{fig:HOGcorrPCsims}
\end{figure}

\subsection{Synthetic observations}

The radiative transfer (RT) post-processing of the simulations was made using the RADMC-3D code\footnote{http://www.ita.uni-heidelberg.de/$\sim$dullemond/software/radmc-3d/} following the procedures described in \cite{clarkInPrep}. 
In brief, the H{\sc i} emission is modelled assuming that the hyperfine energy levels are in local thermodynamic equilibrium (LTE), with a spin temperature $T_{\rm s}$ equal to the local kinetic temperature of the gas.
This is a good approximation for the cold, dense atomic gas that dominates the emission signal in these simulations \citep[e.g.,][]{liszt2001}. 
For the $^{13}$CO, we do not assume LTE, as some of the emission may be coming from regions with densities below the CO critical density. 
Instead, we use the \juan{large} velocity gradient (LVG) module implemented in RADMC-3D by \citet{shetty2011}. 
In addition, as the \cite{clarkInPrep} simulations do not track $^{13}$CO explicitly, it is necessary to compute the $^{13}$CO abundance based on the $^{12}$CO abundance. 
This is done using a fitting function for the $^{13}$CO/$^{12}$CO ratio as a function of the $^{12}$CO column density proposed by \citet{szucs2014}. 
This column-density-dependent conversion factor accounts for the effects of chemical fractionation and selective photodissociation of $^{13}$CO and hence is more accurate than adopting a constant $^{13}$CO/$^{12}$CO ratio.

The {\tt AREPO} results are interpolated onto a regular cartesian grid. 
The grid covers a cubic region of 9.72\,pc with 400 cells per side, corresponding to a spatial resolution of 0.024\,pc.
The synthetic spectra are initially calculated in 500 velocity channels covering the velocity range [$-$5,5]\,\kps.
We resample this original data \juan{into a velocity resolution of} $0.2$\,\kps\ to match the channel width of the GRS data.
The maps of the synthetic observations of H{\sc i} and $^{13}$CO and some selected corresponding spectra are presented in Fig.~\ref{fig:imagesPCsims} and Fig.~\ref{fig:spectraPCsims}, respectively.

It is common at low Galactic latitudes that cold foreground clouds absorb the emission from gas behind. 
This effect is often called H{\sc i} self-absorption (HISA), although it is not self-absorption in the normal radiative transfer sense, 
because the absorbing cloud may be spatially distant from the background H{\sc i} emission, but sharing a common radial velocity \citep{gibson2005a,kavars2005}.
For that reason we use synthetic observations of H{\sc i} that include a 100\,K background emission.
For the sake of completeness and discussion, we present the synthetic observations of H{\sc i} without background emission in Appendix~\ref{app:MHDsims}.

We analyze two configurations of the aforementioned simulation: one with the line of sight parallel to the collision axis (face-on) and one with line of sight perpendicular to the collision axis (edge-on).
Fig.~\ref{fig:imagesPCsims} shows the clear differences between the two configurations.
In the face-on configuration, the H{\sc i} is distributed over the whole map in filamentary structures that appear dark against the bright background while the $^{13}$CO appears more concentrated, but also filamentary in appearance.
In the edge-on configuration, the H{\sc i} appears concentrated in the shocked layer, which is clearly visible against the bright background, and the $^{13}$CO is distributed in a couple of filamentary structures.

The spectra of the face-on and the edge-on synthetic observations, shown in Fig.~\ref{fig:spectraPCsims}, reveal two clear differences between these configurations.
The face-on configuration presents a broad H{\sc i} mean spectrum in absorption against the 100\,K background and clearly centered at \vlos\,$\approx$\,0\,\kps.
The $^{13}$CO is also clearly centered at \vlos\,$\approx$\,0\,\kps.
The edge-on configuration presents a flat H{\sc i} mean spectrum at 100\,K, resulting from the background emission that is dominant in most of the map, and absorption spectra with peaks at \vlos\,$\approx$\,$-2$ and 2\,\kps.
These two peaks are most likely the result of momentum conservation in the shocked layer, as we discuss in more detail in the next section.
The $^{13}$CO is clearly centered at \vlos\,$\approx$\,0\,\kps.

\subsection{HOG analysis results}

We run the HOG analysis of the two sets of synthetic observations (face-on and edge-on) following the procedure described in Sec.~\ref{section:method}.
We use a derivative kernel with a 0.12\,pc (5 pixels) FWHM.
Given that the synthetic observations do not include noise, we consider all non-zero gradients in the synthetic H{\sc i} and $^{13}$CO PPV cubes.

Figure~\ref{fig:HOG_PCsims} shows the HOGs corresponding to a selection of H{\sc i} and $^{13}$CO velocity channels. 
One fact that is evident from the shape of the HOGs is that, at least for some pairs of velocity channels, the distribution of relative orientation angles is not flat and it clearly peaks at $\phi$\,$=$\,0\deg; this indicates that in these channel pairs, the H{\sc i} and $^{13}$CO have contours that are aligned and the two tracers are morphologically correlated.  
This can be visually confirmed in the gradient plots of the velocity-channel pairs with the highest \prs\ values, presented in Fig.~\ref{fig:HOGpanelPCsims}, \juan{where it is evident that the $^{13}$CO emission contours are adjacent to the contours of regions with a relative decrease of H{\sc i} emission}.

The behavior of the relative orientation trend is better visualized in the values of the mean resultant vector length $r$, defined in Eq.~\eqref{eq:mymrv}, and the projected Rayleigh statistic \prs, defined in Eq.~\eqref{eq:myprs}, for all pairs of H{\sc i} and $^{13}$CO channels in the velocity range $-$4.8\,$\leq$\,\vlos\,$\leq$\,4.8\,\kps, presented in Fig.~\ref{fig:HOGcorrPCsims}.
The distribution of \prs\ and $r$ shows that the maximum spatial correlation between the H{\sc i} and $^{13}$CO emission appears at the same velocity in the two tracers, that is, along the diagonal of the correlation plane, where \vhi\,$\approx$\vco.
This observation is not entirely unexpected; if one considers the standard picture of a quiescent MC and its associated atomic envelope and the atomic gas and the molecular gas move together, then, the two tracers should appear approximately at the same velocity.
However, it is worth remarking that this correlation indicates that the contours of the emission of the two tracers match across multiple velocity channels.
This behavior is not exclusive to the case of H{\sc i} with an emission background and $^{13}$CO, it can also be seen when applying HOG to the analysis of synthetic observations of H{\sc i} without background emission and $^{13}$CO emission, as shown in Appendix~\ref{app:MHDsims}.

The HOG, however, does not reveal an unambiguous difference between the signals produced by the observations in the face-on and the edge-on configurations.
To zeroth order, HOG is revealing the spatial coincidence of the two tracers, which does seem to be significantly affected by the orientation of the colliding flows with respect to the line of sight.
In more detail, the face-on configuration presents homogenous high \prs\ values along \vhi\,$\approx$\,\vco\ in the velocity range $-$2.5\,$<$\,\vlos\,$<$\,2.5\,\kps, while in the edge-on configuration the high \prs\ values seem group around \vlos\,$\approx$\,$-$2.0 and 2.0\,\kps, but also close to \vhi\,$\approx$\,\vco. 
In both configurations, these trends are produced by approximately 30\% of the gradient pairs, as inferred from the values of $r$.

The difference between \prs\ in the face-on and the edge-on configurations can be understood in terms of the dynamics imposed by the colliding flow.
In the face-on case, the ram pressure constrains both the cold H{\sc i} and $^{13}$CO to remain close to \vlos\,$\approx$\,0\,\kps.
\juan{The molecular gas formed in the shocked interface does not inherit the structure of the colliding atomic clouds; consequently, we do not see high spatial correlation between the $^{13}$CO at \vlos\,$\approx$\,0\,\kps\ and the H{\sc i} at the velocities of the colliding clouds}.
\juan{Given that the shocked interface is relatively thin, there is not much overlap of structures along the line of sight, which most likely explains the tight correlation around \vhi\,$\approx$\,\vco\ shown in the top panels of Fig.~\ref{fig:HOGcorrPCsims}}.

\juan{In the edge-on case, we are looking at the shocked interface in the direction that is not directly constrained by the ram pressure, where the parcels of H{\sc i} and $^{13}$CO have developed line-of-sight motions that are independent from the proper motion of the parental atomic clouds.}
\juan{In contrast with the face-on case, the larger values of $V$ are centered on \vlos\,$\approx$\,2\,\kps, most likely due to the proper motion of the most dominant parcel of $^{13}$CO formed in the shocked interface.}
\juan{The overlap of structures along the line of sight in the edge-on shocked interface is most likely producing the dispersion of high $V$ values across velocity channels, however, it is still closely concentrated around \vhi\,$\approx$\,\vco.}

The HOG analysis reveals that the H{\sc i} and the $^{13}$CO emission from a colliding flow simulation appear morphologically correlated at roughly the same velocity, independently of the orientation of the primary flow with respect to the line of sight, which is both disappointing and encouraging.
On the one hand, this implies that the \juan{results of the HOG analysis cannot unambiguously differentiate orientations of the cloud collision with respect to the line of sight}. 
On the other hand, this implies that the HOG signal produced by the \juan{atomic cloud collision} is not greatly affected by the orientation of the primary flow with respect to the line of sight and the HOG can be \juan{used} to quantify any departures from this simple scenario.
These departures are evident when we apply HOG to the observations.

\section{Observations}\label{section:observations}

\begin{figure*}[ht!]
\vspace{-0.1cm}
\centerline{
\includegraphics[height=2.9in,angle=0,origin=c]{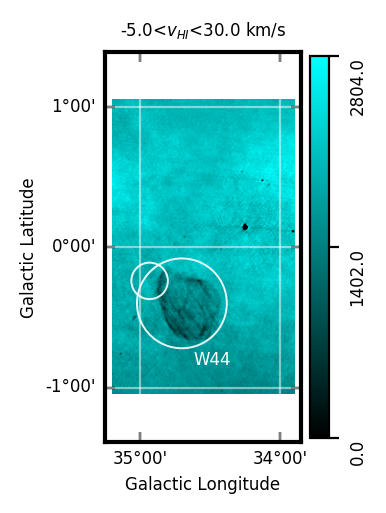}
\hspace{-0.2cm}
\includegraphics[height=2.9in,angle=0,origin=c]{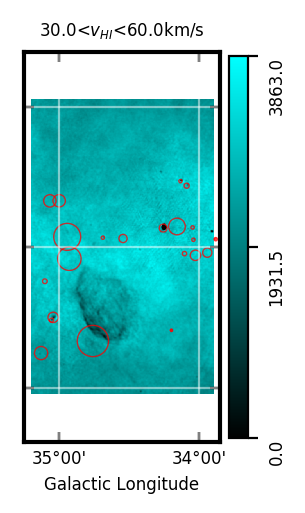}
\hspace{-0.2cm}
\includegraphics[height=2.9in,angle=0,origin=c]{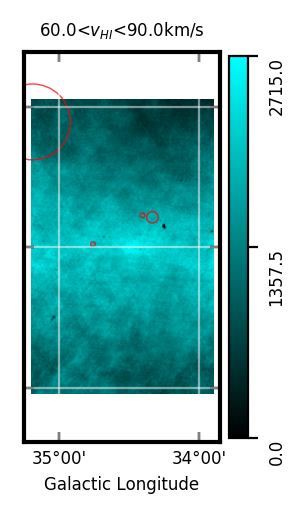}
\hspace{-0.2cm}
\includegraphics[height=2.9in,angle=0,origin=c]{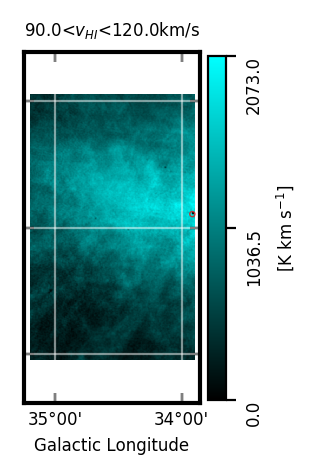}
}
\vspace{-0.1cm}
\centerline{
\includegraphics[height=2.9in,angle=0,origin=c]{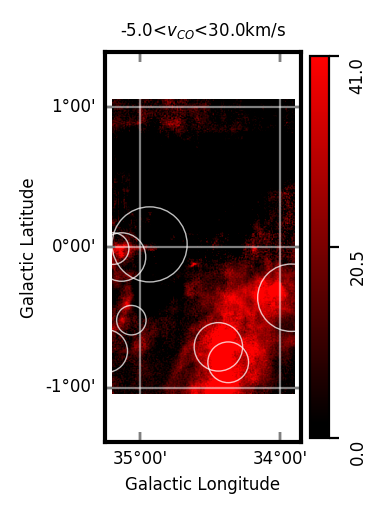}
\hspace{-0.2cm}
\includegraphics[height=2.9in,angle=0,origin=c]{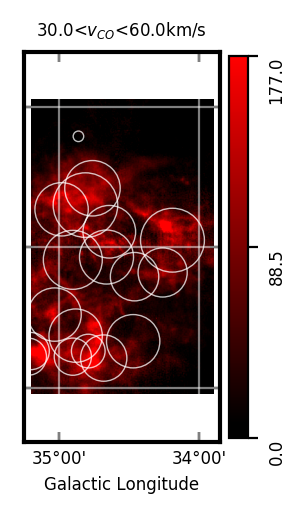}
\hspace{-0.2cm}
\includegraphics[height=2.9in,angle=0,origin=c]{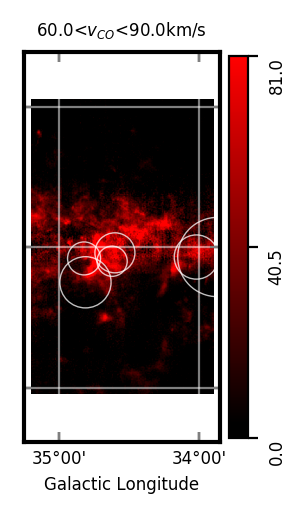}
\hspace{-0.2cm}
\includegraphics[height=2.9in,angle=0,origin=c]{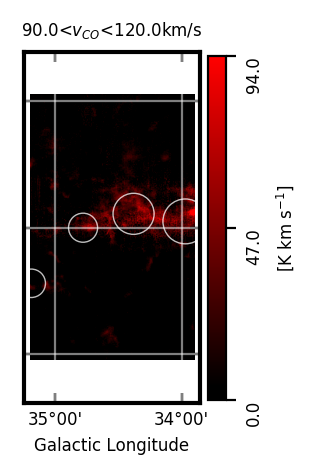}
}
\vspace{-0.1cm}
\caption{
Maps of the THOR H{\sc i} and GRS $^{13}$CO observations.
\emph{Top}. H{\sc i} integrated emission from the THOR+VGPS+GBT observations \citep{beuther2016} in the indicated velocity ranges.
The white circles correspond to the positions and effective sizes of identified supernovae remnants \citep{anderson2017,green2014} in each velocity range.
The red circles correspond to the positions and effective sizes of identified H{\sc ii} regions \citep{anderson2014} in each velocity range.
\emph{Bottom}. $^{13}$CO integrated emission from the GRS observations \citep{jackson2006} in the indicated velocity ranges.
The white circles correspond to the positions and effective sizes of the molecular cloud candidates from the \cite{rathborne2009} catalog in the indicated velocity ranges.
}
\label{fig:HIandCOmaps}
\end{figure*}

\begin{figure}[ht!]
\vspace{-0.1cm}
\centerline{
\includegraphics[height=3in,angle=0,origin=c]{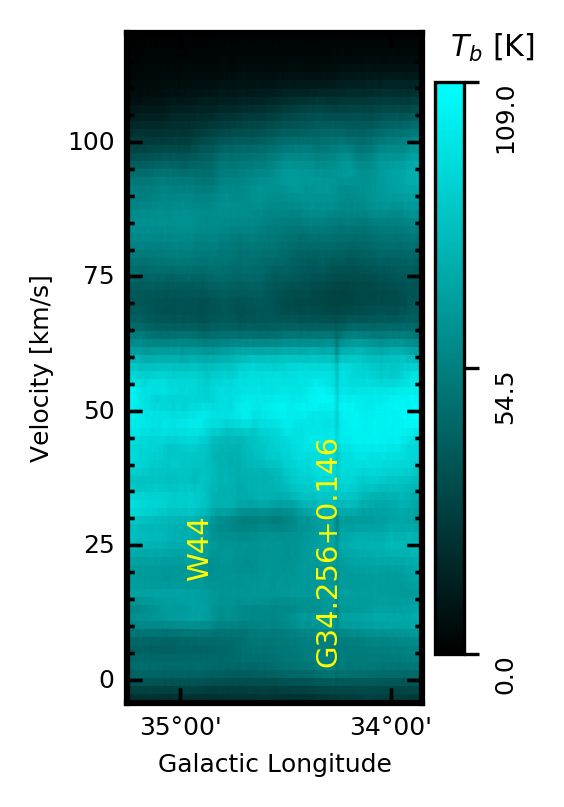}
\hspace{-0.1cm}
\includegraphics[height=3in,angle=0,origin=c]{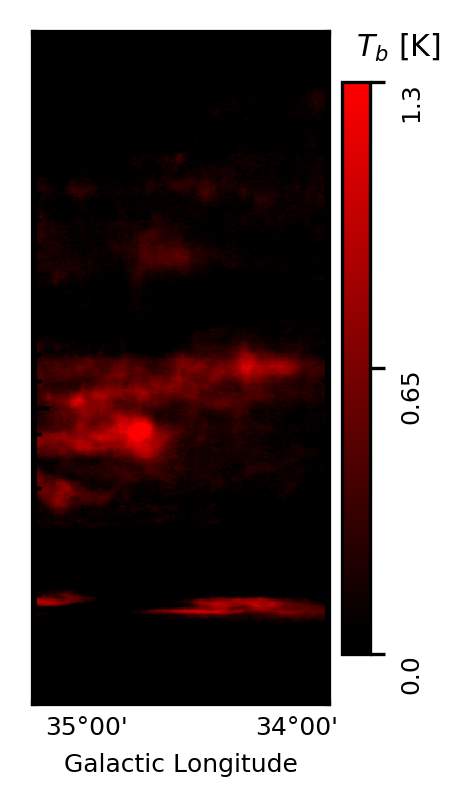}
}
\vspace{-0.1cm}
\caption{
Longitude-velocity (LV) diagrams of the H{\sc i} emission (from THOR, left) and $^{13}$CO emission (from GRS, right) toward the region presented in Fig.~\ref{fig:HIandCOmaps}.
In the H{\sc i} LV diagram, the shadow around 0\,$\leq$\,\vlsr\,$\leq$\,45\,\kps\ and the vertical line around 0\,$\leq$\,\vlsr\,$\leq$\,75\,\kps\ correspond to absorption toward the supernova remnant (SNR) G34.7-0.4 (W44) and the H{\sc ii} region G34.256+0.146, respectively.
}
\label{fig:HIandCOlvdiagrams}
\end{figure}

For this first application of the histogram of oriented gradients (HOG) method, we choose the THOR H{\sc i} and GRS $^{13}$CO observations toward the portion of the Galactic plane defined by 33\pdeg75\,$<$\,$l$\,$<$\,35\pdeg25 and $|b|$\,$<$\,1\pdeg25, which are shown in Fig.~\ref{fig:HIandCOmaps} and Fig.~\ref{fig:HIandCOlvdiagrams}.
Given the need to describe the method in detail, we focus our analysis on this region because it contains a large diversity of objects, such as supernova remnants (SNR), H{\sc ii} regions, H{\sc i} Self-Absorption (HISA) features \citep{bihrPhDT2016,wangInPrep}, and a wealth of MCs that have been identified in emission from $^{12}$CO and $^{13}$CO \citep[][respectively]{miville-deschenes2017,rathborne2009}.  
We reserve the application of the HOG technique to the whole extent of both surveys for a subsequent publication \citep{solerInPrep}.

The selected region includes two SNRs that we identify using the catalogs presented in \cite{green2014} and \cite{anderson2017}.
The most conspicuous of these two SNRs is Westerhout 44 \citep[W44,][]{westerhout1958}, located around $[l, b]$\,$=$\,$[34\pdeg7,-0\pdeg4]$, which is shown in Fig.~\ref{fig:HIandCOmaps}.
Multi-wavelength observations of W44 show the presence of an elongated shell-like structure with a remarkable network of filaments and arcs across the face of this remnant suggesting the presence of shocked gas \citep{giacani1997,reach2006}.

The region also contains a plethora of H{\sc ii} regions, which we identify using the catalog produced using the {\emph WISE} observations \citep{anderson2014}.
One of the most interesting objects in this catalog is the ultra-compact H{\sc ii} (UCH{\sc ii}) region G34.256+0.146, which produces a significant absorption feature that is clearly distiguishable in the H{\sc i} longitude-velocity (LV) diagram presented in Fig.~\ref{fig:HIandCOlvdiagrams}.

This region also includes portions of two giant molecular filaments (GMFs) in the sample presented in \cite{ragan2014}.
First, 38.1-32.4a, a structure that extends across 33\pdeg4\,$\leq$\,$l$\,$\leq$\,37\pdeg1 and $-$0\pdeg4\,$\leq$\,$b$\,$\leq$\,0\pdeg6 and is associated to $^{13}$CO emission in the range 50\,$\leq$\,\vlsr\,$\leq$\,60\,\kps.
Second, GMF38.1-32.4b, a structure that extends across 34\pdeg6\,$\leq$\,$l$\,$\leq$\,35\pdeg6 and $-$1\pdeg0\,$\leq$\,$b$\,$\leq$\,0\pdeg2 and is associated to $^{13}$CO emission in the range 43\,$\leq$\,\vlsr\,$\leq$\,46\,\kps.

\subsection{Atomic hydrogen emission at 21\,cm}\label{sec:HIdata}

We use the H{\sc i} emission observations from The H{\sc i}/OH/Recombination line survey of the inner Milky Way \citep[THOR,][]{beuther2016}.
THOR comprises observations in eight continuum bands between 1 and 2\,GHz made with Karl G. Jansky Very Large Array (VLA) in the C-array configuration covering the portion of the Galactic plane defined by 14\pdeg0\,$\leq$\,$l$\,$\leq$\,67\pdeg4 and $|b|$\,$\leq$\,1\pdeg25 at approximately 20\arcsec\ resolution.
As the survey name implies, the THOR frequency range includes the H{\sc i} 21-cm emission line, four OH lines, and 19 H$\alpha$ recombination lines.

The THOR H{\sc i} data that are taken in C-array configuration are crucial for the study of absorption profiles against Galactic and extragalactic background sources.
However, they do not recover the large-scale emission. 
In the present study, we use the data set resulting from the combination of  the H{\sc i} observations from THOR and the D-configuration VLA Galactic Plane Survey, \citep[VGPS,][]{stil2006} combined with single-dish observations from the Green Bank Telescope (GBT).

The C-array configuration H{\sc i} visibility data from the THOR survey were calibrated with the {\tt CASA}\footnote{https://casa.nrao.edu/} software package as described in \cite{beuther2016}. 
We used the multi-scale {\tt CLEAN} routine in {\tt CASA} to image the continuum-subtracted C-array configuration H{\sc i} visibility together with the D-array configuration visibility from VGPS \citep{stil2006}. 
We chose a pixel size of 4\arcsec, robust\,=\,0.45, and a velocity resolution of 1.5\,\kps\ in the velocity range $-$50\,$\leq$\,\vlsr\,$\leq$\,150\,\kps.
The resulting images were smoothed into a resolution of 40\arcsec\ and feathered with the VGPS images (D+GBT) to recover the large-scale structure. 
Further details on the data reduction and imaging procedure are described in \cite{beuther2016}.
The public release of this new H{\sc i} data product is forthcoming \citep{wangInPrep}.

\subsection{Carbon monoxide (CO) emission}

We compare the H{\sc i} emission observations with the $^{13}$CO($J$\,$=$\,1\,$\rightarrow$\,0) observations from The Boston University-Five College Radio Astronomy Observatory Galactic Ring Survey \citep[GRS,][]{jackson2006}.
The GRS survey has 46\arcsec\ angular resolution with an angular sampling of 22\arcsec.
In this particular region, it covers the range $-$5\,$\leq$\,\vlsr\,$\leq$\,135\,\kps\ at a resolution of 0.21\,\kps.
It has a typical root mean square (RMS) sensitivity of 0.13\,K.
We also make use of the catalog of MC and clump candidates identified in the GRS data \citep{rathborne2009}.

We use $^{13}$CO rather than $^{12}$CO to minimize optical depth effects and facilitate the interpretation of the HOG analysis.
The $^{12}$CO emission is widespread towards the Galactic plane, just like H{\sc i}, and only around 14\% of the molecular gas mass traced by $^{12}$CO emission is identified as part of molecular clouds in $^{13}$CO \citep{roman-duval2016}.
Compared to $^{12}$CO, the $^{13}$CO molecule is approximately 50 times less abundant and, thus, has a much lower optical depth \citep{wilsonANDrood1994}.
As a result, $^{13}$CO is a much better tracer of column density and suffers less from line blending and self-absorption.

\section{HOG analysis of observations}\label{section:results}

\begin{figure*}[ht!]
\centerline{
\includegraphics[width=0.4\textwidth,angle=0,origin=c]{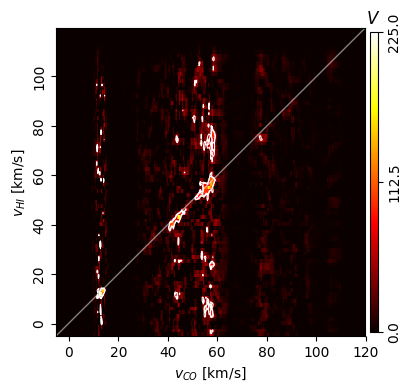}
\includegraphics[width=0.4\textwidth,angle=0,origin=c]{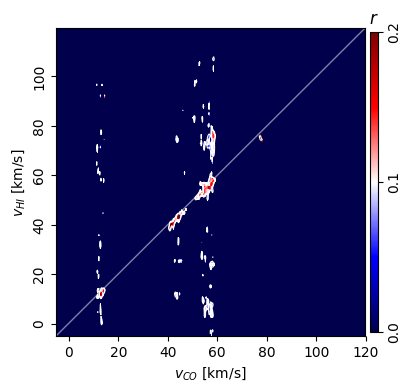}
}
\vspace{-0.1cm}
\caption{
Results of the HOG analysis of the THOR H{\sc i} and GRS $^{13}$CO observations. 
\emph{Left.} Projected Rayleigh statistic, $V(v_{\rm 13CO},v_{\rm HI})$, the HOG statistical test of spatial correlation between H{\sc i} and $^{13}$CO velocity-channel maps, defined in Eq.~\eqref{eq:myprs}.
\juan{The contours indicate the 3$\varsigma_{V}$, 4$\varsigma_{V}$, and 5$\varsigma_{V}$ levels in the corresponding velocity range.}
\emph{Right.} Mean resultant vector length, $r(v_{\rm 13CO},v_{\rm HI})$, with the 3$\varsigma_{V}$ confidence interval, a HOG metric that is roughly equivalent to the percentage of gradient pairs that imply the spatial correlation between the velocity-channel maps, defined in Eq.~\eqref{eq:mymrv}.
}\label{fig:corrPlane}
\end{figure*}

We apply the HOG analysis to the data products described in Sec.~\ref{section:observations} using the method described in Sec.~\ref{section:method}.
\juan{We compute HOG exclusively using gradients that satisfy $I_{ij,k} \geq 5I^{\rm N}$ and $|\nabla I_{ij,k}| \geq 5|\nabla I^{\rm N}|$, where the noise intensity, $I^{\rm N}$, and the noise gradient norm, $|\nabla I^{\rm N}|$, are estimated following the procedure presented in Appendix~\ref{sec:gradselection}.
Here we present and discuss the results obtained using a derivative kernel with a 90\arcsec\ FWHM.
This selection does not imply any loss of generality as described in Appendix~\ref{app:multiscaleHOG}, where we discuss the results of using different derivative kernel sizes.}
\juan{The selection of $I_{ij,k} \geq 5I^{\rm N}$ instead of $I_{ij,k} \geq 3I^{\rm N}$ does not critically change the results of this analysis, as illustrated in Appendix~\ref{app:HOGstatsMCnoise}.}

Figure~\ref{fig:corrPlane} presents the values of the projected Rayleigh statistic, \prs, and the mean resultant vector length, $r$, corresponding to the H{\sc i} and $^{13}$CO emission for the velocity range $-$5\,$\leq$\,\vlsr\,$\leq$\,120\,\kps.
It is clear from Fig.~\ref{fig:corrPlane} that the spatial correlation between the H{\sc i} and $^{13}$CO emission is significant at the same velocity in the two tracers, that is, at \vhi\,$\approx$\,\vco\, or equivalently, along the diagonal of the \prs- and $r$-plane.
As discussed in the previous section, if one considers a toy quiescent MC and its respective atomic envelope, the atomic gas and the molecular gas move together, then, the two tracers should appear approximately at the same \vlsr.
However, this result confirms the prediction from the analysis of the synthetic observations: there is a morphological correlation in the spatial distribution of H{\sc i} and $^{13}$CO.
This spatial correlation is not the result of the concentration of emission around particular velocity channels nor the product of chance correlation, as we prove through the statistical tests presented in Appendix~\ref{app:HOGstatsJacknives}.
We discuss in detail this correlation around \vhi\,$\approx$\,\vco\ in Sec.~\ref{subsection:diag} and particularly focus on the 47.5\,$\leq$\,\vlsr\,$\leq$\,62.5\,\kps\ range in Sec.~\ref{sec:SpecialRange}.

Figure~\ref{fig:corrPlane} also shows some less-dominant correlation in velocity channels that are not necessary around \vhi\,$\approx$\,\vco, such as that seen around \vhi\,$\approx$\,10 and \vco\,$\approx$\,55\,\kps\ or less significantly around \vhi\,$\approx$\,70 and \vco\,$\approx$\,10\,\kps.
This correlation appears associated to some vertical stripes in the $V$-plane, which can be interpreted as the spatial distribution of the $^{13}$CO being correlated with the H{\sc i} in many channels.
We discuss this off-diagonal signal, in terms of its position in the \prs-plane, in Sec.~\ref{subsection:offdiag}.

\subsection{Interesting velocity ranges}\label{subsection:diag}

Figure~\ref{fig:corrPlane} reveals that the largest \prs\ values are grouped around roughly four values of \vlsr; explicitly, \vlsr\,$\approx$\,12, 43, 55, and 75\,\kps.
These velocities are related to the radial velocities of the individual parcels of H{\sc i} and $^{13}$CO that are morphologically correlated, thus, they are most likely associated with the rotation of the Galaxy and its spiral arm structure.
Visual inspection of the spiral arm model presented in \cite{reid2014} suggests that the $^{13}$CO emission at 12\,\kps\ might be associated with the Perseus arm, at 43 and 55\,\kps\ with the far side of the Sagittarius arm, and at 75\,\kps\ with the Aquila spur. 
However, establishing the association between the central velocities of this emission and the spiral arm structure is not straightforward and it is beyond the scope of this work.

In what follows we detail the HOG analysis around each of these central velocities to establish if the morphological correlation can be associated to a particular set of objects.
For that purpose we focus our analysis both in the velocity ranges identified using the values of \prs\ and the MC candidates identified in catalogs presented in \cite{rathborne2009} and \cite{miville-deschenes2017}.
For the sake of simplicity, we also identify the region with maximum \prs\ values in a Galactic longitude and latitude grid of 3$\times$7 elements, which we call blocks following the vocabulary introduced in machine vision studies \citep[for example,][]{ml:zhu2006}. 
This selection of grid is arbitrary and just aims to guide the eye to the areas of the maps where the $\phi$ distribution is more significantly peaked around 0\deg. 

\begin{figure*}[ht!]
\centerline{
\includegraphics[height=3.0in,angle=0,origin=c]{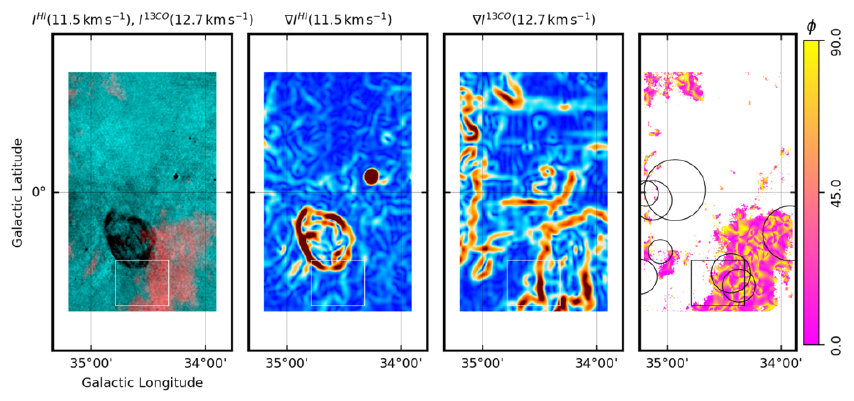}
}
\caption{
Intensity, intensity gradients, and relative orientation angle maps from the THOR H{\sc i} and GRS $^{13}$CO observations presented in Fig,~\ref{fig:HIandCOmaps}.
\emph{Left.} H{\sc i} (teal) and $^{13}$CO emission (red) in the velocity channels with the largest spatial correlation in the velocity range $-$5\,$\leq$\,\vlsr\,$\leq$\,30\,\kps, as inferred from the \prs\ values shown in Fig.~\ref{fig:corrPlane}.
\emph{Middle left.} Norm of the gradient of the H{\sc i} intensity map in the indicated velocity channel.
\emph{Middle right.} Norm of the gradient of the $^{13}$CO intensity map in the indicated velocity channel.
\emph{Right.} Relative orientation angle $\phi$, Eq.~\eqref{eq:phi}, between the gradients of the H{\sc i} and $^{13}$CO intensity maps in the indicated velocity channels.
The white color in the $\phi$ map corresponds to areas where the gradient is not significant in either tracer, as estimated using the rejection criteria described in App.~\ref{sec:gradselection}.
The rectangle shows the block, selected from a 7\,$\times$\,3 spatial grid, with the largest values of \prs.
The black circles in the rightmost panel correspond to the positions and effective sizes of the MC candidates from the \cite{rathborne2009} catalog in the aforementioned velocity range.
}
\label{fig:HOGanalysis-5to30}
\end{figure*}

In the $-$5\,$\leq$\,\vlsr\,$\leq$\,30\,\kps\ velocity range, the most conspicuous feature in \prs\ is centered on \vlsr\,$\approx$\,12\,\kps.
Figure~\ref{fig:HOGanalysis-5to30} reveals that in the pair of H{\sc i} and $^{13}$CO velocity channels with the largest values of \prs, the gradients in the H{\sc i} map are dominated by W44, but these do not have a particular correspondence with the $^{13}$CO gradients.
The maximum values of \prs\ correspond to the area in the southeast of W44, around $[l,b]$\,$=$\,[35\pdeg0,0\pdeg6], where an elongated $^{13}$CO emission blob has a clear correspondence with the H{\sc i}.
This $^{13}$CO emission feature is not among the objects identified in the \cite{rathborne2009} cloud catalog or included within the effective radius of the objects identified in \cite{miville-deschenes2017}.

\begin{figure*}[ht!]
\centerline{
\includegraphics[height=3.0in,angle=0,origin=c]{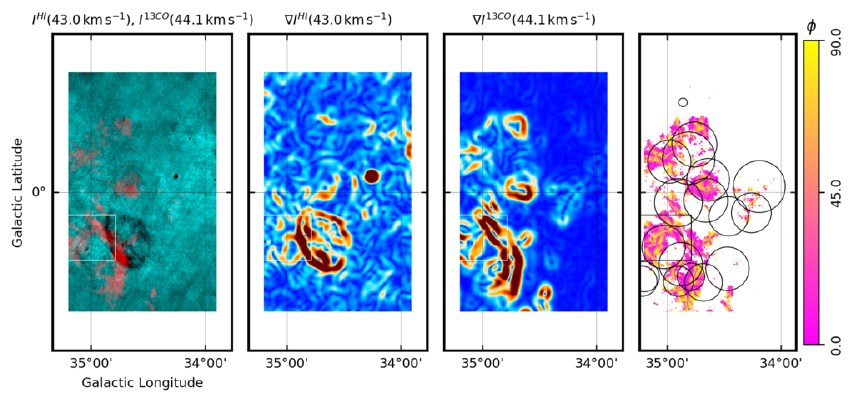}
}
\centerline{
\includegraphics[height=3.0in,angle=0,origin=c]{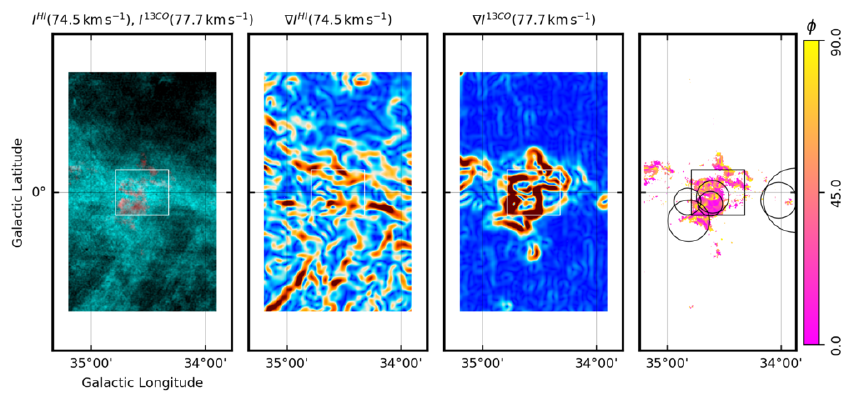}
}
\centerline{
\includegraphics[height=3.0in,angle=0,origin=c]{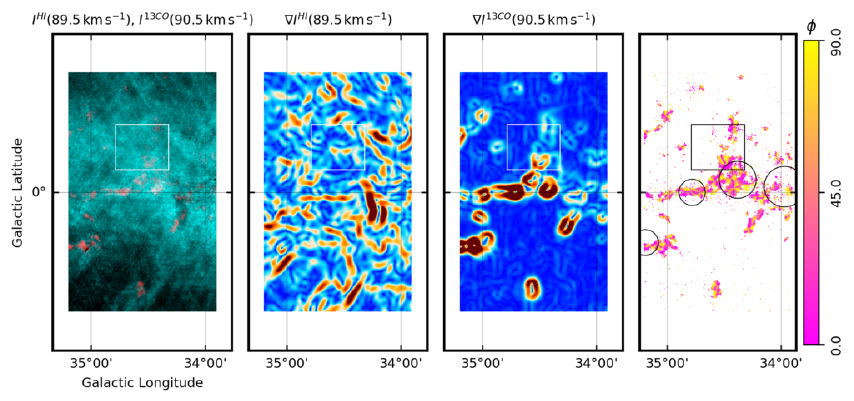}
}
\vspace{-0.1cm}
\caption{
Same as Fig.~\ref{fig:HOGanalysis-5to30} for 30\,$\leq$\,\vlsr\,$\leq$\,60\,\kps\ (top), 60\,$\leq$\,\vlsr\,$\leq$\,90\,\kps\ (middle), and 90\,$\leq$\,\vlsr\,$\leq$\,120\,\kps\ (bottom).
}
\label{fig:HOGanalysis30to60}
\end{figure*}

In the 30\,$\leq$\,\vlsr\,$\leq$\,60\,\kps\ velocity range, the most significant features in \prs\ are centered at \vlsr\,$\approx$\,43 and 55\,\kps.
The two velocities roughly correspond to those of the two giant molecular filaments (GMFs) identified in \cite{ragan2014}.
The top panel of Fig.~\ref{fig:HOGanalysis30to60} shows that in the velocity channel maps corresponding to the largest \prs\ values, the H{\sc i} gradients are still dominated by W44 and the largest correlation appears around the eastern edge of that SNR, around $[l,b]$\,$=$\,[34\pdeg8,$-$0\pdeg4].

The studied area of the sky contains a large number of MC candidates from the \cite{rathborne2009} and \cite{miville-deschenes2017} catalogs in this velocity range.
One of the objects in the \cite{rathborne2009} catalog, centered at $[l,b]$\,$=$\,[35\pdeg0,$-$0\pdeg5], is coincident with the large-\prs\ region identified in the top panel of Fig.~\ref{fig:HOGanalysis30to60}.
Additionally, there is also large regions of coincident gradients in the \cite{rathborne2009} MC candidates centered at $[l,b]$\,$=$\,[34\pdeg6,0\pdeg25] and [34\pdeg6,$-$0\pdeg25], although there are extended regions with $\phi$\,$\approx$\,0\deg\ that do not correspond to any MC candidate.

In the 60\,$\leq$\,\vlsr\,$\leq$\,90\,\kps\ velocity range, the most significant features in \prs\ are centered at \vlsr\,$\approx$\,75\,\kps.
The middle panel of Figure~\ref{fig:HOGanalysis30to60} shows that in the velocity channel maps corresponding to the largest \prs\ values, the correlation between the gradients is concentrated in the region around $[l,b]$\,$=$\,[34\pdeg5,0\pdeg0], which is coincident with two \cite{rathborne2009} and one \cite{miville-deschenes2017} MC candidates.

There is not \juan{a} significant \juan{spatial} correlation in the 90\,$\leq$\,\vlsr\,$\leq$\,120\,\kps\ velocity range when it is compared to the $V$ values obtained in the full $-$5\,$\leq$\,\vlsr\,$\leq$\,120\,\kps\ range, as illustrated in Fig.~\ref{fig:corrPlane}.
However, when considering the pair of velocity channels with the maximum value of $V$ in the 90\,$\leq$\,\vlsr\,$\leq$\,120\,\kps\ range, we find significant spatial correlation toward the \cite{rathborne2009} and \cite{miville-deschenes2017} MC candidates centered on $[l,b]$\,$=$\,[34\pdeg4,0\pdeg15], as shown in the bottom panel of Fig.~\ref{fig:HOGanalysis30to60}.
There, the regions with $\phi$\,$\approx$\,0\deg\ seem to be less extended than those shown in the 30\,$\leq$\,\vlsr\,$\leq$\,60\,\kps\ and 60\,$\leq$\,\vlsr\,$\leq$\,90\,\kps\ ranges and they cover just a few small patches.

There is some interesting correlation between H{\sc i} and $^{13}$CO around $[l,b]$\,$=$\,[34\pdeg2,$-$0\pdeg2], where there is a clear HISA feature correlated with a small patch of $^{13}$CO emission, as it is evident in the gradients and the relative orientation angles presented in the bottom panel of Fig.~\ref{fig:HOGanalysis30to60}.
Nevertheless, this region is not coincident with any of the MC candidates in the \cite{rathborne2009} and \cite{miville-deschenes2017} catalogs.

\subsection{HOG in the 47.5\,$\leq$\,\vlsr\,$\leq$62.5\,\kps\ range}\label{sec:SpecialRange}

\begin{figure}[ht!]
\centerline{
\includegraphics[height=0.275\textheight,angle=0,origin=c]{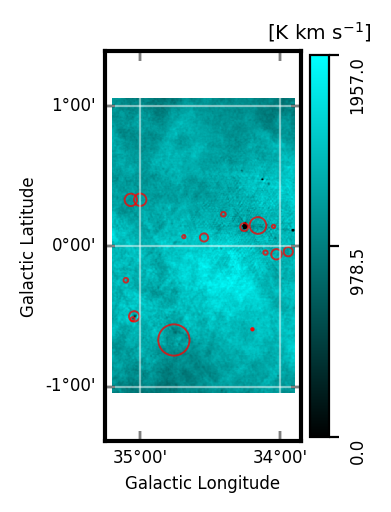}
\includegraphics[height=0.275\textheight,angle=0,origin=c]{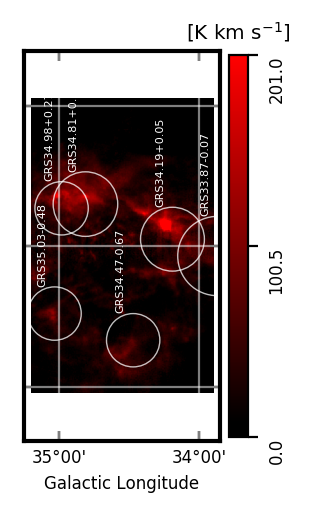}
}
\vspace{-0.1cm}
\caption{Integrated H{\sc i} (left) and $^{13}$CO (right) emission in the range 47.5\,$\leq$\,\vlsr\,$\leq$\,62.5\,\kps.
The red circles in the left-hand-side panel correspond to the H{\sc ii} regions from the \cite{anderson2014} catalog in the aforementioned velocity range.
The white circles in the right-hand-side panel correspond to the positions and effective sizes of the MC candidates from the \cite{rathborne2009} catalog in the aforementioned velocity range.
}
\label{fig:HIv47.5to62.5}
\end{figure}

\begin{figure}[ht!]
\centerline{
\includegraphics[width=0.25\textwidth,angle=0,origin=c]{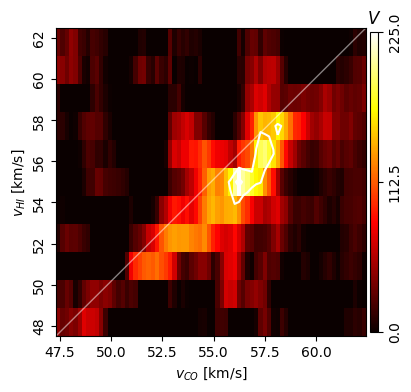}
\includegraphics[width=0.25\textwidth,angle=0,origin=c]{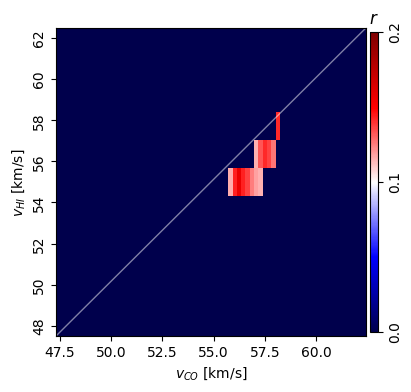}
}
\centerline{
\includegraphics[width=0.25\textwidth,angle=0,origin=c]{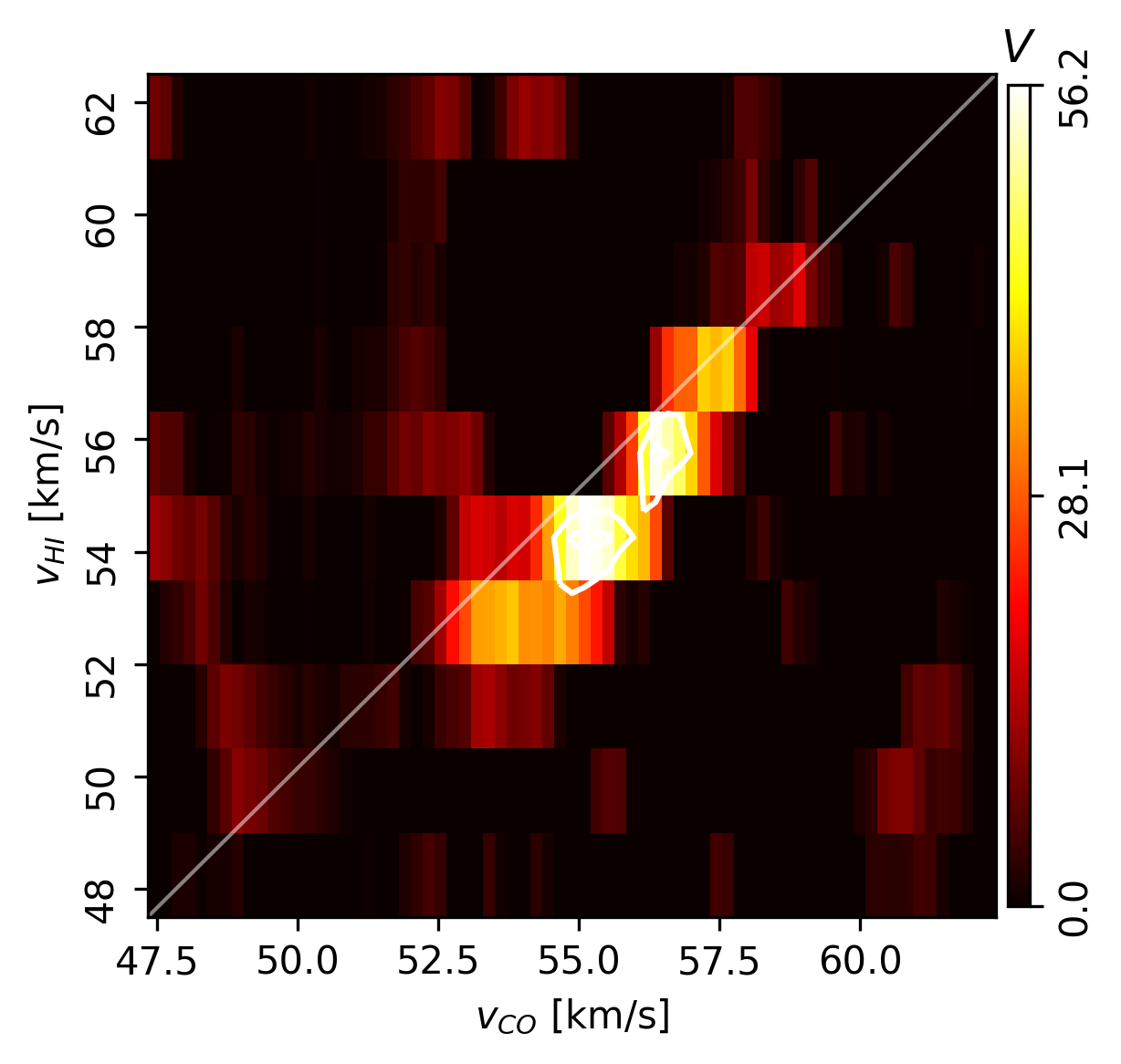}
\includegraphics[width=0.25\textwidth,angle=0,origin=c]{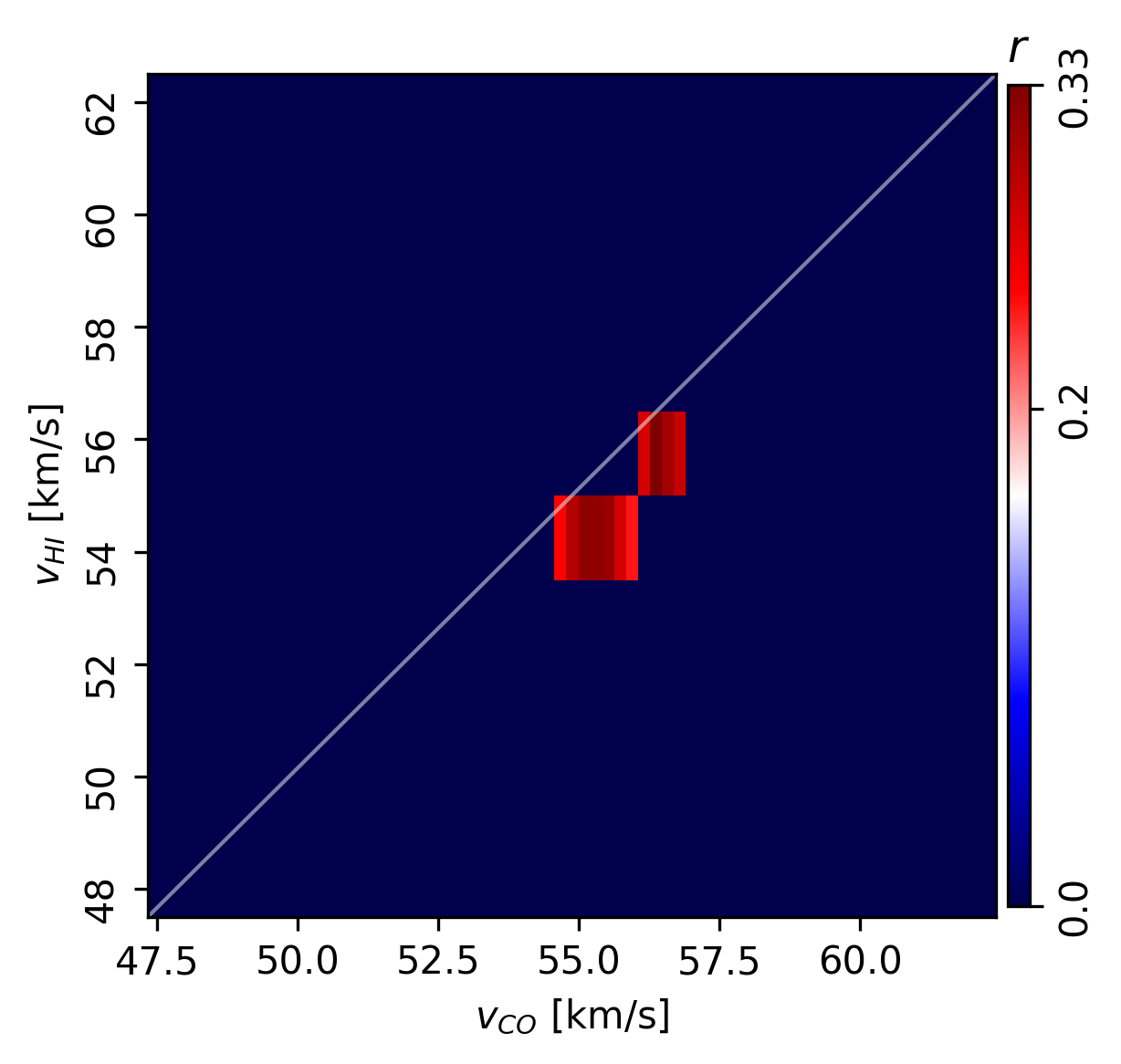}
}
\vspace{-0.1cm}
\caption{
\emph{Top.} Results of the HOG analysis of the THOR H{\sc i} and GRS $^{13}$CO observations in the velocity range 47.5\,$\leq$\,\vlsr\,$\leq$\,62.5\,\kps\ over the region presented in Fig.~\ref{fig:HIv47.5to62.5}. 
\emph{Left.} Projected Rayleigh statistic, $V(v_{\rm 13CO},v_{\rm HI})$, the HOG statistical test of spatial correlation between H{\sc i} and $^{13}$CO velocity-channel maps, defined in Eq.~\eqref{eq:myprs}.
\juan{The contours indicate the 3$\varsigma_{V}$, 4$\varsigma_{V}$, and 5$\varsigma_{V}$ levels in the corresponding velocity range.}
\emph{Right.} Mean resultant vector length, $r(v_{\rm 13CO},v_{\rm HI})$, with the 3$\varsigma_{V}$ confidence interval, a HOG metric that is roughly equivalent to the percentage of gradient pairs that imply the spatial correlation between the velocity-channel maps, defined in Eq.~\eqref{eq:mymrv}.
\emph{Bottom.} Same for the maximum-\prs\ block shown in Fig.~\ref{fig:HOGanalysis47to62}
}\label{fig:corrPlane47to62}
\end{figure}

Due to its large \prs\ values and the relatively low number of MC candidates, which facilitates our analysis, we devote special attention to velocity range around \vlsr\,$\approx$\,55\,\kps.
The distribution of the H{\sc i} and $^{13}$CO emission in this velocity range, shown in Fig.~\ref{fig:HIv47.5to62.5}, suggests at first glimpse the correlation between a large scale HISA ring, seen as the shadows in the H{\sc i} emission maps, and a $^{13}$CO ring where the \cite{rathborne2009} MC candidates are located.
The detailed values of \prs\ in this velocity range, presented in Fig.~\ref{fig:corrPlane47to62}, show a departure from the maxima along the \vhi\,$\approx$\,\vco\ range, although this behavior is below the 3$\varsigma_{V}$ level in the 47.5\,$\leq$\,\vlsr\,$\leq$62.5\,\kps.
We detail the individual behavior toward different portions of this region by making use of the objects identified in the \cite{rathborne2009} MC catalog.
Note that \cite{rathborne2009} employs just one of the multiple methods for producing MC catalogs from emission observations and the MC candidates identified there are not indisputable.
Here we use it just as a guide for our analysis of different portions of the studied area.

\begin{figure*}[ht!]
\centerline{
\includegraphics[height=3.0in,angle=0,origin=c]{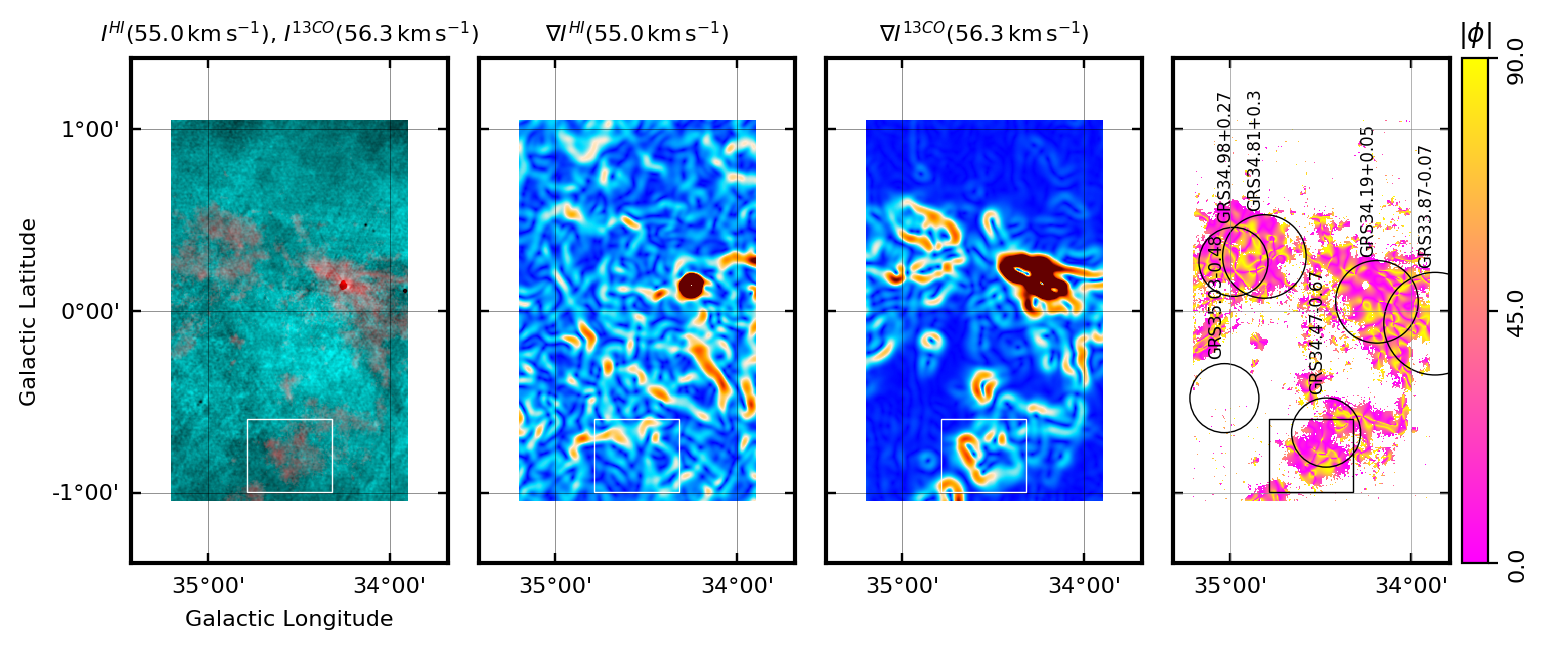}
}
\caption{
Same as Fig.~\ref{fig:HOGanalysis-5to30} for 47.5\,$\leq$\,\vlsr\,$\leq$\,62.5\,\kps.
The labels in the rightmost panel correspond to the MC candidates from the \cite{rathborne2009} catalog in this velocity range.
}
\label{fig:HOGanalysis47to62}
\end{figure*}

The rightmost panel of Figure~\ref{fig:HOGanalysis47to62} reveals that in the velocity channel maps with the largest \prs\ values, the spatial correlation between the H{\sc i} and the $^{13}$CO emission is located in extended patches.
We further study these regions by estimating the \prs\ values in the block with the highest \prs\ values and in the effective area covered by four of the \cite{rathborne2009} MC candidates; namely, GRS34.19$+$0.05, GRS34.47$-$0.67, GRS34.81$+$0.3, GRS34.98$+$0.27.
We exclude from this analysis MC candidates GRS33.87$-$0.07 and GRS35.03$-$0.48, also found in the selected velocity range, given the partial coverage of GRS33.87$-$0.07 and the low \prs\ values found toward GRS35.03$-$0.48. 

Figure~\ref{fig:corrPlane47to62} shows the correlation plane toward the block with the highest \prs\ values, indicated by the box in Fig.~\ref{fig:HOGanalysis47to62}.
Toward that portion of the map, the \prs\ values around \vlsr\,$\approx$\,55\,\kps\  are maximum for \vhi\,$\approx$\,\vco\ in the range 47.5\,$\leq$\,\vlsr\,$\leq$62.5\,\kps.
\juan{This behavior} is similar to that observed in the synthetic observations presented in Sec.~\ref{section:mhd}. 
\juan{However}, it does not necessarily imply the presence of colliding clouds toward this region.
Note that the block \juan{with the largest \prs\ value does} not contain any identified H{\sc ii} regions.

\subsubsection{$V$ values toward MC candidates}

\begin{figure}[ht!]
\centerline{
\includegraphics[width=0.25\textwidth,angle=0,origin=c]{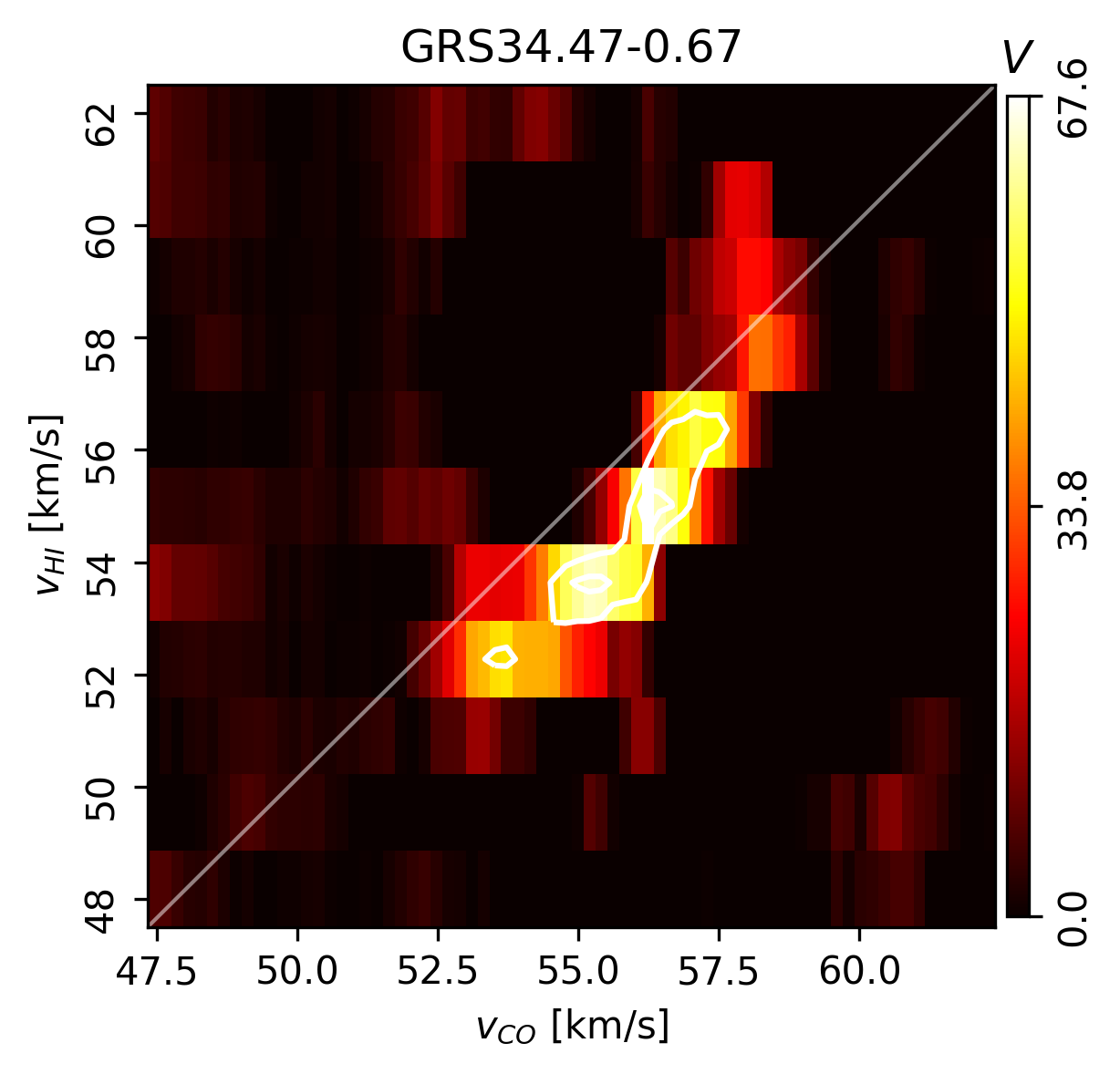}
\includegraphics[width=0.25\textwidth,angle=0,origin=c]{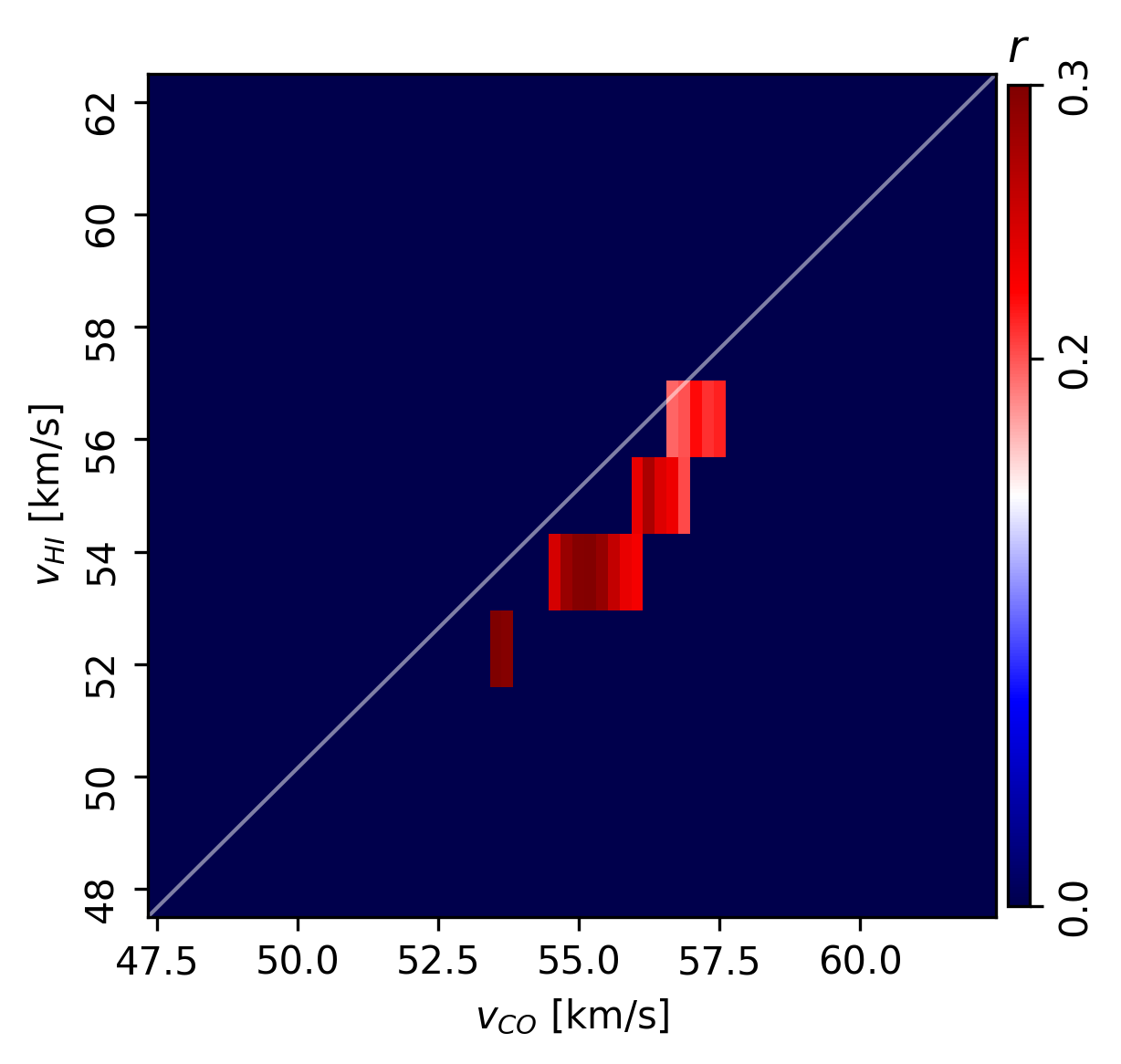}
}
\centerline{
\includegraphics[width=0.25\textwidth,angle=0,origin=c]{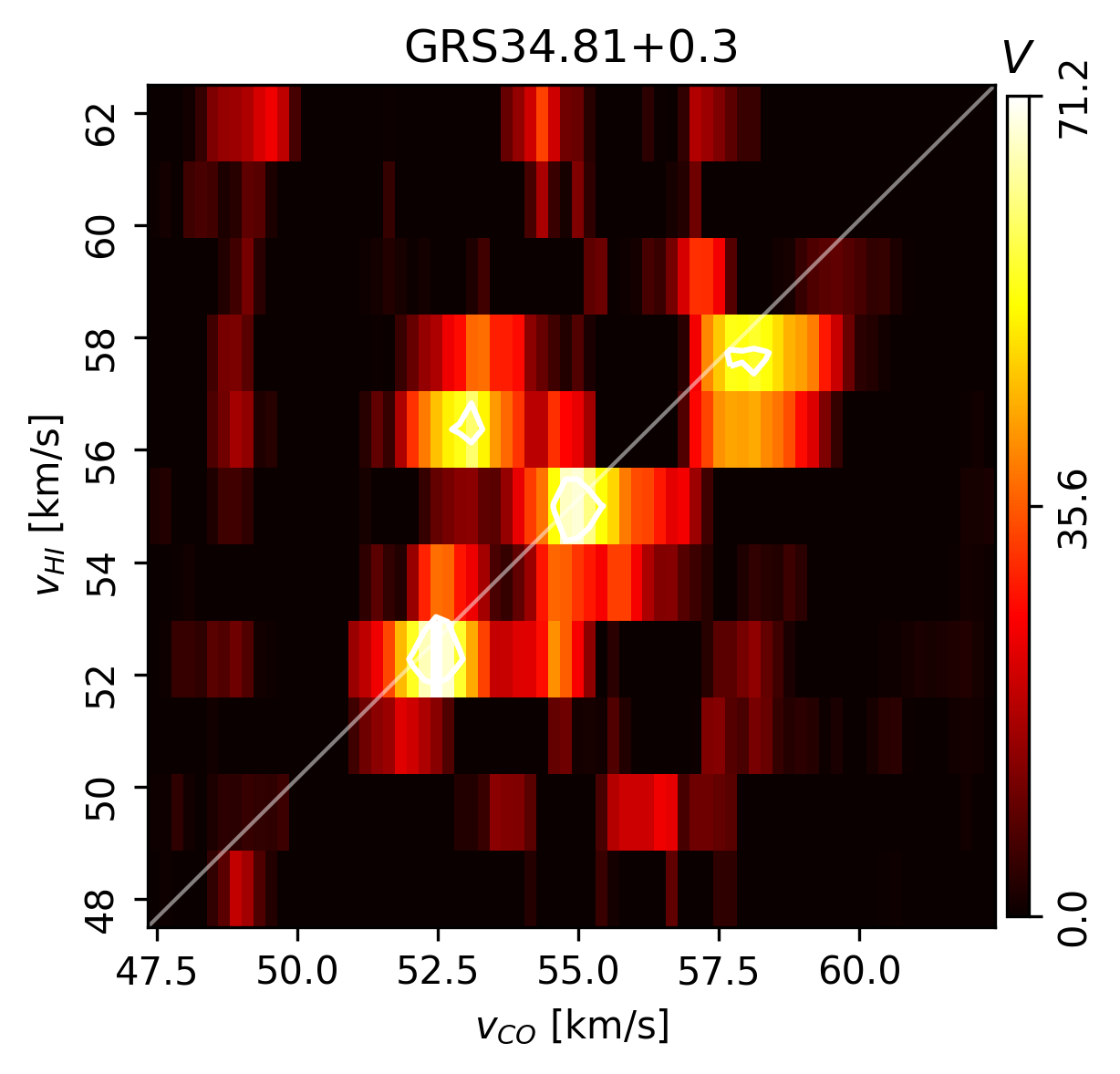}
\includegraphics[width=0.25\textwidth,angle=0,origin=c]{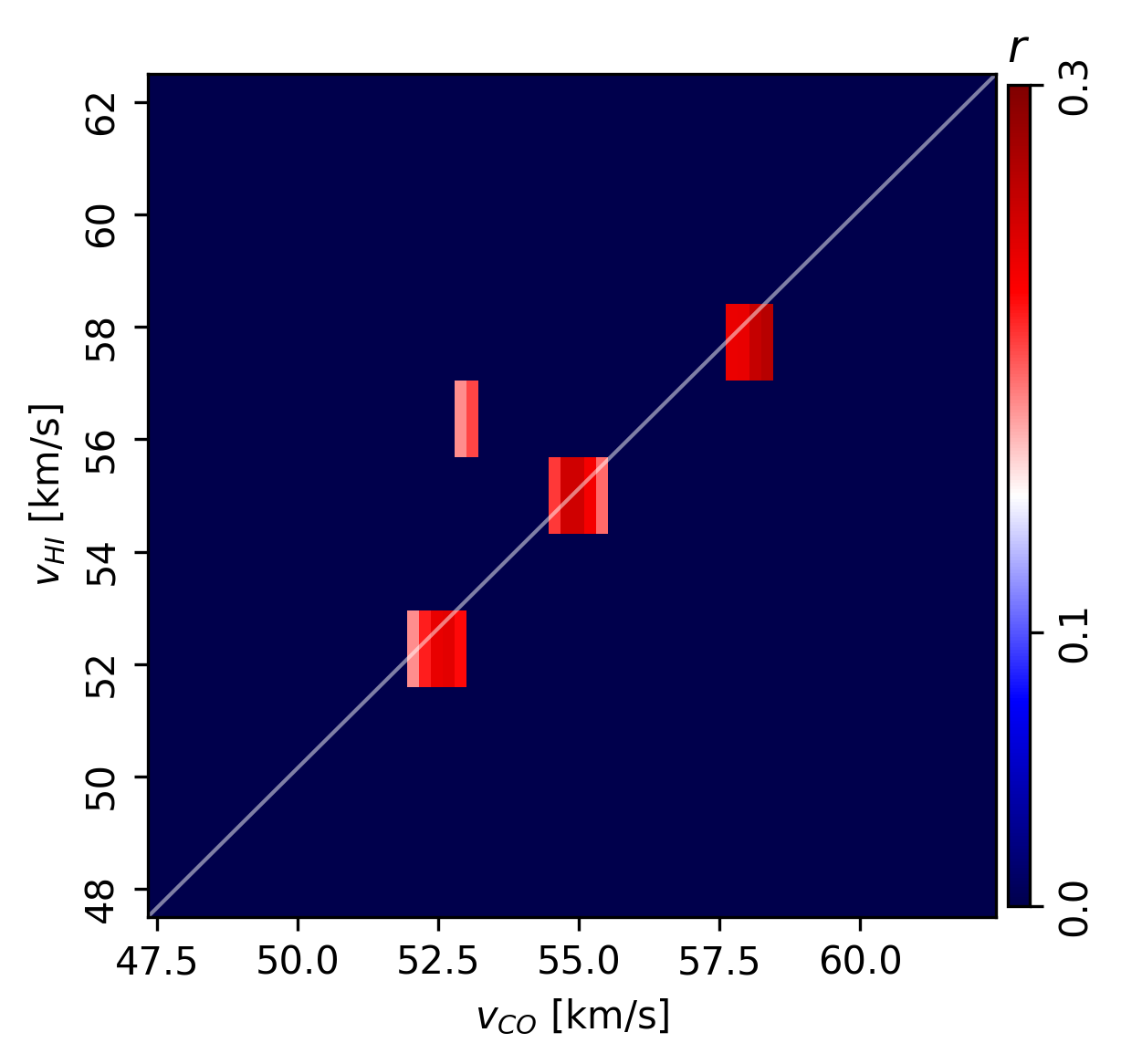}
}
\centerline{
\includegraphics[width=0.25\textwidth,angle=0,origin=c]{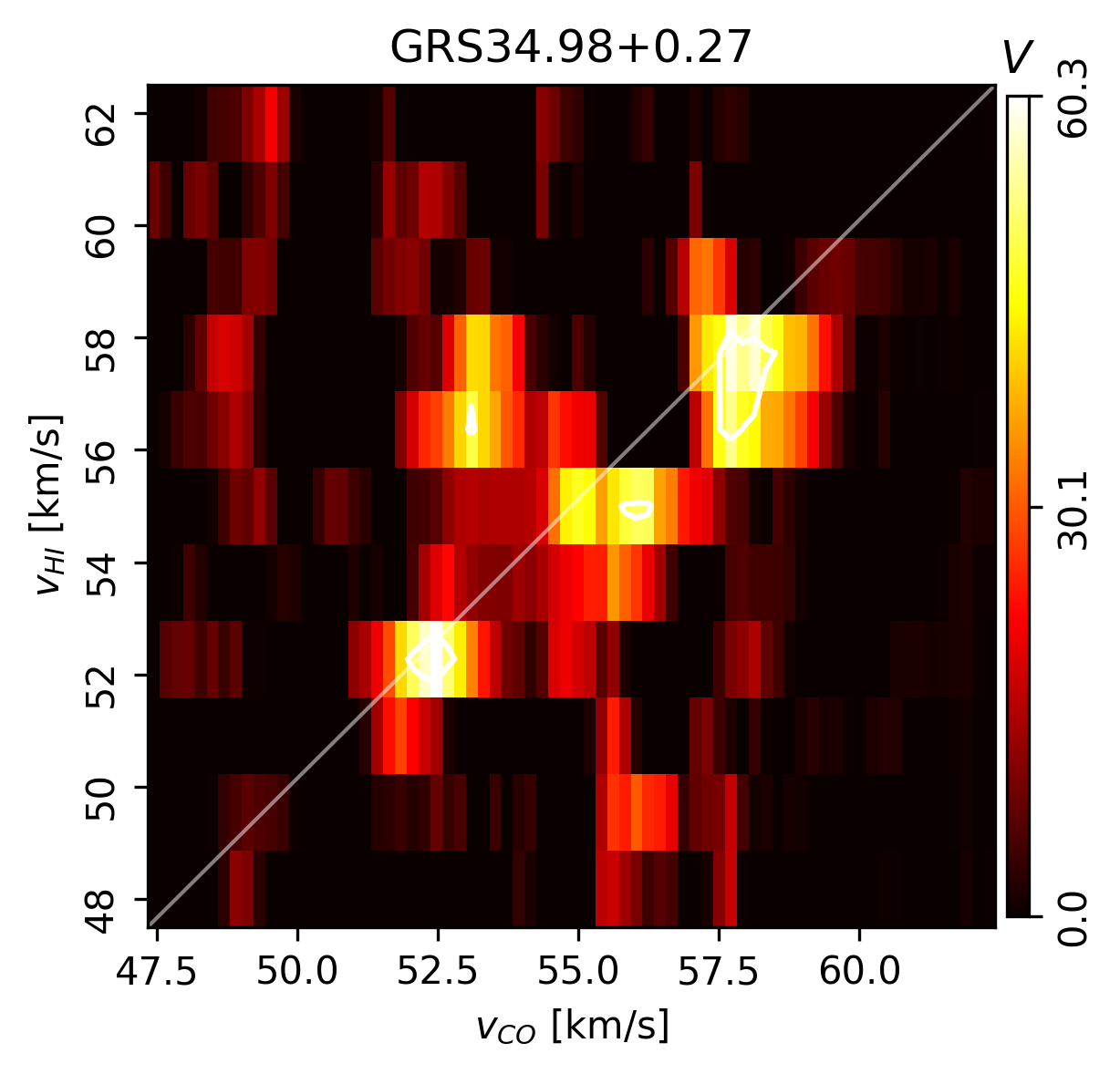}
\includegraphics[width=0.25\textwidth,angle=0,origin=c]{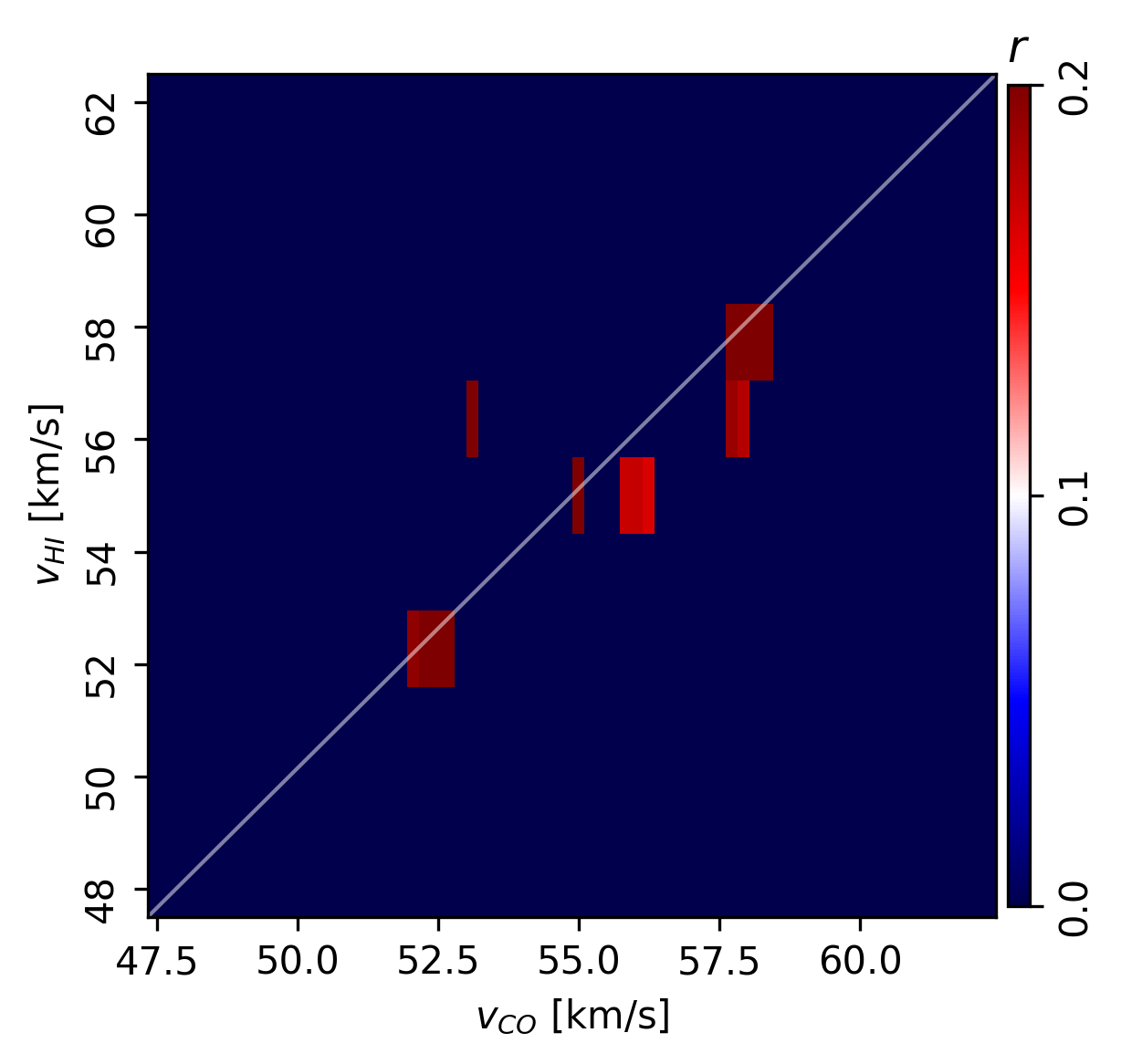}
}
\centerline{
\includegraphics[width=0.25\textwidth,angle=0,origin=c]{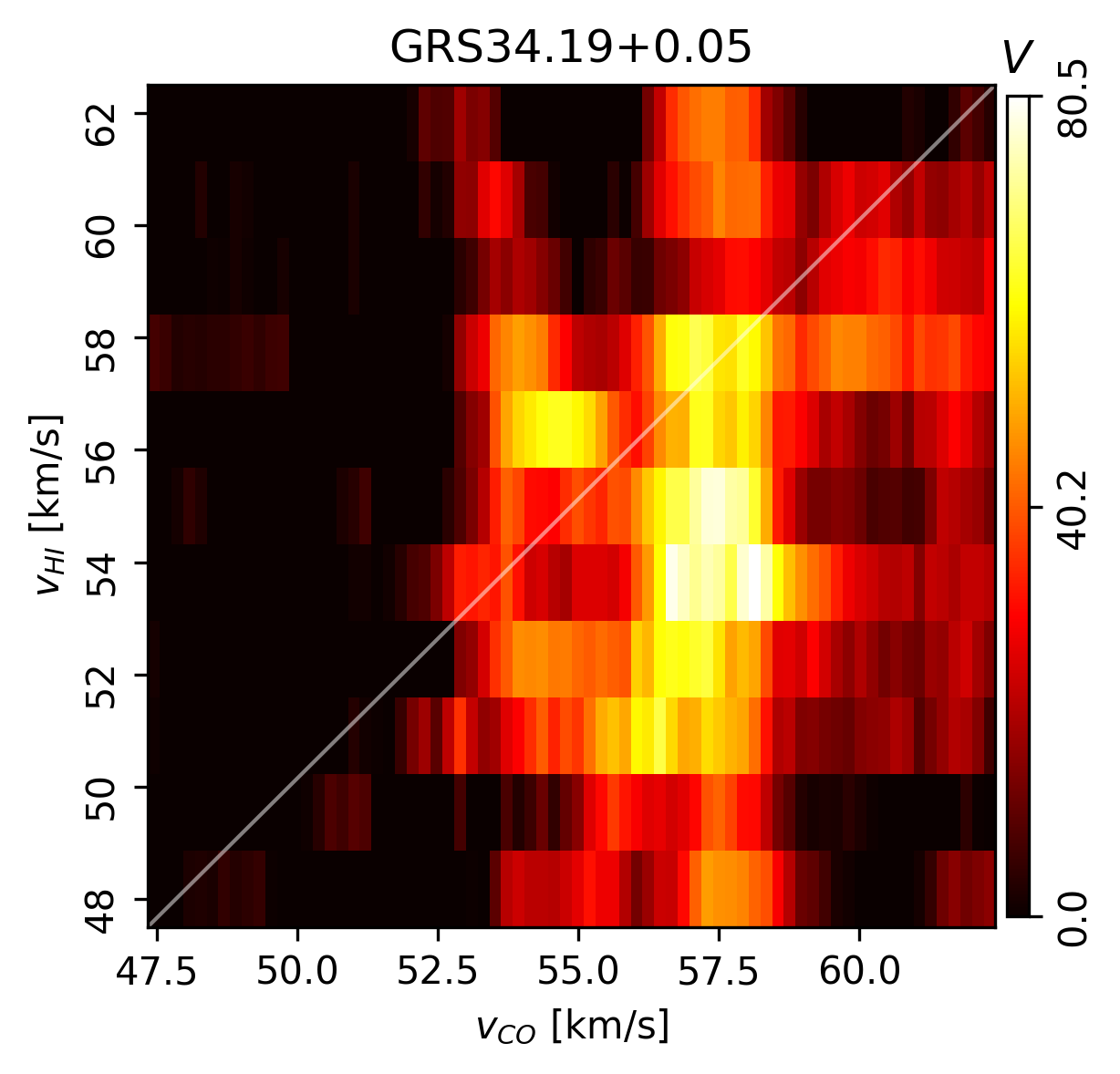}
\includegraphics[width=0.25\textwidth,angle=0,origin=c]{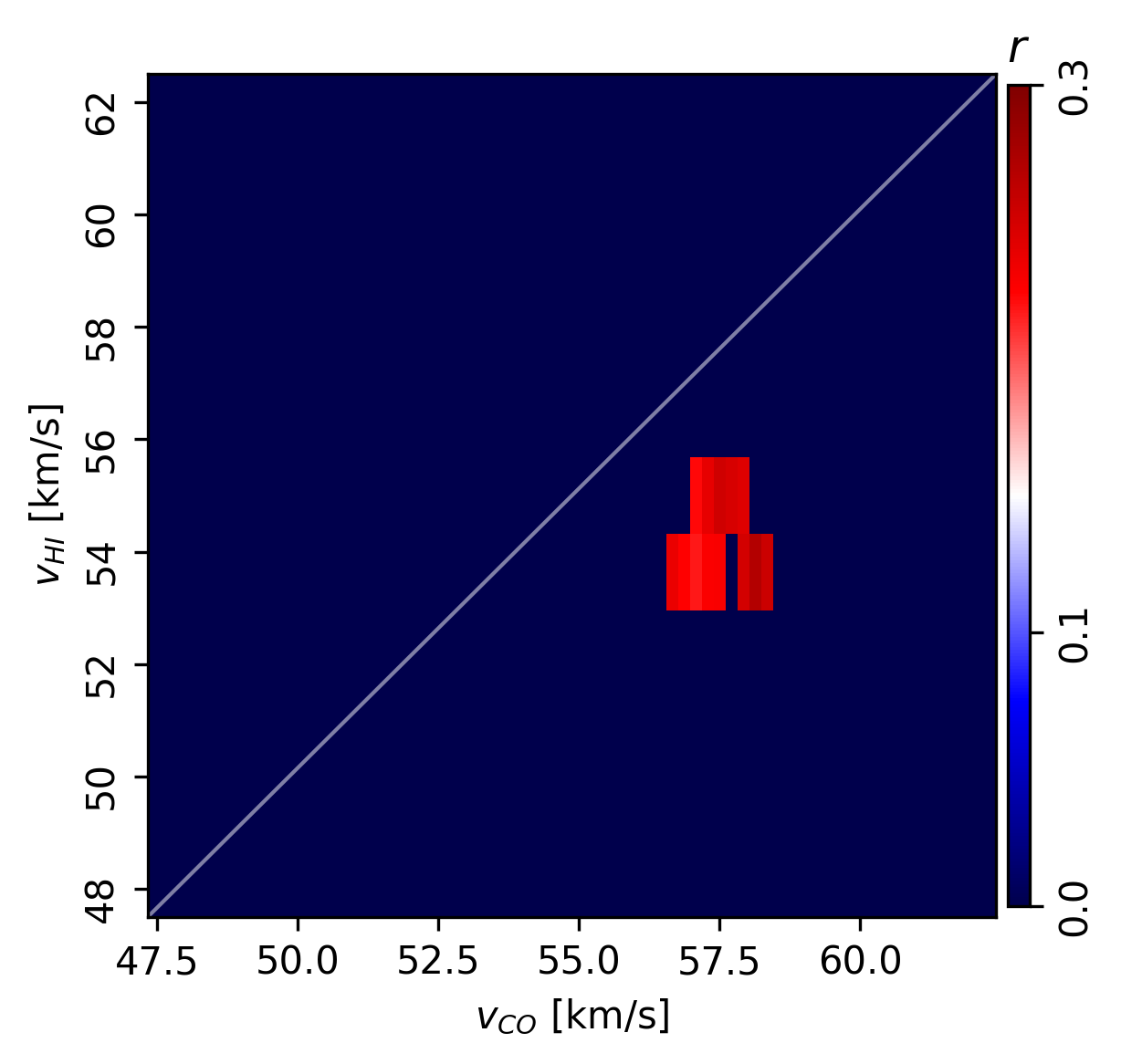}
}
\vspace{-0.1cm}
\caption{
Same as Fig.~\ref{fig:corrPlane47to62} for MC candidates GRS34.47$-$0.67, GRS34.81$+$0.3, GRS34.98$+$0.27, and GRS34.19$+$0.05.
}
\label{fig:corrPlane47to62HOGtowardsMCs}
\end{figure}

The correlation plane corresponding to the MC candidates GRS34.81$+$0.3, GRS34.98$+$0.27, and GRS34.19$+$0.05 show that the concentration of significantly high \prs\ values along \vhi\,$\approx$\,\vco\ is not a general trend.
For example, toward G34.81$+$0.3 and G34.98$+$0.27 the \prs\ values, presented in Fig.~\ref{fig:corrPlane47to62HOGtowardsMCs}, are large around \vhi\,$\approx$\,\vco, but also around \vco\,$\approx$\,52.5 and \vhi\,$\approx$\,57\,\kps.
The latter implies morphological correlation in the distribution of the emission in channels maps with a velocity offset of few kilometers per second.
This velocity offset does not necessarily imply the flow of one tracer with respect to the other, as discussed in Sec.~\ref{section:mhd}, but it does suggest a dynamic behavior beyond that described by the colliding clouds.

Even more interestingly, the HOG analysis toward GRS34.19$+$0.05 presented in Fig.~\ref{fig:corrPlane47to62HOGtowardsMCs} shows large \prs\ values distributed across a broad range of velocities, thus implying morphological correlations in velocity channels separated by up to a few kilometers per second.
This behavior is not entirely unexpected if we consider that GRS34.19$+$0.05 contains the G34.256+0.136 H{\sc ii} region at \vlsr\,$\approx$\,54\,\kps\ that extends across an area of approximately 3.4\arcmin\ in diameter \citep{kuchar1997,kolpak2003,anderson2014}. 
At glance, one could explain it by considering the H{\sc i} absorption toward the H{\sc ii} region that is present over a range of velocities, but this would only produce a vertical stripe in the distribution of $V$, that is high $V$ values for a broad range of \vhi\ and a narrow range of \vco.

It is plausible that the energy injection from the H{\sc ii} region into the surrounding $^{13}$CO and H{\sc i} can produce the high $V$ values in a broad range of \vhi\ and \vco, by contrast, a region like GRS34.47$-$0.67 lacks an embedded energy source and shows high $V$ values only around \vhi\,$\approx$\,\vco.
Molecular candidates GRS34.81$+$0.30 and GRS34.90$+$0.28 are also in the vicinity of H{\sc ii} regions in the right velocity range, in this case G035.0528$-$00.5180 and G035.1992-01.7424 \citep{lumsden2013}, yet their distribution of $V$ values across \vhi\ and \vco\ is not as broad as in GRS34.19$+$0.05. 
The study of dedicated MHD simulations of the impact of H{\sc ii} regions in a MC \citep[see for example,][]{geen2017,kimANDostriker2018} is necessary to unambiguously describe the imprint of this kind of feedback in the HOG correlation. 

\subsubsection{Is the spatial correlation between H{\sc i} and CO related to H{\sc i} self-absorption?}\label{sec:HOGandHISAs}

The distribution of H{\sc i} and $^{13}$CO intensities shown in Fig.~\ref{fig:HIv47.5to62.5} and Fig.~\ref{fig:HOGanalysis47to62} suggests that the high $V$ values mostly correspond to the spatial correlation between $^{13}$CO emission and the contours of regions with a relative decrease in the H{\sc i} intensity, which would be produced by HISA.
To further explore this possibility, we consider the H{\sc i} and $^{13}$CO spectra toward the MC candidates GRS34.47$-$0.67, GRS34.19$+$0.05, GRS34.81$+$0.3, and GRS34.98$+$0.27.

\begin{figure}[ht!]
\centerline{
\includegraphics[width=0.25\textwidth,angle=0,origin=c]{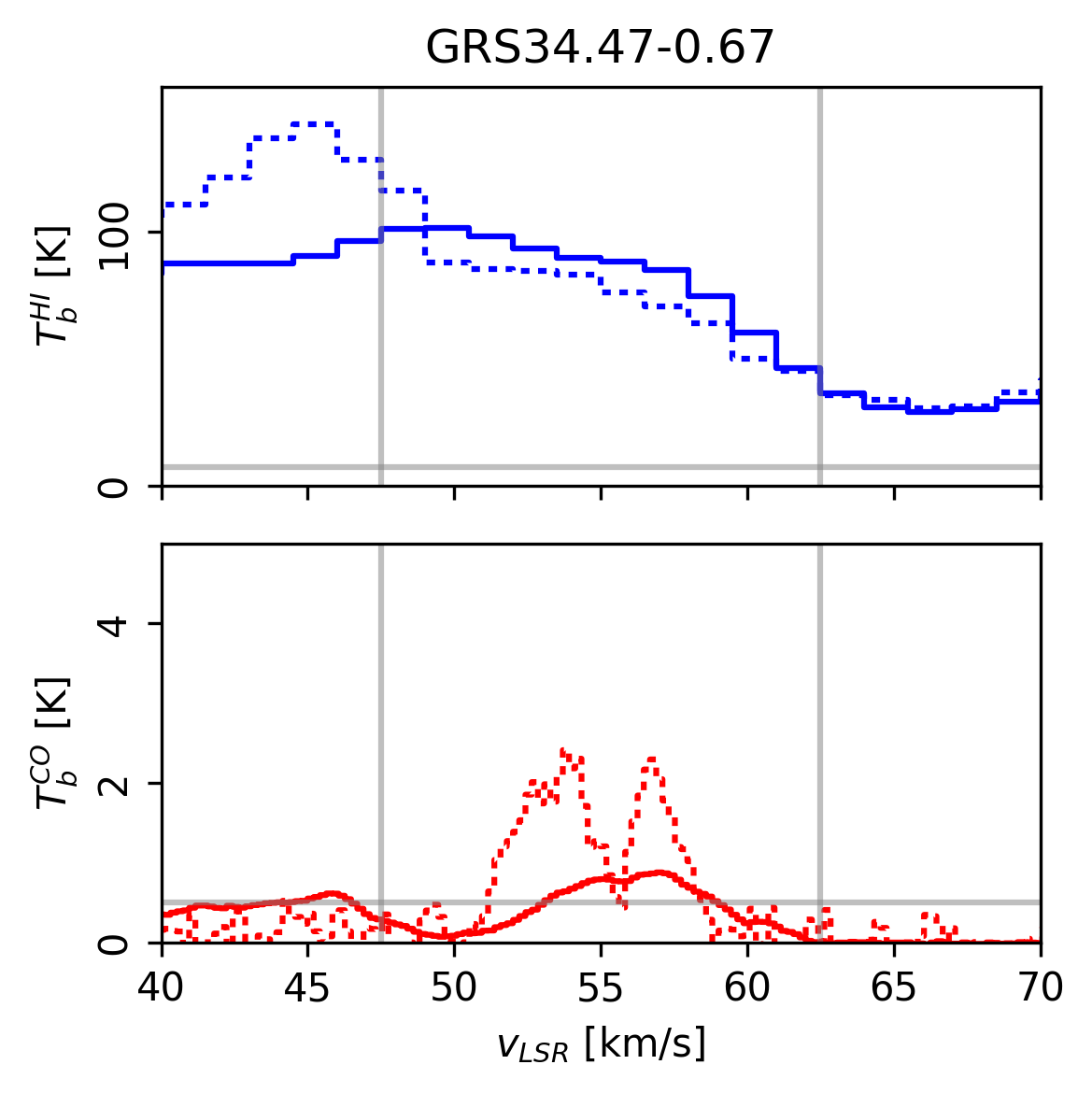}
\includegraphics[width=0.25\textwidth,angle=0,origin=c]{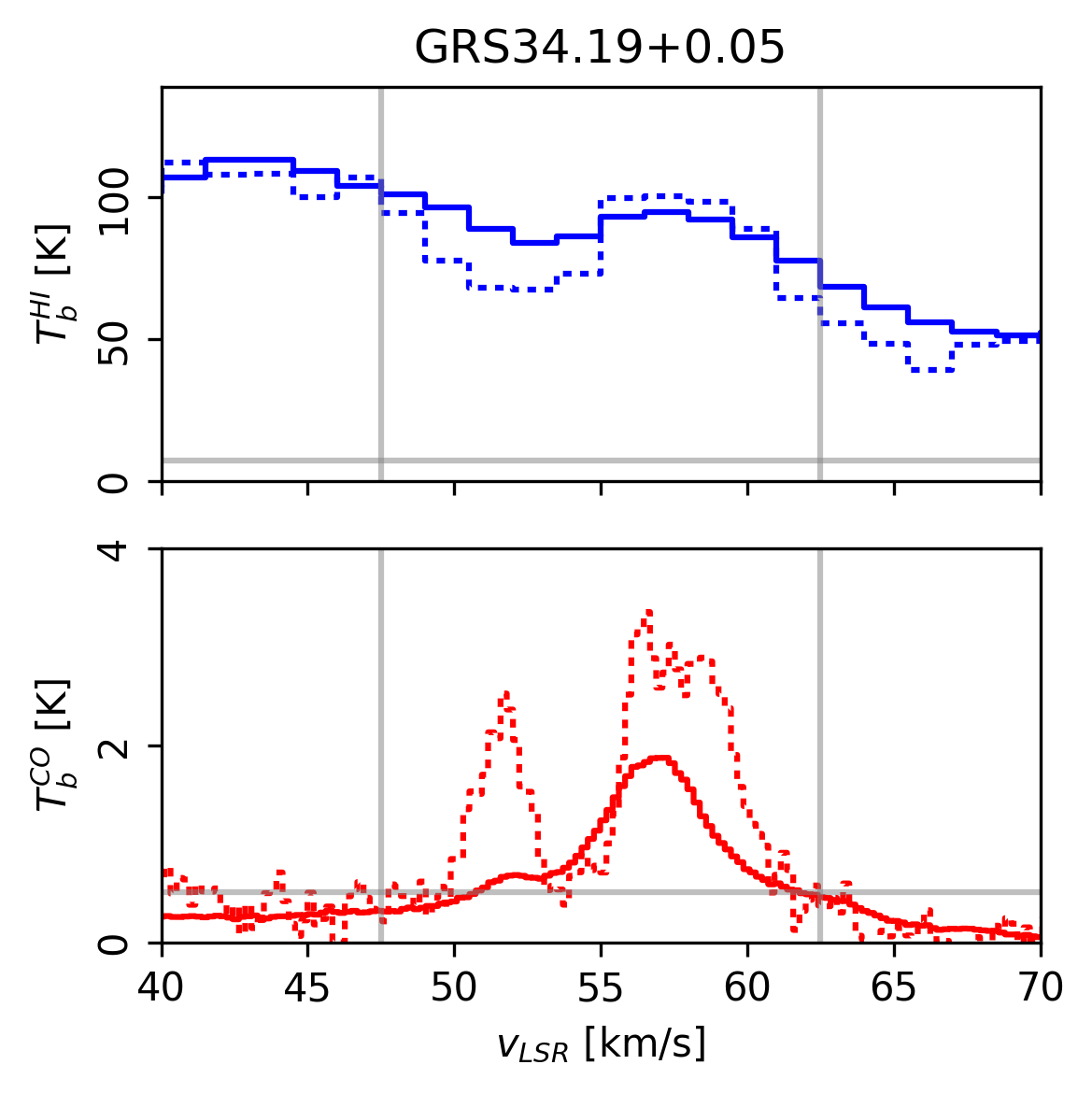}
}
\centerline{
\includegraphics[width=0.25\textwidth,angle=0,origin=c]{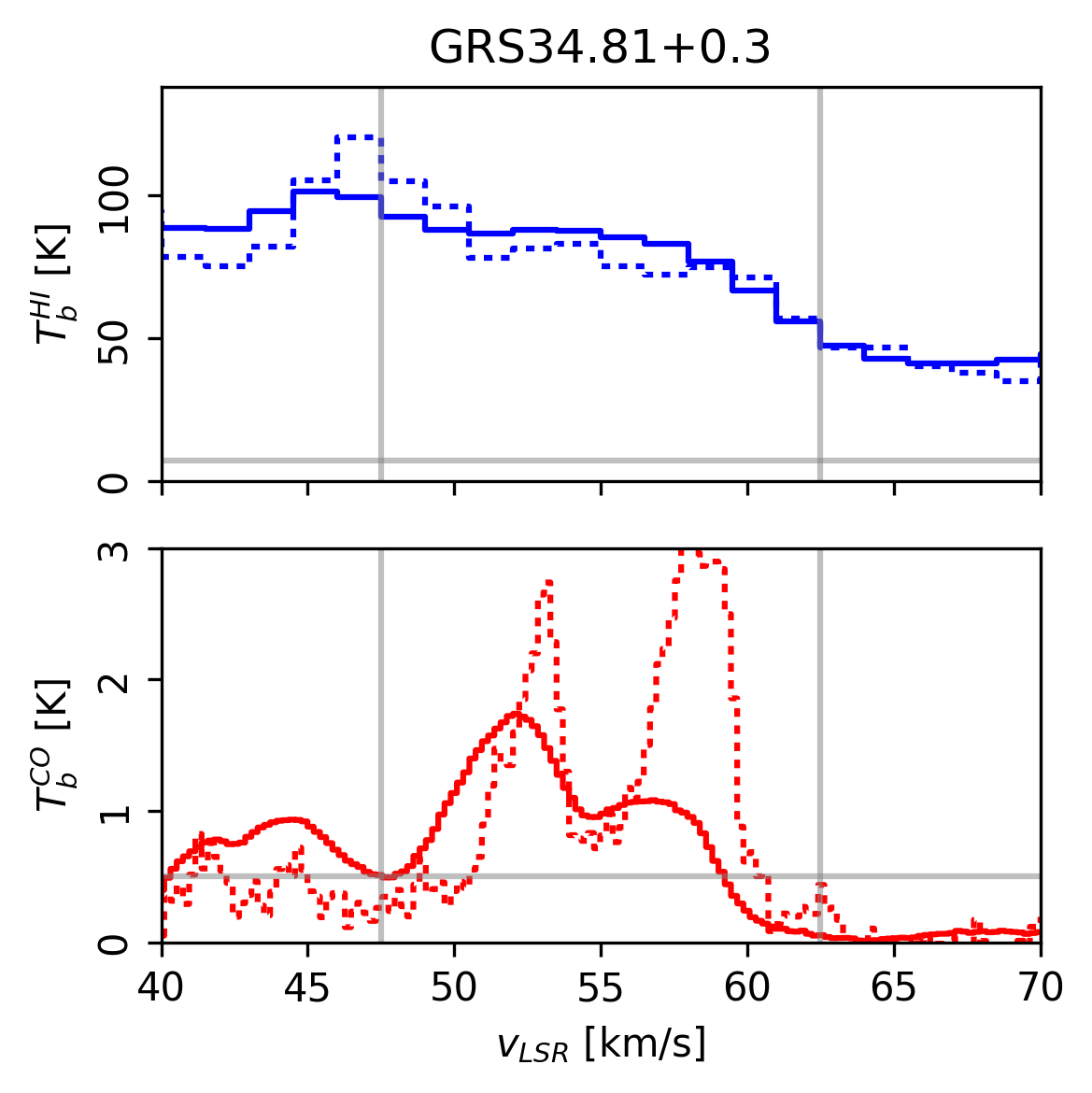}
\includegraphics[width=0.25\textwidth,angle=0,origin=c]{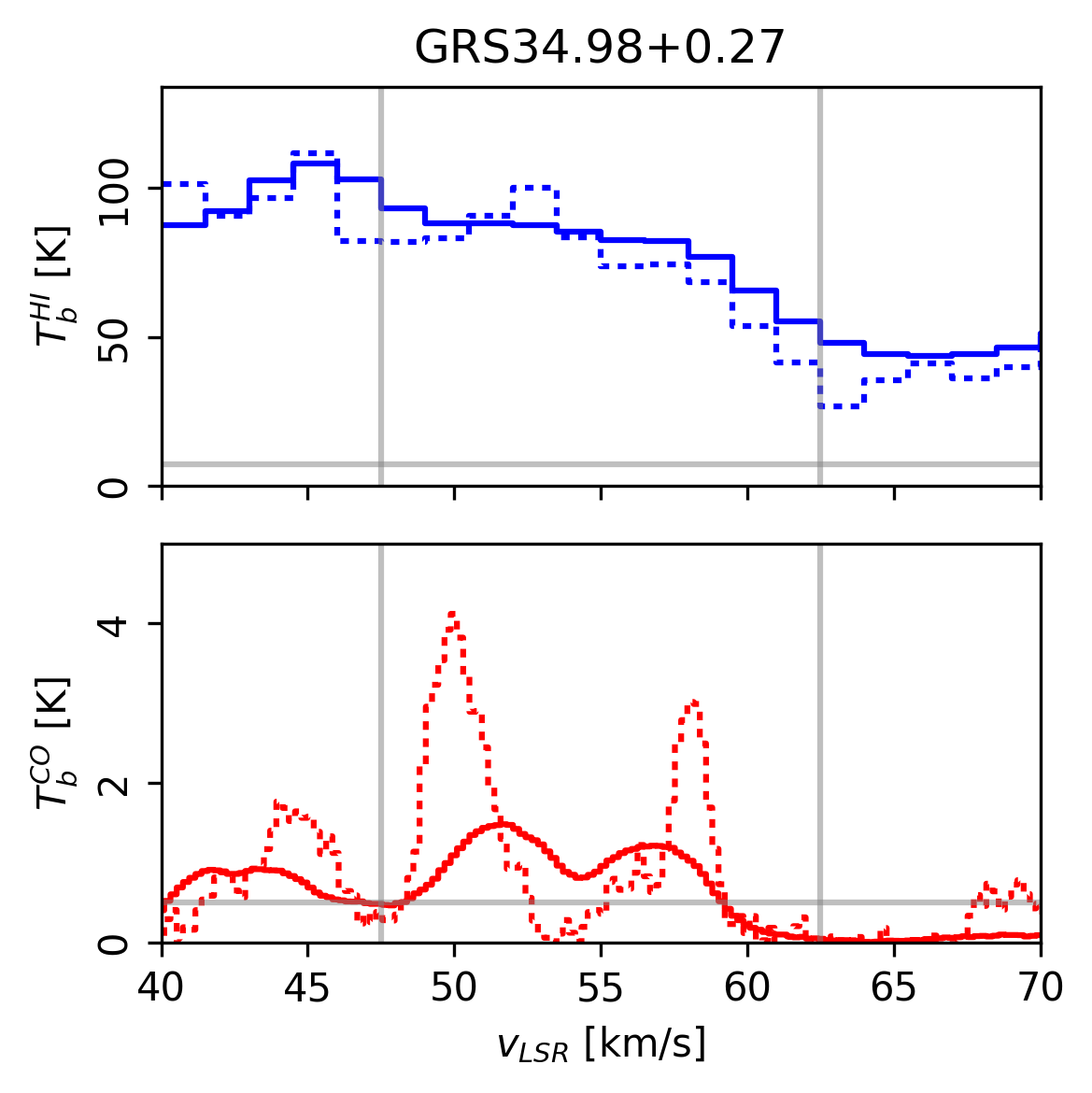}
}
\caption{H{\sc i} and $^{13}$CO spectra toward \cite{rathborne2009} MC candidates GRS34.47$-$0.67, GRS34.19$+$0.05, GRS34.81$+$0.3, and GRS34.98$+$0.27.
The solid and dotted lines represent the mean spectra over their effective area and toward the central position of the MC candidate, respectively.
}\label{fig:corrSpectra}
\end{figure}

These spectra, presented in Fig.~\ref{fig:corrSpectra}, suggest that toward the GRS34.19$+$0.05 and GRS34.98$+$0.27 MC candidates there are dips in the H{\sc i} emission around 50\,\kps\ that can be associated with the $^{13}$CO emission.
\juan{Closer evaluation of the spectra toward these regions indicates that they correspond to HISA \citep{bihrPhDT2016,wangInPrep}.}
However, the same is not true for GRS34.47$-$0.67 and GRS34.81$+$0.3, where the peaks in $^{13}$CO spectra do not seem associated with a decrease in the H{\sc i} that can be \juan{readily} identified as HISA.
\juan{It is possible that the CNM, which can be spatially correlated with the $^{13}$CO towards those two regions, does not have enough contrast with the hotter H{\sc i} background to produce a clearly identifiable HISA feature in the spectra.
But it is also possible that there is a spatial correlation between the $^{13}$CO and the thermally unstable H{\sc i}, which does not produce HISA features, as it is shown in the synthetic observations presented in App.~\ref{app:MHDsimsClark}.}

\subsection{HOG correlation at large separations between \vhi\ and \vco}\label{subsection:offdiag}


Fig.~\ref{fig:corrPlane} shows that the most significant spatial correlation revealed by the HOG technique appears at \vhi\,$\approx$\,\vco. 
However, there is a substantial signal in both $V$ and $r$ in velocity channels separated by tens of kilometers per second, for example, around \vco\,\,$\approx$\,10\,\kps\ and 
60\,$\lesssim$\,\vhi\,$\lesssim$\,100\,\kps\ and \vco\,\,$\approx$\,50\,\kps\ and 0\,$\lesssim$\,\vhi\,$\lesssim$\,40\,\kps.
To explore the origin of these features, we consider the distribution of the gradients and relative orientation angles in H{\sc i} and $^{13}$CO velocity-channel pairs with high $V$ that are separated by a few tens of kilometers per second.

\begin{figure*}[ht!]
\centerline{
\includegraphics[height=3.0in,angle=0,origin=c]{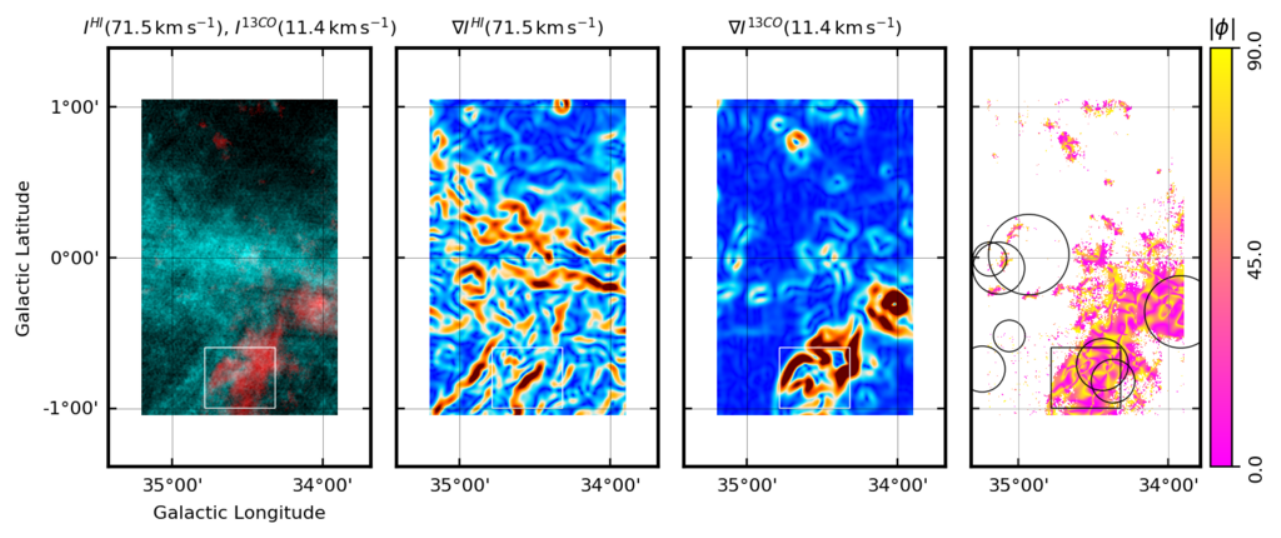}
}
\vspace{-0.2cm}
\centerline{
\includegraphics[height=3.0in,angle=0,origin=c]{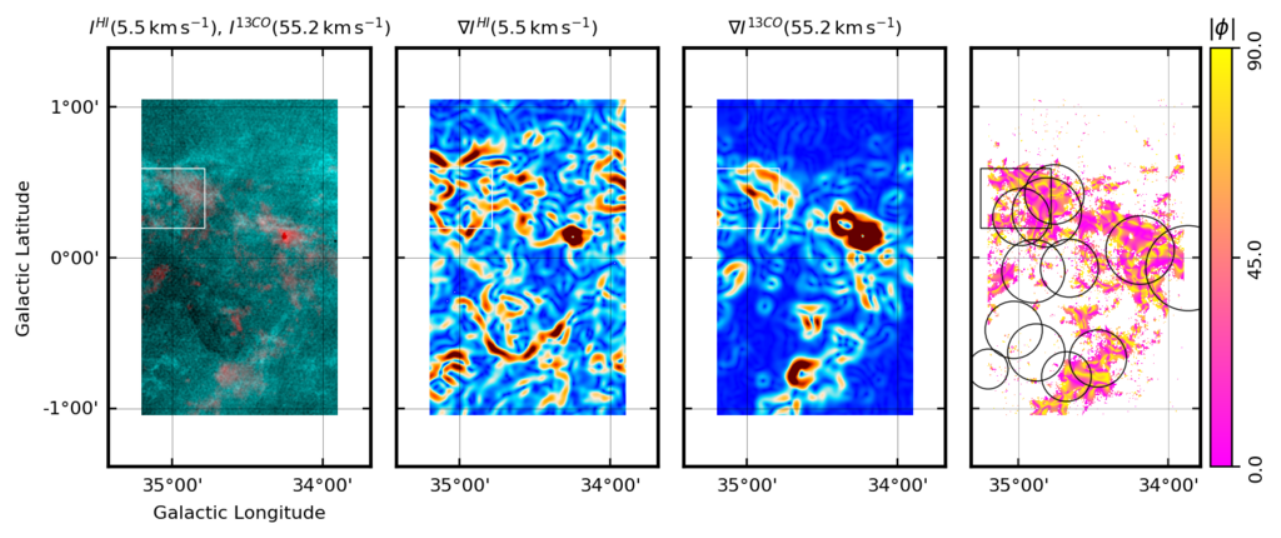}
}
\caption{
Same as Fig.~\ref{fig:HOGanalysis-5to30} for two pairs of H{\sc i} and $^{13}$CO velocity channels maps with high spatial correlation, as inferred from the $V$ values in Fig.~\ref{fig:corrPlane}, but large difference between the velocities \vhi\ and \vco.
}
\label{fig:HOGanalysisOffDiagonal}
\end{figure*}

The gradients in the velocity-channel maps corresponding to \vhi\,$=$\,71.5 and \vco\,\,$=$\,11.4\,\kps, presented in the top panel in Fig.~\ref{fig:HOGanalysisOffDiagonal}, indicate that there is indeed some extended correlation in the spatial distribution of both tracers around $l$\,$\approx$\,34\pdeg5 and $b$\,$\approx$\,$-$1\pdeg0.
In this particular case, the $^{13}$CO distribution seems to be associated with some elongated H{\sc i} features oriented at roughly 45\deg\ with respect to the vertical direction.
Similarly, the velocity-channel maps corresponding to \vhi\,$=$\,5.5 and \vco\,\,$=$\,55.2\,\kps, presented in the bottom panel in Fig.~\ref{fig:HOGanalysisOffDiagonal}, also indicate some extended correlation around $l$\,$\approx$\,35\pdeg0 and $b$\,$\approx$\,1\pdeg0.
What distinguishes this correlation from that found around \vhi\,$\approx$\,\vco\ is that in the former the high $V$ values, $V$\,$>$\,5$\varsigma_{V}$, appear just in a few scattered pairs of velocity channels.
In contrast the high $V$ values around \vhi\,$\approx$\,\vco\ appear distributed in several pairs of consecutive velocity channels. 

The presence of the vertical stripes in the distribution of $V$ indicates that there is some degree of chance correlation wherever there is significant $^{13}$CO emission, although in most cases it is below the 5$\varsigma_{V}$ confidence level.
This correlation is distributed over a broad range of H{\sc i} velocity channels due to the fact that there is H{\sc i} extended structure in all of them, thus increasing the amount of chance correlation with the $^{13}$CO emission.
This conclusion is confirmed by the presence of similar vertical stripes in the null tests introduced in App.~\ref{app:StatSignificance}, where the values of $V$ can only be the result of chance correlation.

\section{Discussion}\label{section:discussion}

The analysis of the H{\sc i} and $^{13}$CO observations using the histogram of oriented gradients (HOG) technique produces three main results that we discuss here.
\begin{enumerate}
\item There is a significant spatial correlation between the two tracers in extended portions of the \juan{region studied}.
\item When considering the spatial correlation revealed by the HOG technique toward particular MC candidates, we find that different clouds present substantial differences in the velocity ranges over which the HOG correlation is distributed.
\item Toward some of the MC candidates the HOG results imply a morphological correlation in the emission of the two tracers in velocity channels separated by \juan{up to} a few kilometers per second.
\end{enumerate}

\subsection{Spatial correlation of H{\sc i} and $^{13}$CO}

Using HOG, we find evidences of the spatial correlation of H{\sc i} and $^{13}$CO, or more explicitly, we find that the two tracers have coincident intensity contours traced by the orientation of their gradients.
We quantify this spatial correlation using the tools of circular statistics, namely, the projected Rayleigh statistic, \prs. 
Previous studies of the association between H{\sc i} absorption features and molecular gas have been based on the agreement between the velocities, the close agreement of non-thermal line widths, and the matching of the inferred temperatures \citep[e.g.,][]{kavars2003,liANDgoldsmith2003,barriault2010}.

In an overly simplistic model of the ISM, a spherical cloud of diffuse gas and dust in axisymmetric collapse immersed in a bath of isotropic interstellar radiation begins to form a MC when the column density gets sufficiently high that the gas/dust can self-shield, the H{\sc i} converts to H$_{2}$, and the $^{13}$CO appears toward the center.
In this toy model, the HOG correlation indicates that some of the contours of the H{\sc i} emission match with the contours of the $^{13}$CO emission, even if \juan{they} do not share a boundary in 3D.
Given that we are comparing the gradients, the HOG correlation is not directly related to the correlation or anti-correlation between the amount of atomic and molecular gas, but rather to their spatial \juan{distributions}.
For this toy model cloud it is expected that the gradients of the H{\sc i} and $^{13}$CO emission match, but this is not necessarily the case for a real MC, where the density structure is much more complex and the spectra of both tracers are affected by optical depth and self-absorption, such that even a perfect correlation between atomic and molecular hydrogen would not necessarily result in a good correlation of the H{\sc i} and $^{13}$CO maps.
However, the results of the HOG analysis reveal that this spatial correlation is present in the observations.

\subsubsection{H{\sc i} self-absorption and $^{13}$CO}

The observation of spatial correlation between H{\sc i} and $^{13}$CO has been reported in previous studies of the association of molecular gas and H{\sc i} self-absorption (HISA) features \citep{gibson2005a} and narrow H{\sc i} self-absorption (HINSA) features \citep{goldsmith2005,krco2008}.
However, it was limited by the process of identification and extraction of HISA features, which entails a particular level of complexity.
In our blind approach, the H{\sc i} contours are not particularly associated to the cold gas producing the HISAs, but are rather any contour features that characterize the map.
Then, it is convenient to discuss how an object that in principle has no defined edges, such as a cloud of gas in the ISM, can produce structures that can be identified in two different tracers.

\cite{heilesANDtroland2003b} indicate that a model of CNM cores contained in WNM envelopes, as suggested in \cite{mckeeANDostriker1977}, provides a good description of the data toward many sources.
Additionally, some of \juan{these} H{\sc i} envelopes are identified around MCs \citep[e.g.,][]{wannier1983,stanimirovic2014}.
In the turbulent ISM these different phases are not contained within each other like a matryoshka doll; there \juan{are no} clearly defined boundaries \juan{but rather} gradients that depend on the distribution of column density structure, radiation field, and spin temperature.
Those are the gradients that we consider as potentially responsible for the signal that is found using the HOG technique.

For the particular case of the comparison of H{\sc i} and $^{13}$CO, the conditions of the transition between H{\sc i} and H$_{2}$ and the relation between H$_{2}$ and $^{13}$CO that are ultimately responsible for the observed emission gradients are very hard to determine for a random MC candidate.
The HOG technique does not address the physical and chemical phenomena that produce those gradients, but rather embraces their complexity following a phenomenological and statistical approach to find out where are they coincident and what do they reveal about the MC formation process.

It is unexpected that the H{\sc i} and $^{13}$CO have a tendency to have coincident intensity contours unless these arise from regions of H{\sc i} self-absorption, as supported by the simulation analysis presented in Sec.~\ref{section:mhd}.
However, the observed spatial correlation is not exclusively related to HISA features, as \juan{shown} in Sec.~\ref{sec:HOGandHISAs}.
This indicates two possibilities: either the spatial correlations are related to self-absorption that is not evident in the central and average spectra presented in Fig.~\ref{fig:corrSpectra}, or the spatial correlation is produced by the general H{\sc i} emission.
The first possibility calls for the combination of HOG and the dedicated identification of HISA, which we will address in a subsequent publication \citep{wangInPrep}.
The second possibility implies that the interpretation of the HOG results is less simple than what is inferred from the study of the \juan{atomic-cloud-collision MHD simulations presented in Sec.~\ref{section:mhd}}.
For a given velocity channel, the H{\sc i} signal is contributed from gas parcels both within the cloud/cloud envelope and material not physically associated with the cloud but with one with broad velocity dispersion that leaks into the cloud velocity interval.
Although the study of the MHD simulations without H{\sc i} emission background, presented in App.~\ref{app:MHDsims}, shows that there is a significant level of spatial correlation between H{\sc i} and $^{13}$CO even without the explicit presence of H{\sc i} self-absorption, the general interpretation of the spatial correlation between the two tracers will have to be supported by further study of MHD simulations and synthetic observations that reproduce the HISAs better.

\subsubsection{Emission background}\label{section:discussionEmissionBackground}

\juan{In contrast to the continuum emission maps in the application of the HOG to the \emph{Planck} data \citep{soler2013,planck2015-XXXV}, the velocity channels in this analysis include a background component that is not simply the result of the integration of the emission along the line of sight.
A particular velocity channel map potentially includes contributions from structures that are not physically connected but produce emission at the same velocity, for example, emission from locations of the Galaxy that have the same \vlsr\ or from portions of unconnected expanding shells, spiral shocks, or non-circular motions near the Galactic bar.
The HOG technique evaluates the morphological correlation between the intensity maps of two tracers, independent of the physical conditions producing the observed intensity distribution in a particular velocity channel map.
In principle, it is sensitive to the chance correlation introduced by this emission background.
However, it is unlikely that this background emission from disconnected regions has a similar structure and would produce singularly high spatial correlation between the considered tracers.}

\juan{In the \cite{reid2014} spiral arm model A5 around Galactic longitude $l$\,$\approx$\,34\pdeg5, the velocities \vlsr\,$\approx$\,12, 42, and 54\,\kps\ correspond to kinematic distances of roughly 0.78\,$\pm$\,0.45, 2.57\,$\pm$\,0.37, and 3.21\,$\pm$\,0.36\,kpc in the near side of the Galaxy and approximately 12.67\,$\pm$\,0.46, 10.97\,$\pm$\,0.37 and 10.36\,$\pm$\,0.35\,kpc in the far side\footnote{BeSSeL survey revised kinematic distance calculator http://bessel.vlbi-astrometry.org}. 
These large differences between the near and far distances make it unlikely that the morphological correlations between H{\sc i} and $^{13}$CO structures identified in the HOG analysis around those velocities are significantly affected by emission from the other side of the Galaxy.
For the same Galactic longitude, the estimated gap between the near and far kinematic distances is lower for larger \vlsr, for example, it is around 1.45\,kpc for \vlsr\,$\approx$\,100\,\kps\ and close to zero close to the tangent point, at roughly \vlsr\,$\approx$\,120\,\kps.
However, it is difficult to assess if the lack of HOG correlation at \vlsr\,$>$\,90\,\kps\ can be entirely attributed to the blending of density structures into the same velocity range.}

\juan{If we consider a CO cloud with line-of-sight velocity (LOS) $v_{0}$ located directly in front of an expanding H{\sc i} shell with mean LOS velocity $v_{1}$ and expansion velocity $v_{e}$, the value of $V$ corresponding to the spatial correlation between the emission of the two tracers at $v_{0}$ would not exclusively be that of the CO cloud and its atomic envelope, but would also include the emission from the portion of the shell moving at $v_{0}$\,=\,$v_{1}+v_{e}$.
If the H{\sc i} shell is spatially disconnected from the CO cloud, there is no reason why the spatial distribution of the its H{\sc i} emission at $v_{0}$ should be correlated with the CO emission and its contribution to the estimated values of $V$ is that of chance correlation.
This chance correlation is well exemplified toward W44, where the expansion of the supernova remnant potentially contributes to the H{\sc i} emission over a broad range of velocity channels, approximately 10\,$<$\,\vlsr\,$<$\,45\,\kps\ as inferred from Fig.~\ref{fig:HIandCOlvdiagrams}, but there is not an exceptionally high spatial correlation with the $^{13}$CO emission in that velocity range, as shown in Fig.~\ref{fig:corrPlane}.}

\subsection{The H{\sc i} and $^{13}$CO correlation in different environments}

When separating the region in individual MC candidates we find three interesting cases in terms of the spatial correlation inferred from $V$, all illustrated in Fig.~\ref{fig:corrPlane47to62HOGtowardsMCs}.
First, MCs where the H{\sc i} and $^{13}$CO emission appear correlated at roughly the same velocities.
Second, clouds that show correlation around \vhi\,$\approx$\,\vco\ and also correlation in some H{\sc i} and $^{13}$CO velocity channels separated by a few kilometers per second.
Third, clouds that \juan{show} correlation \juan{between} H{\sc i} and $^{13}$CO in many velocity channels distributed on a broad velocity range.
Only the first case is arguably consistent with the synthetic observations of the \cite{clarkInPrep} colliding flows simulation.

The GRS34.19$+$0.05 MC candidate presents high \prs\ values close to \vhi\,$\approx$\,\vco. 
This trend is very similar to that found in the synthetic observations presented in Sec.~\ref{section:mhd}, however, it does not necessarily imply that this specific configuration corresponds to the physics responsible for the observed values.
\juan{In principle, this spatial correlation is expected if the atomic and molecular gases are both cospatial and comoving.}
\juan{The spatial correlation, illustrated in Fig.~\ref{fig:HOGanalysis47to62}, corresponds to the $^{13}$CO emission associated with a relative decrease in the H{\sc i} intensity, which is most likely produced by HISA \citep{bihrPhDT2016,wangInPrep}, thus suggesting that the observed correlation corresponds to that between the molecular gas sampled by $^{13}$CO and the CNM.}

The MC candidates GRS34.81$+$0.3 and GRS34.98$+$0.27 present high \prs\ values close to \vhi\,$\approx$\,\vco, but they also show significant \prs\ values around \vco\,$=$\,52.5 and \vhi\,$=$\,56\,\kps.
This significant velocity offset is not reproduced by the synthetic observations presented in Sec.~\ref{section:mhd}, although velocity offsets have been traditionally associated with relative motions between the tracers \citep[e.g.,][]{motte2014}.
One \juan{possible explanation to this observation is the potential superposition of clouds along the line of sight} \citep{beaumont2013}.
\juan{However, it is unlikely that spatially separated parcels of H{\sc i} can have such a high spatial correlation with the same $^{13}$CO cloud.
Another possibility is that the H{\sc ii} regions introduce a velocity offset between the dense molecular gas and the less dense atomic gas.}
A final possibility is that more general conditions than those in the Sec.~\ref{section:mhd} synthetic observations can produce this trend. 
We discuss the latter two possibilities in more detail in Sec.~\ref{section:discussion3}.

The deviation from the clustering of high \prs\ values around \vhi\,$\approx$\,\vco\ is more evident toward GRS34.19$+$0.05.
There, the broad range of velocities with large \prs\ can in principle be \juan{related to} the effects of the H{\sc ii} regions G34.256+0.136 and G34.172+0.175 and the presence of infrared bubbles \citep{churchwell2006,xu2016}.
It is worth noting that observationally, the MCs are arbitrarily defined identities, the spatial and velocity associations of $^{13}$CO that we call MCs may not correspond to an individual objects with well defined boundaries.
So, in general terms, what we are finding with the HOG is that the proximity of H{\sc ii} regions or the relative isolation is related to different behaviors of the spatial correlation sampled by $V$, and not necessarily that there are two types of MCs in the catalog.

\subsection{Potential causes of the H{\sc i} and $^{13}$CO velocity offsets}\label{section:discussion3}

\subsubsection{Cloud evolution}

\juan{Ionizing radiation from high-mass stars creates H{\sc ii} regions, while stellar winds and supernovae drive the matter in star-forming MCs into thin shells.
These shells are accelerated by the combined effect of winds, radiation pressure, and supernova explosions \cite[see][and references there in]{rahner2017}.
Under the influence of the wind responsible for the shell expansion, the surrounding gas is accelerated, but the less dense atomic gas is accelerated more so that over time, a velocity difference is accumulated between it and the molecular gas \citep{pound1997,pellegrini2007}}.

\juan{We considered this scenario of cloud evolution in the study of the THOR data towards the W49A region, where we found that cloud structure and dynamics of the region are in agreement with a feedback-driven shell that is re-collapsing due to the gravitational attraction \cite{rugel2018}.} 
\juan{However, this is the first study where we include the atomic gas that is associated to the star-forming cloud}.
\juan{Potentially, the velocity separation between spatially-correlated H{\sc i} and $^{13}$CO channel maps can be used to study the energy input from H{\sc ii}, but fully exploring that possibility requires additional analysis of models and MHD simulations that are beyond the scope of this work}.

\subsubsection{Cloud formation}

Our analysis of MHD simulations, presented in Sec.~\ref{section:mhd}, suggests that the ideal head-on collision of atomic clouds does not reproduce the velocity offset between the velocity channels with high spatial correlation revealed by the HOG analysis.
However, it is expected that more general \juan{MC-formation} conditions; such as not-head-on collisions, Galactic shear, and different mean magnetic field orientation with respect to the collision axis; could produce different correlations between the atomic and the molecular emission.
Indeed, numerical studies of the thermally bistable and turbulent atomic gas show that once formed, the CNM gas is dynamically stable and individual CNM structures have supersonic relative motions that are related to the dynamic of the WNM \citep{heitsch2006,hennebelle2007,saury2014}.
For example, the presence of the magnetic field would impose an anisotropy in the flows and if two fronts of atomic gas were not directed parallel to the magnetic field lines, they would have to re-orient themselves and the accumulation of dense gas can appear at a different velocity with respect to the flow of gas that is producing it \citep{hennebelle2000,hartmann2001, soler2017b}.
In a similar way, the Galactic shear, the spiral-arm gravitational potential, or simply the angle between the shock fronts of gas pushed by the ram pressure of supernovae can produce anisotropies that could potentially lead to velocity offsets observed between the atomic and molecular tracers.

In order to test the aforementioned hypothesis, we performed a quick experiment in a segment of one of the stratified, supernova-regulated, 1\,kpc-scale, magnetized ISM magneto-hydrodynamical simulations presented in \cite{hennebelle2018}.
These simulations trace the evolution of the supernova-regulated multi-phase ISM and, although they do not explicitly estimate the formation of molecular gas, they provide self-consistent initial conditions for the dynamics of the bistable atomic gas.
In this simulation, the presence of multiple shock fronts produced by the supernovae explosions makes it extremely unlikely that the accumulation of the dense gas, which can potentially become a MC, is the result of just one collision of atomic flows or the isotropic collapse into one gravitational potential well.
In that sense, the accumulation of dense gas in this simulated volume represents a MC formation scenario that is less dependent on the initial conditions of the simulation.

We selected a (20\,pc)$^{3}$ volume around a density structure identified using a friend-of-friends (FoFs) algorithm with a threshold density $n_{C}$\,$=$\,$10^{3}$\,cm$^{-3}$.
Although the FoFs algorithm is not optimal for the general selection of connected structures, in this case we simply use it to identify a reference parcel of gas.
\juan{We applied the HOG technique to synthetic observations of H{\sc i} and $^{13}$CO emission produced using simple density and temperature thresholds, which is not an optimal approach but it is sufficient for our quick experiment}.
We refer to App.~\ref{app:FRIGGsims} for further details on these synthetic observations. 

\begin{figure}[ht!]
\centerline{
\includegraphics[width=0.25\textwidth,angle=0,origin=c]{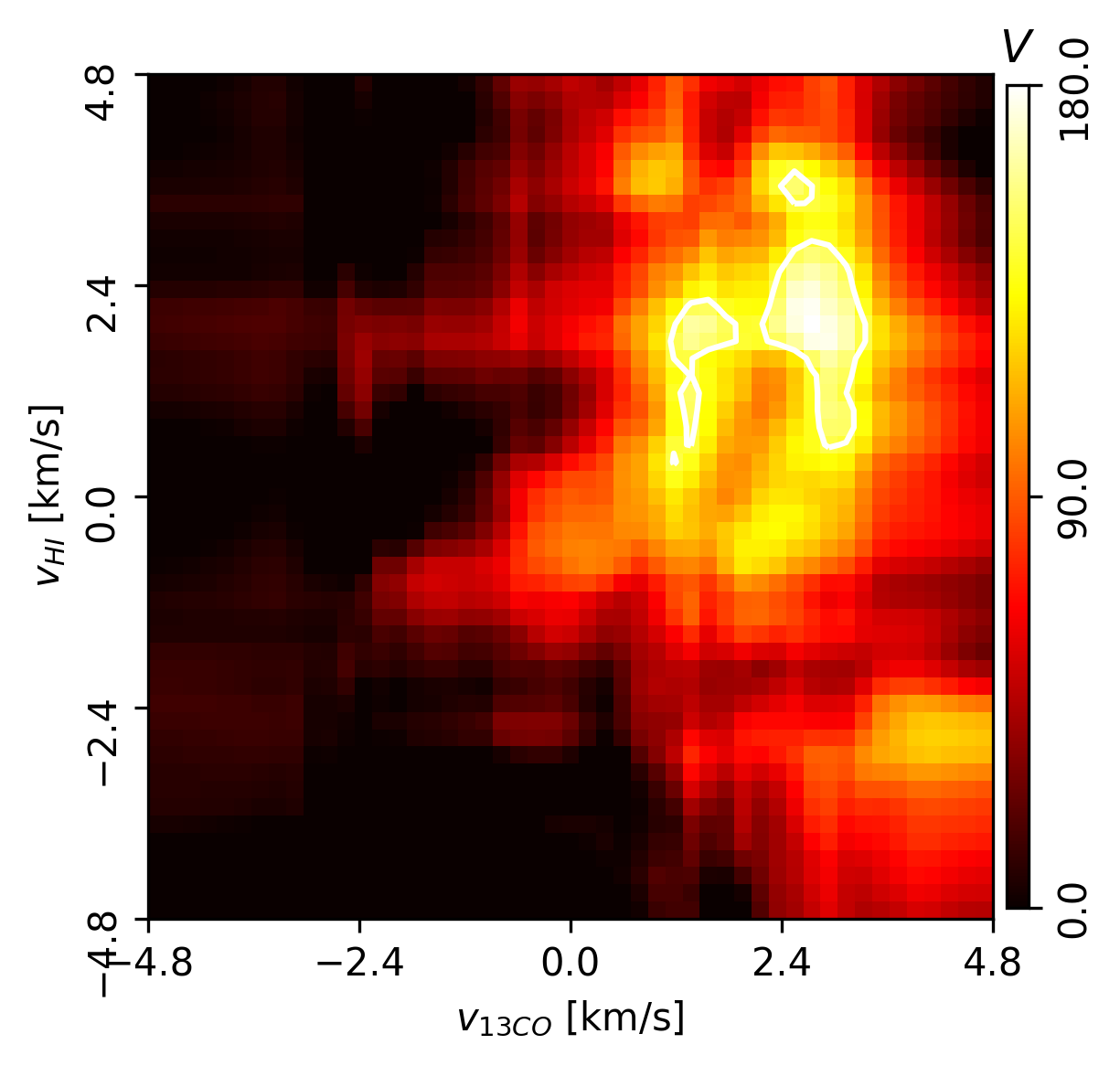}
\includegraphics[width=0.25\textwidth,angle=0,origin=c]{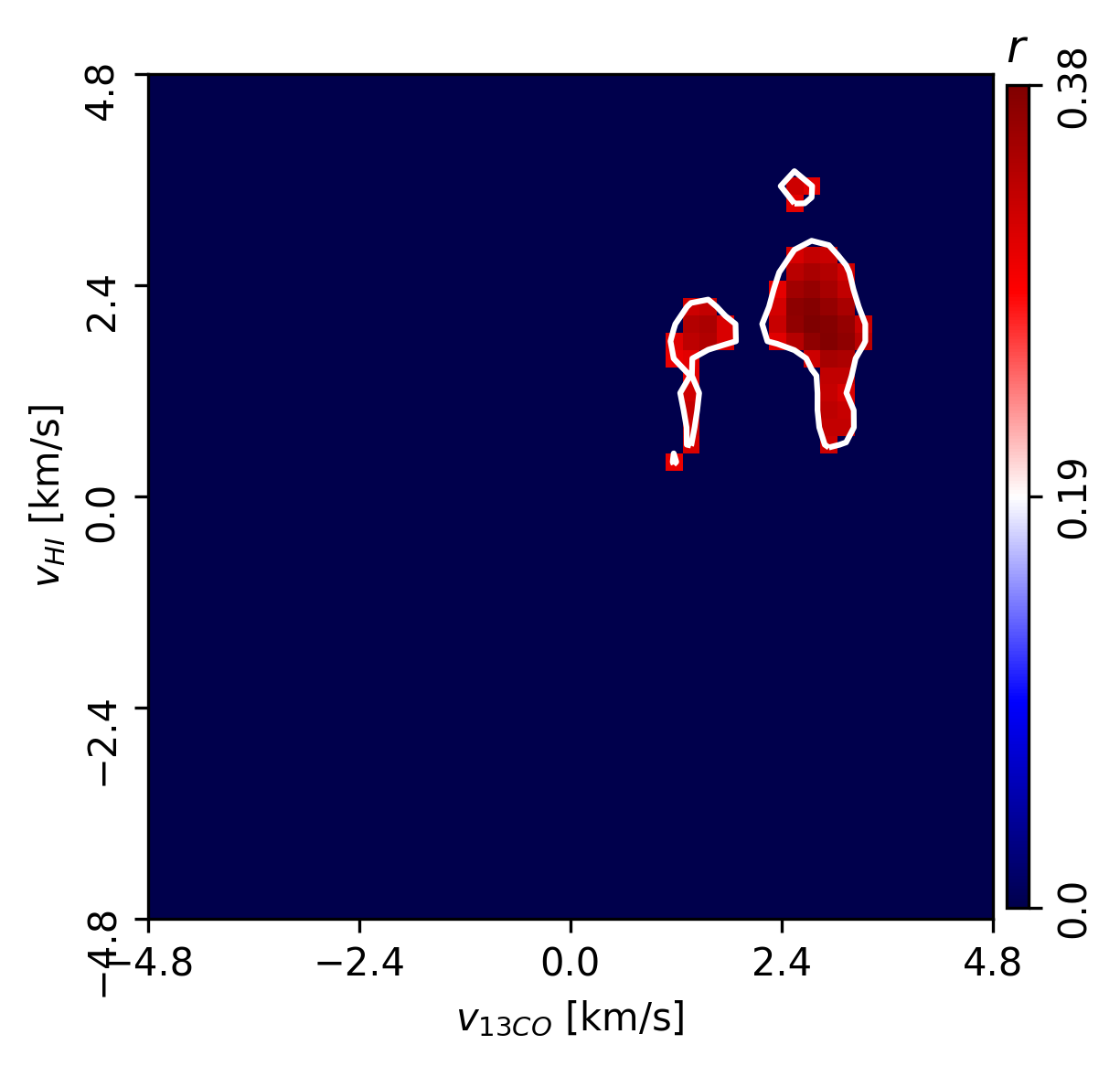}
}
\vspace{-0.1cm}
\caption{
Results of the HOG analysis of the H{\sc i} (left) and the $^{13}$CO synthetic observations of a segment of the 1-kpc stratified box MHD simulations presented in \cite{hennebelle2018} and detailed in Appendix~\ref{app:FRIGGsims}.
\emph{Left.} Projected Rayleigh statistic, $V(v_{\rm 13CO},v_{\rm HI})$, the HOG statistical test of spatial correlation between H{\sc i} and $^{13}$CO velocity-channel maps, defined in Eq.~\eqref{eq:myprs}.
\juan{The contours indicate the 3$\varsigma_{V}$, 4$\varsigma_{V}$, and 5$\varsigma_{V}$ levels in the corresponding velocity range.}
\emph{Right.} Mean resultant vector length, $r(v_{\rm 13CO},v_{\rm HI})$, with the 3$\varsigma_{V}$ confidence interval, a HOG metric that is roughly equivalent to the percentage of gradient pairs that imply the spatial correlation between the velocity-channel maps, defined in Eq.~\eqref{eq:mymrv}.
}
\label{fig:corrPlaneFRIGG}
\end{figure}

The results of the HOG analysis, \juan{shown} in Fig.~\ref{fig:corrPlaneFRIGG}, indicate not only the spatial correlation in velocity channels \vhi\,$\approx$\vco, but also a significant correlation in H{\sc i} and $^{13}$CO velocity channels separated by a few kilometers per second.
These offsets are persistent for roughly 10$^{5}$ years in the simulation and change throughout the evolution of the region.
Their presence alone does not clarify the origin of the offsets seen in the analysis of the observations, but suggests that HOG can potentially constitute a good metric for the study of the cloud evolution and formation in numerical simulations. 
A detailed study of the prevalence of these trends and the physical conditions that produce it in this particular set of MHD simulations is beyond the scope of this work, but constitutes an obvious step to follow in a forthcoming analysis.
The main goal of such a study is to identify if the spatial correlation obtained with HOG can be related to the gas motions in MCs that have formed self-consistently within the kilo-parsec numerical simulation and compare its results with other complementary techniques used to characterize the MC kinematics \citep[e.g.,][]{lazarianANDpogosyan2000,henshaw2016,chiraInPrep}.

\section{Conclusions and perspectives}\label{section:conclusions}

We characterize the histogram of oriented gradients (HOG), a tool developed for machine vision that we employ in the study of spectral line observations of atomic and molecular gas.
This technique does not assume the organization of the atomic or molecular gas in clouds or complexes. 
In that sense, it constitutes a ``blind'' estimator of the coincidence in the spatial distribution of the two tracers. 

We applied HOG to a set of synthetic H{\sc i} and $^{13}$CO observations from a MHD simulation of MC formation in the collision of two atomic clouds.
There we find significant spatial correlation between the synthetic H{\sc i} and $^{13}$CO emission contours across a broad range of velocity channels.
The highest spatial correlation appears around velocity channel pairs with \vhi\,$\approx$\,\vco\ independently of the cloud-collision direction with respect to the line of sight.

Using HOG, we studied the spatial correlation of H{\sc i} and $^{13}$CO emission observations toward a portion of the Galactic plane.
We significant spatial correlation between the H{\sc i} and $^{13}$CO emission. 
The highest spatial correlation appears around velocity channel pairs with \vhi\,$\approx$\,\vco, although in some regions there is significant correlation in H{\sc i} and $^{13}$CO velocity channels separated by a few kilometers per second.

We used the catalog of MC candidates derived from the $^{13}$CO observations \citep{rathborne2009} to analyze the spatial correlation toward particular objects.
Part of the spatial correlation identified with the HOG technique appears to be associated with these MC candidates, however, there are extended portions of the maps that are spatially correlated and do not correspond to any of them.
The HOG results indicate a different spatial correlation across velocity channels between the two tracers towards MC candidates in the proximity of H{\sc ii} regions.
This observation can be interpreted in two ways: either the H{\sc ii} regions are producing this particular dynamical behavior, by their input of energy that potentially affects the atomic and the molecular medium in different ways, or the regions with this dynamical behavior are the ones producing H{\sc ii} regions, by resulting from efficient accumulation of gas.
Either scenario is worth exploring in the future using dedicated synthetic observations of MHD simulations.

We showed that the significant correlation in H{\sc i} and $^{13}$CO velocity channels separated by a few \kps\ is also found in the synthetic observations of a portion of an MHD simulation with multiple supernovae explosions in a multiphase magnetized medium.
But the identification of the physical conditions that produce this velocity offset and its importance for identifying a particular mechanism of MC formation will be the subject of future work based on MHD simulations.
In the observational front, we will also continue this work by extending the HOG analysis to the full extent of the THOR observations, using of improved MC catalogs to evaluate HOG toward individual objects, and the combining HOG with the identification of the physical properties of the H{\sc i} gas.

We conclude that the HOG is a useful tool to evaluate the spatial correlation between tracers of different regimes of the ISM.
In this particular case, we used the extended H{\sc i} and $^{13}$CO emission to characterize MCs, but HOG can be used for the systematic comparison of extended observations of other tracers in Galactic and extragalactic targets.
The broad range of scales, the diversity of physical conditions, and the large volumes of observed and simulated data make understanding of the dynamical behavior of the ISM a big-data problem.
Hiding within those mounds of data are the trends that reveal what determines where and when stars form.
HOG constitutes just one of the multiple data-driven tools that in the future should pave the way to a more comprehensive picture of the ISM.

\begin{acknowledgements}
JDS, HB, MR, YW, and JCM acknowledge funding from the European Research Council under the Horizon 2020 Framework Program via the ERC Consolidator Grant CSF-648505.
SCOG and RK acknowledge support from the Deutsche Forschungsgemeinschaft via SFB 881, ``The Milky Way System'' (sub-projects B1, B2 and B8), and from the European Research Council under the European Community's Seventh Framework Programme (FP7/2007-2013) via the ERC Advanced Grant STARLIGHT (project number 339177).
FB acknowledges funding from the European Union's Horizon 2020 research and innovation programme under grant agreement No. 726384.
JK has received funding from the European Union's Horizon 2020 research and innovation programme under grant agreement No. 639459 (PROMISE).
SER acknowledges support from the European Union's Horizon 2020 research and innovation programme under the Marie Sk{\l}odowska-Curie grant agreement No. 706390.
NR acknowledges support from the Infosys Foundation through the Infosys Young Investigator grant.
RJS acknowledges support from an STFC ERF.
The National Radio Astronomy Observatory is a facility of the National Science Foundation operated under cooperative agreement by Associated Universities, Inc.
The Galactic Ring Survey is a joint project of Boston University and Five College Radio Astronomy Observatory, funded by the National Science Foundation.
This research was carried out in part at the Jet Propulsion Laboratory, operated for NASA by the California Institute of Technology.
Part of the crucial discussions that lead to this work took part under the program Milky-Way-Gaia of the PSI2 project funded by the IDEX Paris-Saclay, ANR-11-IDEX-0003-02. 

We thank the anonymous referee for his/her thorough review and highly appreciate the comments, which significantly contributed to improving the quality of this paper. 
JDS thanks the following people who helped with their encouragement and conversation: Peter G. Martin, Marc-Antoine Miville-Desch\^{e}nes, Norm Murray, Edith Falgarone, Hans-Walter Rix, Jonathan Henshaw, Shu-ichiro Inutsuka, and Eric Pellegrini.
\end{acknowledgements}

\bibliographystyle{aa}
\bibliography{HIandCO.bbl}

\appendix 


\section{Histograms of oriented gradients}\label{app:HOG}

\subsection{Computation of the gradient}\label{app:HOGgradient}

In the HOG, the computation of the gradient is performed by convolving the individual velocity-channel maps with the derivative of a two-dimensional Gaussian, what is known as a Gaussian derivative \citep{soler2013}.
The size of the Gaussian determines the area of the vicinity over which the gradient will be calculated.
Varying the size of the Gaussian kernel enables the sampling of different scales and reduces the effect of noise in the pixels.

In algebraic terms, we estimate the gradient of the $k$-th velocity channel of the PPV cube $I_{ij,k}$, where the indexes $i$ and $j$ run over the spatial coordinates $x$ and $y$, by computing
\begin{equation}\label{eq:gaussianderivative}
\nabla I_{ij,k} = I_{ij,k} \otimes \nabla \left(A\exp\left[-\frac{x^{2}}{2\sigma_{\rm g}^{2}}-\frac{y^{2}}{2\sigma_{\rm g}^{2}} \right]\right),
\end{equation}
where $\otimes$ is the convolution operator, $\nabla$ represents the standard gradient calculated using forward differences of adjacent pixels, and $A$ is a normalization factor, such that the integral of the Gaussian function equals unity.
For the sake of simplicity, we choose the same variance $\sigma_{\rm g}$ of the Gaussian function in the $x$- and $y$-direction.
In practice, we apply the Gaussian filter routine in the multi-dimensional image processing {\tt ndimage} package of {\tt scipy} with reflecting boundary conditions.

\begin{figure*}[ht!]
\centerline{
\includegraphics[width=\textwidth,angle=0,origin=c]{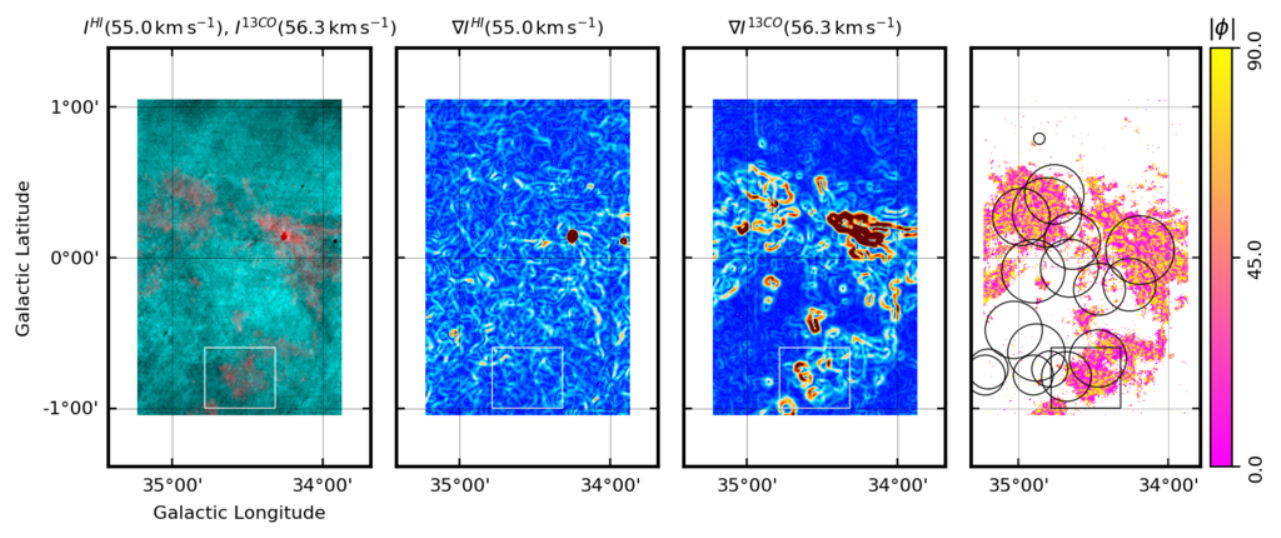}
}
\vspace{-0.2cm}
\centerline{
\includegraphics[height=3.0in,angle=0,origin=c]{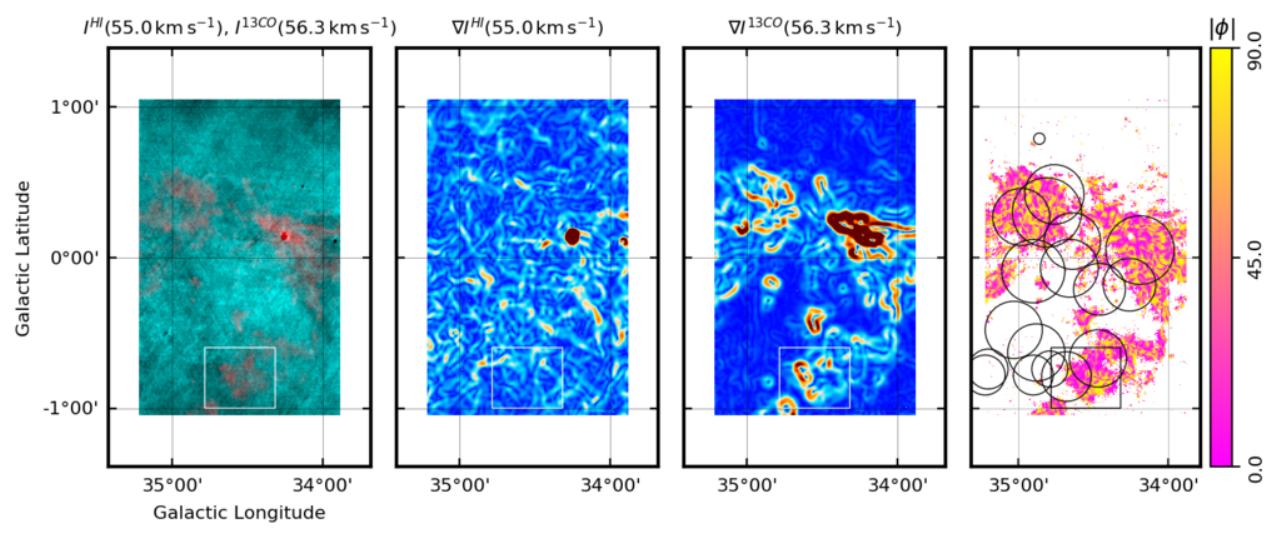}
}
\vspace{-0.2cm}
\centerline{
\includegraphics[height=3.0in,angle=0,origin=c]{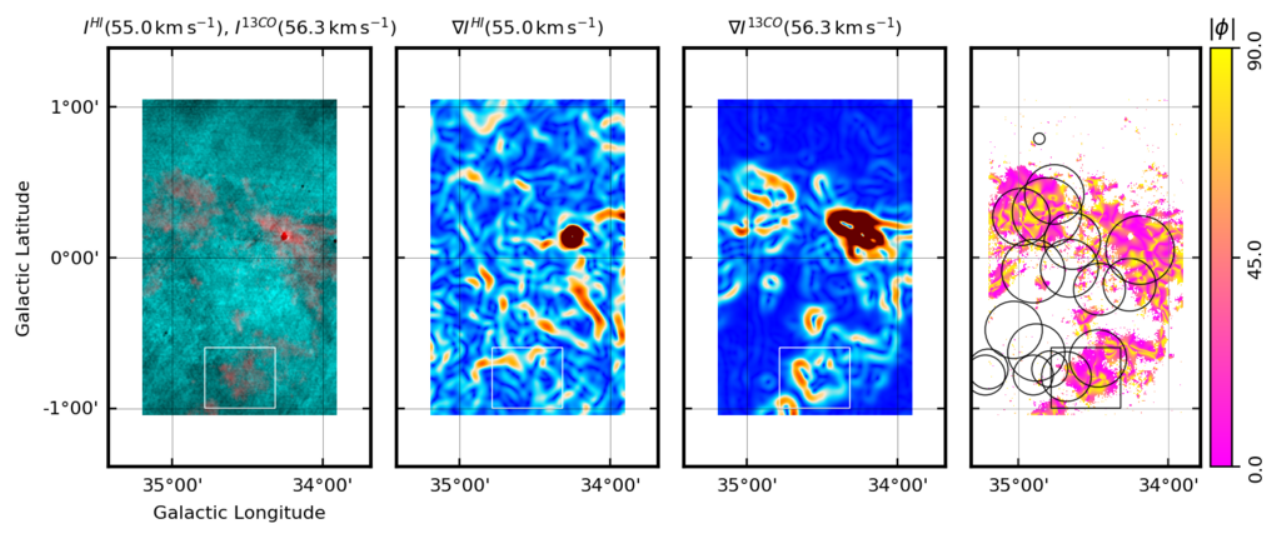}}
\caption{
Same as Fig.~\ref{fig:HOGanalysis47to62} for derivative kernels with 46\arcsec\ (top), 60\arcsec\ (middle), and 105\arcsec\ FWHM (bottom).
}
\label{fig:HOGanalysis-5to30multiKernel}
\end{figure*}

\begin{figure}[ht!]
\centerline{
\includegraphics[width=0.25\textwidth,angle=0,origin=c]{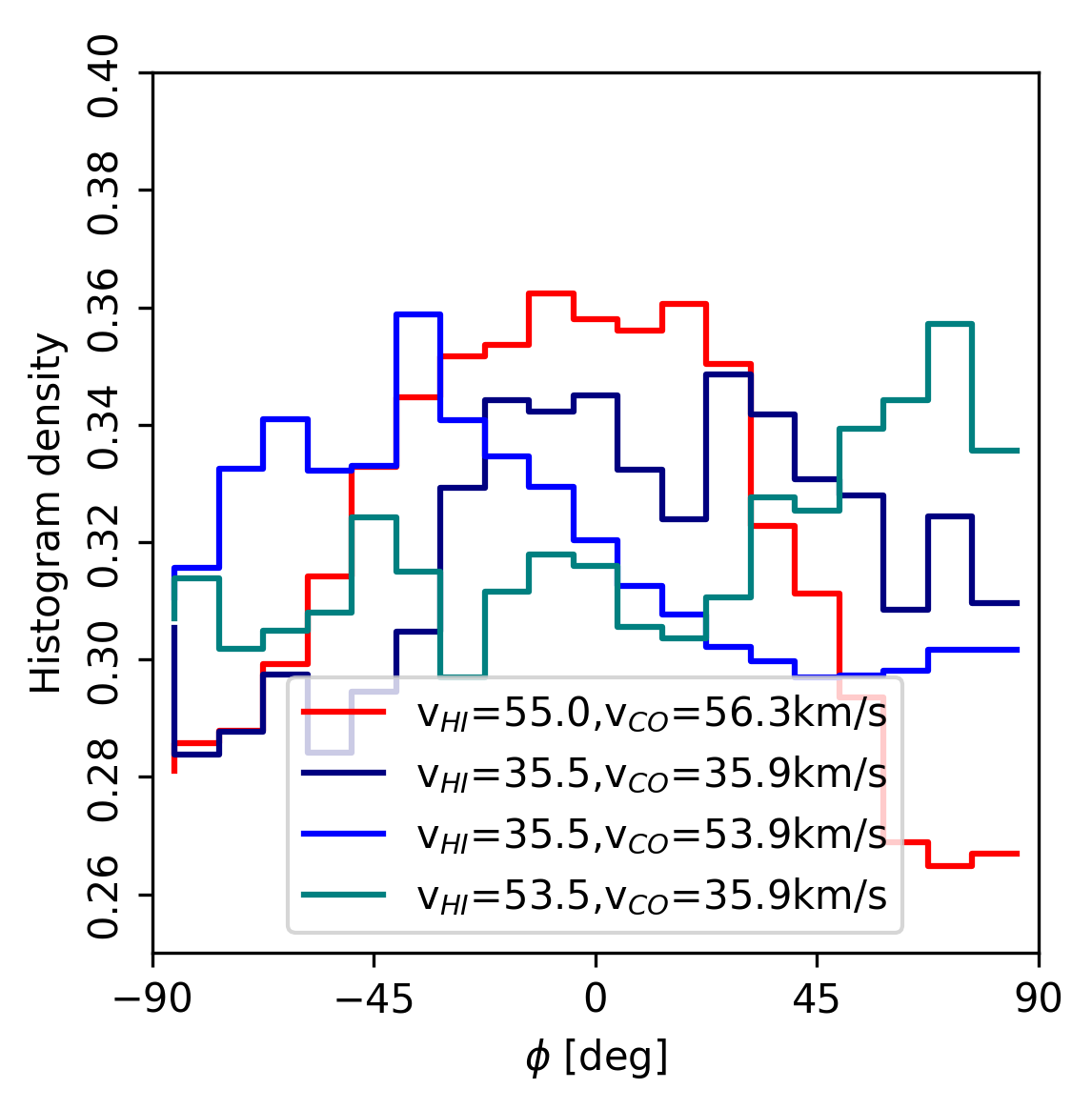}
\includegraphics[width=0.25\textwidth,angle=0,origin=c]{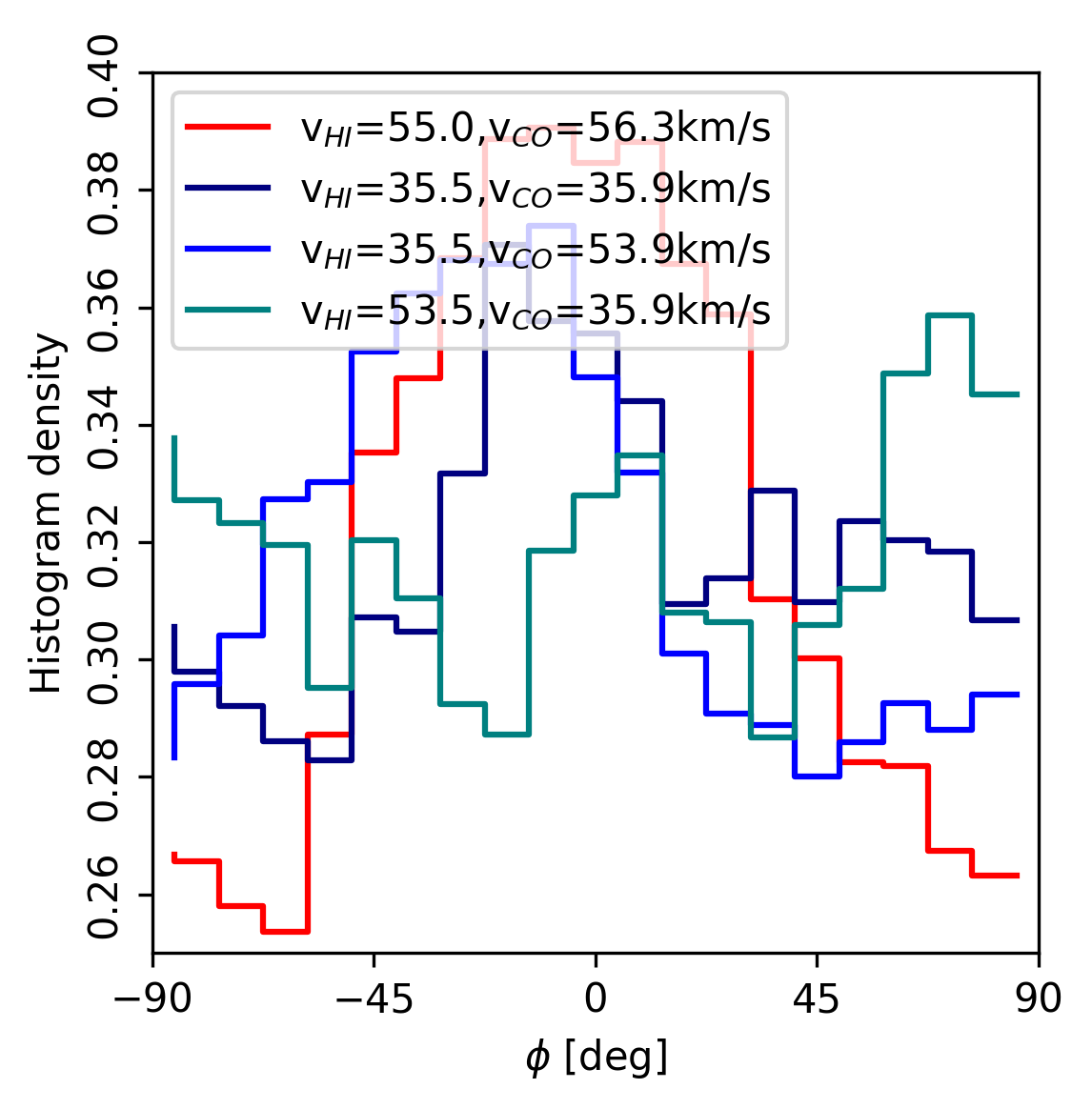}
}
\centerline{
\includegraphics[width=0.25\textwidth,angle=0,origin=c]{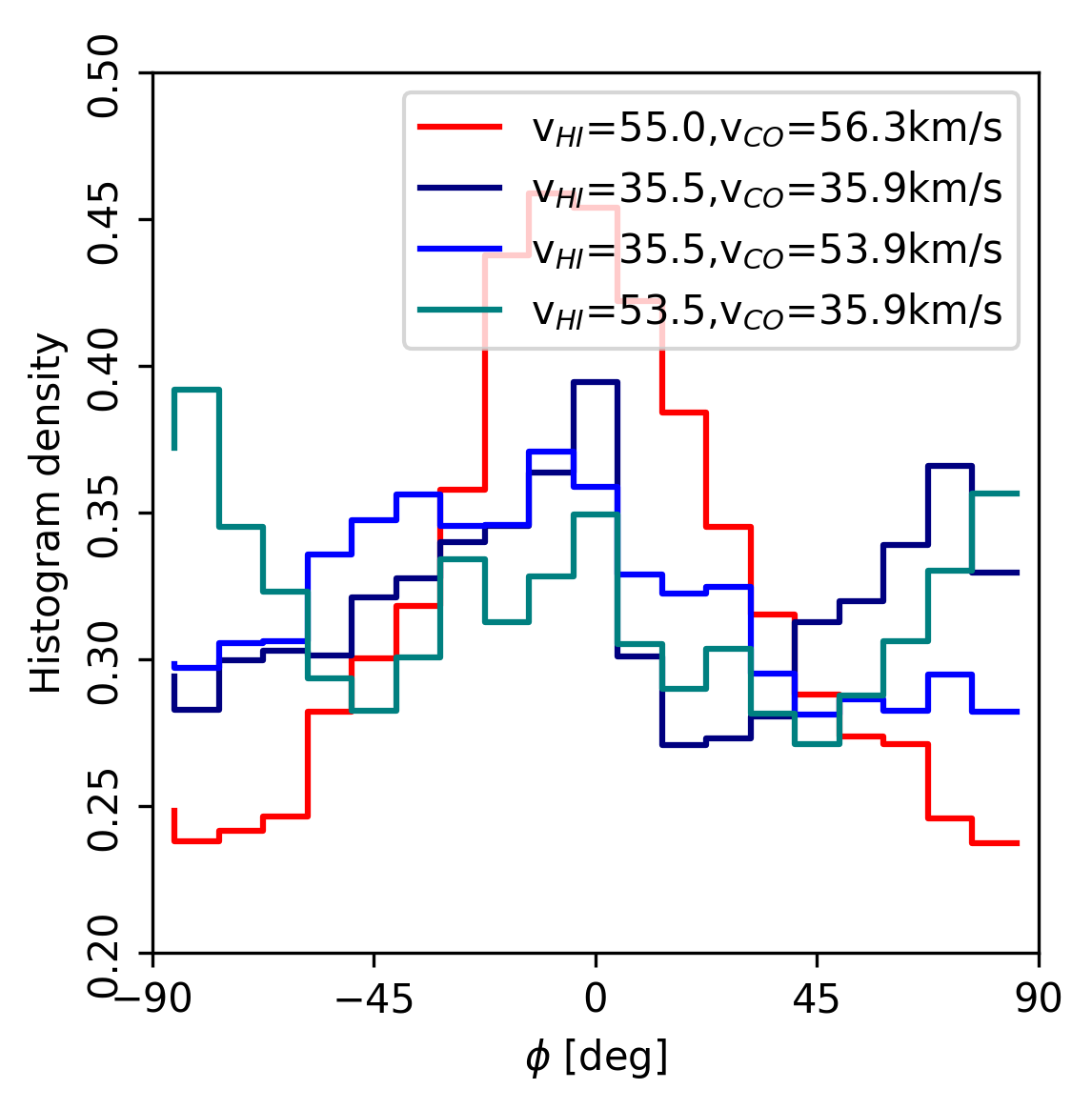}
\includegraphics[width=0.25\textwidth,angle=0,origin=c]{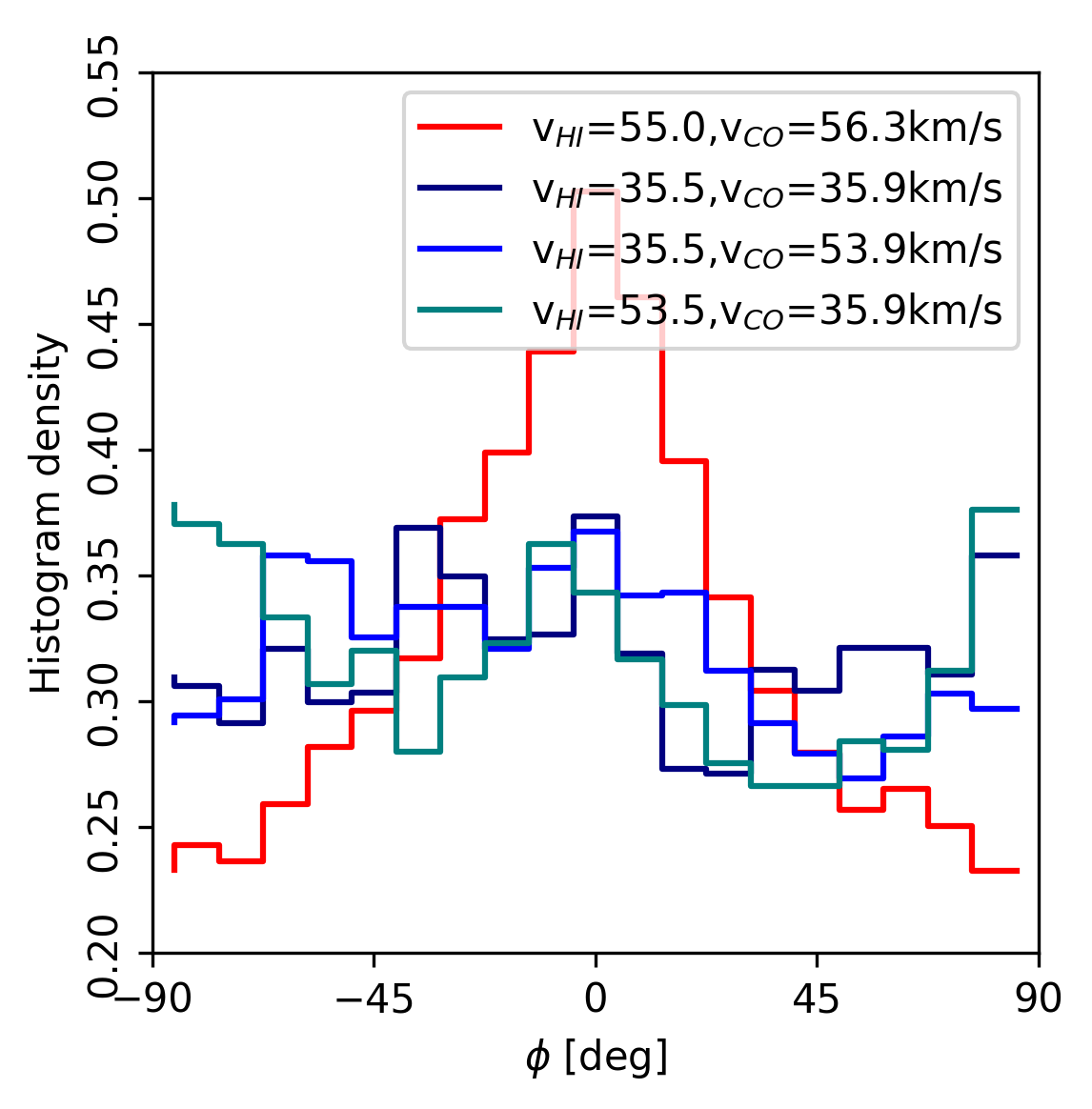}
}
\caption{
Histograms of oriented gradients (HOGs) corresponding to analysis presented in Fig.~\ref{fig:HOGanalysis-5to30multiKernel}.
They correspond to derivative kernels with 46\arcsec\ (top left), 60\arcsec\ (top right), 90\arcsec\ (bottom left) and 105\arcsec\ (bottom right) FWHM.
}\label{fig:HOGcomparison47to62}
\end{figure}

Figure~\ref{fig:HOGanalysis-5to30multiKernel} illustrates the effect of the derivative kernel size in the HOG analysis.
The gradient diameter sets the spatial scale of the intensity contours that we compare using the gradient and consequently, it corresponds to the size of the correlated patches in the maps of $\phi$, shown in the rightmost panel of Fig.~\ref{fig:HOGanalysis-5to30multiKernel}.
However, we note that the regions of the map with $\phi$\,$\approx$\,0\deg\ are persistent across derivative kernels with FWHM\,$=$\,46\arcsec\ (GRS data resolution), 60\arcsec, 90\arcsec\ (shown in Fig.~\ref{fig:HOGanalysis47to62}), and 105\arcsec.
Additionally, the shape of the HOGs is not significantly changed by the selection of the kernel size in the aforementioned range, as shown in Fig.~\ref{fig:HOGcomparison47to62}.
This is not entirely unexpected given that the structures in the velocity-channel maps show spatial correlations across multiple scales \citep[see for example][]{lazarian2000,brunt2003}.

Despite the fact that the HOG results are persistent for the selected derivative kernel sizes, this may not always be the case.
On the one hand, the selection of a very large derivative gradient would wash out the signal and will not profit from the angular resolution of the observations. 
On the other hand, using a very small derivative gradient will make more evident the features produce by noise and non-ideal telescope beams.
Given that to first order the selection of the kernel size between 46\arcsec\ and 105\arcsec\ does not significantly change the correlation that we report in this paper, as we further discuss in Sec.~\ref{app:multiscaleHOG}, we have chosen to report the results of the analysis using the 90\arcsec\ kernel, which is roughly twice the angular resolution of the THOR and GRS data (40 and 46\arcsec, respectively) and reduces most of the interferometer features in the THOR H{\sc i} observations.
In the future we will explore in detail further improvements that could be obtained with the selection of the derivative kernel sizes.

\subsection{Gradient selection}\label{sec:gradselection}

The observed signal in the $k$-th velocity channel can be interpreted as
\begin{equation}
I_{ij,k} = I^{0}_{ij,k} + \delta_{ij,k}, 
\end{equation}
where $I^{0}_{ij,k}$ is the emission and $\delta_{ij,k}$ the noise in the map.
The gradient of the observed velocity-channel maps can be written as
\begin{equation}
\nabla I_{ij,k} = \nabla I^{0}_{ij,k} + \nabla(\delta_{ij,k}).
\end{equation}

We quantify the contribution of the noise to the gradient by evaluating $\nabla I_{ij,k}$ in velocity-channel maps without signal.
For that purpose, we identify the velocity-channel map, $I_{ij,k'}$, with the lowest average emission and compute the reference noise intensity
\begin{equation}
I^{\rm N} \equiv \sqrt{\left<I^{2}_{ij,k'}\right>_{ij}},
 \end{equation}
where $\left<\cdots\right>_{ij}$ denotes the average over the spatial coordinates.
Then, we compute the average intensity gradient in that velocity-channel map,
\begin{equation}
|\nabla I^{\rm N}|\equiv\left<|\nabla(I_{ij,k'})|\right>_{ij},
\end{equation}
where $|\cdots|$ denotes the norm of the vector.

Assuming that $I^{\rm N}$ is a good approximation for $\delta_{ij,k}$ and $|\nabla I^{\rm N}|$ is representative of $\nabla(\delta_{ij,k})$, we compute the HOG using only the gradients in regions of each velocity map where $I_{ij,k} \geq 5I^{\rm N}$ and $|\nabla I_{ij,k}| \geq 5|\nabla I^{\rm N}|$.  
The first criterion guarantees that the gradients are not coming from a region of the map with low signal-to-noise ratio (S/N).
The second criterion guarantees that the gradients are larger than those produced by the noise in the velocity-channel maps.
Additionally, we exclude the gradients that are within a distance $\sigma_{\rm g}$, as defined in Eq.~\eqref{eq:gaussianderivative}, from the map edge.

A potential source of noise in the gradients is the presence of features from the image reconstruction of interferometric data, as is the case in the THOR H{\sc i} observations.
The noise in the THOR maps is highly non-uniform and non-Gaussian, so using the simple spatial average of the gradient of the noise to judge the significance of the intensity gradients may be insufficient.
However, given that we are comparing the interferometric observations with single-dish observations, these spatial features would only be present in the H{\sc i} velocity-channel maps and they would only contribute to the chance correlation between them.
Additionally, the smoothing implied in the Gaussian derivative operation mitigates the effect of non-uniform noise.
\juan{Further studies of the effect of noise on the intensity gradients in the specific context of the GRS $^{13}$CO and THOR H{\sc i} analysis are presented in Sec.~\ref{app:HOGstatsMCnoise} and Sec.~\ref{app:HOGstatsTHORnoise}}.

\subsection{Statistical evaluation of the HOG results.}\label{app:StatSignificance}

In the histograms of relative orientation (HOG) method we have a set of orientation angles $\phi_{k}$ in the range $[-\pi/2,\pi/2]$, estimated from Eq.~\eqref{eq:phi}.
To test for uniformity of these data we map each angle into twice itself, i.e. $\theta_{k} \rightarrow \theta$\,$=$\,$2\phi_{k}$. 
This method of angle doubling is a common technique for converting axial data, which carries information about the orientation and not the direction, to circular data, in order to utilize the tools of circular statistics \citep[see][and references therein]{jow2018}.

We present the results of the HOG analysis using the resultant vector length, $r$, which we define in Eq.~\eqref{eq:mv}.
This is a normalized quantity that, to zeroth order, can be interpreted as the fraction of parallel gradient vectors and encapsulates the information in the HOGs.
However, the value of $r$ is purely descriptive.
To quantify the statistical significance of $r$ values we apply the projected Rayleigh test: a test that the angle distribution is peaked at 0\deg, that is, that it represents mostly parallel gradient vectors.

\subsubsection{The mean resultant vector length}\label{app:R}

The first step in the analysis of relative orientations is the definition of the resultant vector length \citep{batschelet1981}.
Given a set $\theta_{k}$ of $N$ angles, one can associate them with a set of unitary vectors.
In a rectangular $x,y$-coordinate system, the components of the unit vectors are
\begin{equation}\label{eq:rvlxy}
x_{k}=\cos\theta_{k} \mbox{ and } y_{k}=\sin\theta_{k}.
\end{equation}

The sum of vectors, also called the resultant vector, has components
\begin{equation}\label{eq:rvlvw}
X=\sum^{N}_{k}w_{k}\cos\theta_{k} \mbox{ and } Y=\sum^{N}_{k}w_{k}\sin\theta_{k},
\end{equation}
where $w_{k}$ is the statistical weight associated to the angle $\phi_{k}$.

The length $r$ of the mean vector is simply
\begin{equation}\label{eq:mv}
r=\frac{1}{\sum^{N}_{k}w_{k}}(X^{2}+Y^{2})^{1/2}.
\end{equation}
If the resultant vector length is close to zero, then no single preferred direction exists.
This may be the case where all directions are equally likely, that is, a uniform distribution of angles.
This may also be the case with certain multi-modal distributions, for example, when two opposite directions are equally probable.
The statistical significance of a preferential relative orientation is evaluated by testing whether $r$ differs from zero significantly.

\begin{figure}[ht!]
\centerline{
\includegraphics[width=0.3\textwidth,angle=0,origin=c]{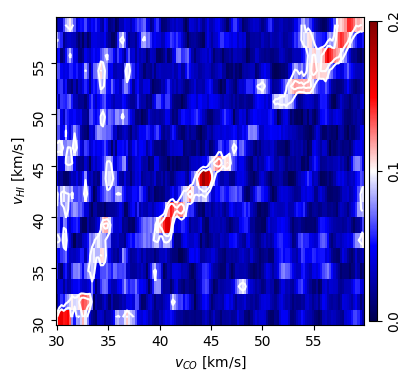}
}
\vspace{-0.2cm}
\caption{Mean resultant vector length, $r(v_{\rm 13CO},v_{\rm HI})$, as defined in Eq.~\eqref{eq:mv}, for the relative orientation angles $\phi$, Eq.~\eqref{eq:phi}, calculated from the H{\sc i} and $^{13}$CO velocity-channel maps in the range $30$\,$<$\,\vlsr\,$<$\,$60$\,\kps.
The contours \juan{are at} the 3$\varsigma_{V}$ and 4$\sigma_{r}$ \juan{levels}, where $\sigma_{r}$ represents the standard deviation of $r(v_{\rm 13CO},v_{\rm HI})$ in the indicated velocity range.
}\label{fig:RplaneExample}
\end{figure}

The mean resultant vector length $r$ is the normalized quantity that we use to systematically characterize the histograms of oriented gradients (HOGs), which we show for reference in Fig.~\ref{fig:HOGcomparison47to62}.
If all the angles $\theta_{k}$ are identical, $r$\,=\,1.
If the angles $\theta_{k}$ are uniformly distributed, $r$ is close to zero. 
If the angles $\theta_{k}$ are not uniformly distributed, $r$ is larger than zero and roughly corresponds to the percentage of angles that represent a preferential orientation. 
Figure~\ref{fig:RplaneExample} shows the values of $r$ obtained from the relative orientation angles estimated using Eq.~\eqref{eq:phi} in the H{\sc i} and $^{13}$CO emission maps at the indicated velocities.

\subsubsection{The projected Rayleigh statistic}\label{app:PRS}

The Rayleigh test is used to determine whether or not a set of angles are uniformly distributed \citep{rayleigh1879}.
The Rayleigh test assumes that the sample is generated from a von Mises distribution, that is, a continuous probability distribution on the circle, a close approximation to the wrapped normal distribution, which is the circular analogue of the normal distribution.
That means that it should be used only when the distribution is unimodal.
In the Rayleigh test, under the null hypothesis $H_{0}$ the population is uniformly distributed around the circle. 
Under the alternative hypothesis $H_{A}$ the population is not distributed uniformly around the circle. 

In the Rayleigh test, if the magnitude of the mean resultant length in the most common instance of unspecified mean direction, defined as
\begin{equation}\label{eq:rs}
Z \equiv (X^{2}+Y^{2})^{1/2} = \left[r^{2}\sum^{N}_{k}w_{k}\right]^{1/2},
\end{equation}
is large, the null hypothesis $H_{0}$ is rejected.
This can be interpreted as the net displacement of a random walk with $N$ steps, each with a corresponding length $w_{k}$.  
If the angles $\theta_{k}$ are uniformly distributed, the net displacement is close to zero.
If the angles $\theta_{k}$ are not uniformly distributed, the net displacement is larger than zero.

\begin{figure}[ht!]
\centerline{
\includegraphics[width=0.3\textwidth,angle=0,origin=c]{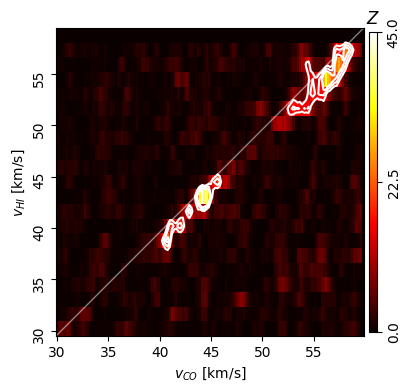}
}
\vspace{-0.2cm}
\caption{Rayleigh statistic $Z$, as defined in Eq.~\eqref{eq:rs}, for the relative orientation angles $\phi$ calculated between H{\sc i} and $^{13}$CO velocity-channel maps in the range $30$\,$<$\,\vlsr\,$<$\,$60$\,\kps.
The contours represent the 3$\sigma_{Z}$ and 4$\sigma_{Z}$ contours, where $\sigma_{Z}$ represents the standard deviation of $Z$ within the indicated velocity range.
}\label{fig:ZplaneExample}
\end{figure}

Figure~\ref{fig:ZplaneExample} shows the results of the Rayleigh test applied to the same observations presented in Fig.~\ref{fig:RplaneExample}.
The values of $Z$ clearly show that many of the points where the values of $r$ were significant do not pass the non-uniformity test.  
Most of the high-$Z$ values are located along the diagonal of the plot, that is, in emission maps corresponding to velocity channels where \vhi\,$\approx$\,\vco.  
To account for the correlation by random chance, we consider only the values which are above three times the standard deviation of $Z$ within the selected velocity range.

In the HOG application, we are interested in testing whether the relative orientation is preferentially parallel (corresponding to $\phi_i=0$) and what is the statistical strength of that trend.
To perform that test we use the projected Rayleigh statistic \citep[PRS, ][]{jow2018}, also known in its general form as the $V$ statistic \citep{durand1958,mardia1972statistics}.

The null hypothesis $H_{0}$ that we test is randomness, which means that the angles of the sample are independent observations from a uniform circular distribution.
The $V$ test for circular uniformity is similar to the Rayleigh test with the difference that under the alternative hypothesis $H_{A}$ is assumed to have a known mean direction, which in our case corresponds to $\theta_{A}=0$\deg.
In that particular case, the PRS is
\begin{equation}\label{eq:prs}
V = \frac{\sum^{N}_{k}w_{k}\cos2\phi_{k}}{\sqrt{\sum^{N}_{k}(w_{k})^{2}/2}},
\end{equation}
where $w_{k}$ is the statistical weight assigned to the angle $\phi_{k}$.

In our application, $V$\,$>$\,$0$ indicates mostly parallel relative orientation between the gradient vectors.  
The values $V$\,$<$\,$0$ correspond to the mostly perpendicular relative orientation between the gradient vectors, which does not carry any particular significance when comparing two images, but it is important when this test is used in the study of the relative orientation between column density structures and the magnetic field \citep{soler2017,jow2018}.

\begin{figure}[ht!]
\centerline{
\includegraphics[width=0.3\textwidth,angle=0,origin=c]{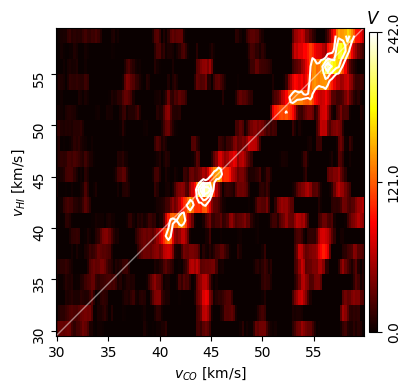}
}
\vspace{-0.2cm}
\centerline{
\includegraphics[width=0.3\textwidth,angle=0,origin=c]{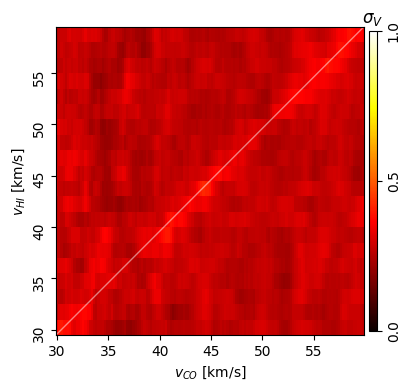}
}
\vspace{-0.2cm}
\centerline{
\includegraphics[width=0.3\textwidth,angle=0,origin=c]{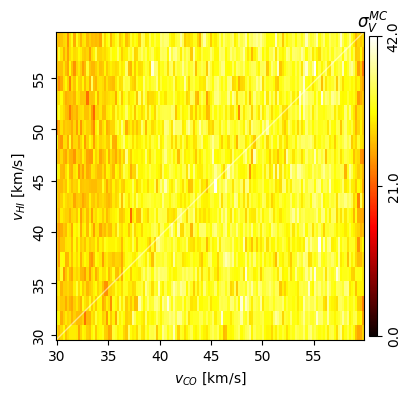}
}
\vspace{-0.2cm}
\centerline{
\includegraphics[width=0.3\textwidth,angle=0,origin=c]{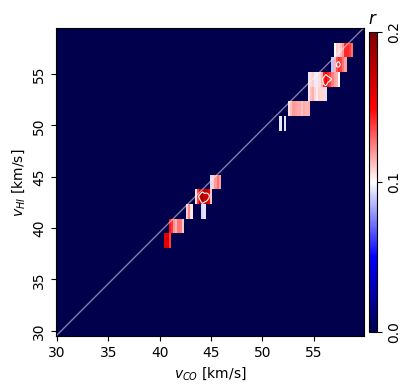}
}
\caption{
From top to bottom: projected Rayleigh statistic ($V$, Eq.~\ref{eq:prs}), its variance ($\sigma_{V}$, Eq.~\ref{eq:sigmaVpair}), \juan{its variance estimated using the Monte Carlo sampling introduced in Sec.~\ref{app:HOGstatsMCnoise},} \juan{and $r$ values from channel pairs with} $V>3\varsigma_{V}$, where $\varsigma_{V}$ represents the population variance of $V$ in the indicated velocity range, as estimated from Eq.~\ref{eq:sigmav}.
These values correspond to the relative orientation angles $\phi$ calculated between H{\sc i} and $^{13}$CO velocity-channel maps in the range $30$\,$<$\,\vlsr\,$<$\,$60$\,\kps.
}\label{fig:VplaneExample}
\end{figure}

The top panel of Fig.~\ref{fig:VplaneExample} shows the results of the project Rayleigh test applied to the same observations considered in Fig.~\ref{fig:ZplaneExample}.
As in the case of $Z$, most of the high-$V$ values are located along the diagonal of the plot, that is, in emission maps corresponding to velocity channels with \vhi\,$\approx$\,\vco.  
The asymptotic limit of the $V$ distribution, in the large $N$ limit, is the standard normal distribution \citep{jow2018}. 
Thus, for a general distribution of angles, the variance in $V$ is simply the variance of each $\cos (2\phi)/\sqrt{1/2}$ and can be estimated as
\begin{equation}\label{eq:sigmaVpair}
\sigma_{V}=\frac{2\sum^{N}_{k}(\cos 2\phi_{k})^{2} - V^{2}}{N}.
\end{equation}
Note that this value corresponds to the variance in the distribution of $V$ for a particular pair of velocity channels maps, shown in the \juan{top-middle} panel of Fig.~\ref{fig:VplaneExample}.
In general, the values derived from Eq.~\ref{eq:sigmaVpair} are smaller than those that arise from the random correlation between velocity channels, $\varsigma_{V}$, which is estimated using Eq.~\ref{eq:sigmav}, hence we use the latter to report the statistical significance of the $V$ values.
\juan{For the sake of completeness, we also report in Fig.~\ref{fig:VplaneExample} the values of $\sigma^{\rm MC}_{V}$ derived from the Monte Carlo sampling introduced in Sec.~\ref{app:HOGstatsMCnoise}}.
\juan{The large difference between the $\sigma_{V}$ values derived from Eq.~\eqref{eq:sigmaVpair} and $\sigma^{\rm MC}_{V}$ confirms the inadequacy of the assumption of statistically independent angles in the analysis of the HOG results.}
Once pre-selected using the values of $\varsigma_{V}$, the values of $r$ roughly corresponds to the fraction of parallel gradient vectors.
This selection leads to the values presented in the bottom panel of Fig.~\ref{fig:VplaneExample}.

\begin{figure}[ht!]
\centerline{
\includegraphics[width=0.3\textwidth,angle=0,origin=c]{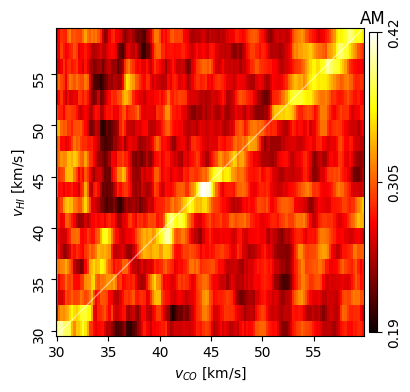}
}
\caption{
Alignment measurement (AM), the method for estimating the degree of alignment between vectors introduced in \citep{lazarian2007}, for the relative orientation angles $\phi$ calculated between H{\sc i} and $^{13}$CO velocity-channel maps in the range $30$\,$<$\,\vlsr\,$<$\,$60$\,\kps.
}\label{fig:AM}
\end{figure}

\juan{We also calculated the alignment measurement (AM), an alternative method for estimating the degree of alignment between vectors.
The alignment measurement is widely used in the study of dust grain alignment with magnetic fields \citep[see for example,][]{lazarian2007} and also in the family of papers represented by \cite{lazarian2018}}.
The calculation of this quantity is made using
\begin{equation}
{\rm AM} = \left<2\cos\phi_{ij} - 1 \right>_{ij},
\end{equation}
where $\left<\cdots\right>_{ij}$ represents the average over the relative orientation angles.
The values of AM, shown in Fig.~\ref{fig:AM}, present a very similar trend to that found with $V$, although \juan{the} AM has a shorter dynamic range than $V$, that is, the contrast between regions with low and high correlation is much lower. 

\subsection{Derivative kernel size}\label{app:multiscaleHOG}

\begin{figure}[ht!]
\centerline{
\includegraphics[width=0.25\textwidth,angle=0,origin=c]{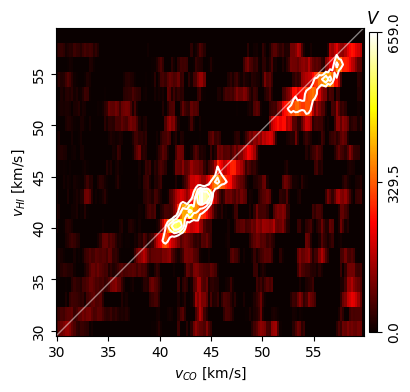}
\includegraphics[width=0.25\textwidth,angle=0,origin=c]{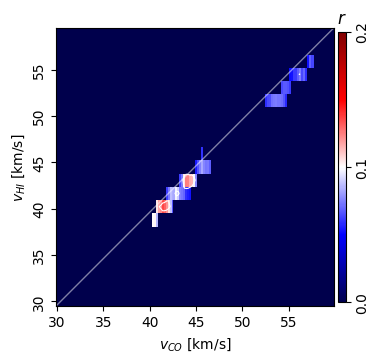}
}
\centerline{
\includegraphics[width=0.25\textwidth,angle=0,origin=c]{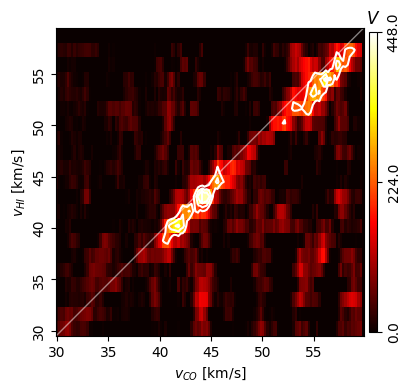}\includegraphics[width=0.25\textwidth,angle=0,origin=c]{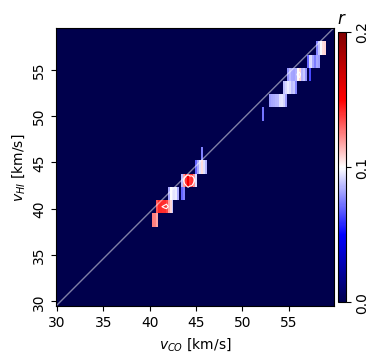}
}
\centerline{
\includegraphics[width=0.25\textwidth,angle=0,origin=c]{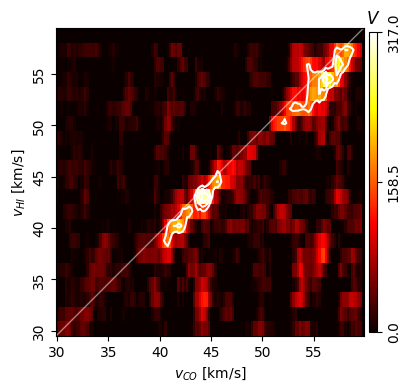}
\includegraphics[width=0.25\textwidth,angle=0,origin=c]{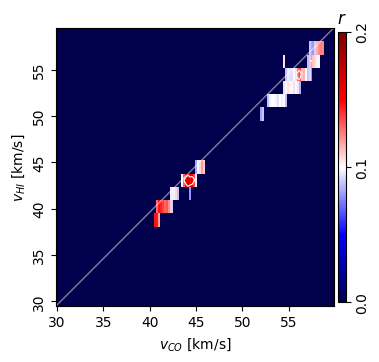}
}
\centerline{
\includegraphics[width=0.25\textwidth,angle=0,origin=c]{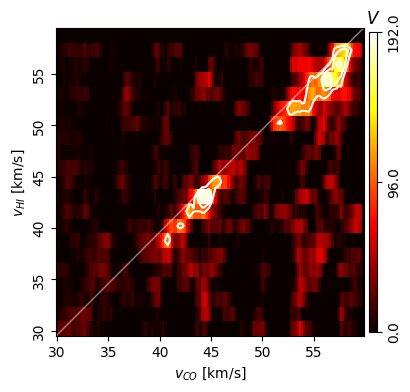}
\includegraphics[width=0.25\textwidth,angle=0,origin=c]{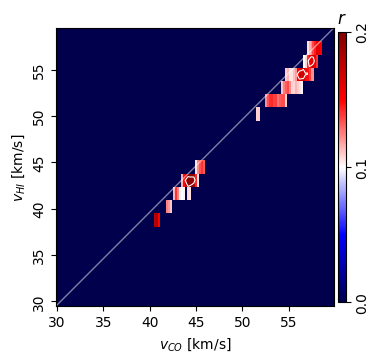}
}
\caption{
Results of the HOG analysis of the THOR H{\sc i} and GRS $^{13}$CO observations between 30\,$\leq$\,\vlsr\,$\leq$\,60\,\kps\ using derivative kernels with 46\arcsec\ (top), 60\arcsec\ (middle top), 75\arcsec\ (middle bottom), and 105\arcsec\ (bottom) FWMH.
\emph{Left}. Projected Rayleigh statistic, $V$.
\emph{Right}. Mean resultant vector length, $r$.
The contours correspond to the 3$\varsigma_{V}$, 4$\varsigma_{V}$, and 5$\varsigma_{V}$ values in the corresponding velocity ranges.
}\label{fig:corrPlaneMultiKernel}
\end{figure}

We discussed the effect of the derivative kernel size in the relative orientation angle, $\phi$, maps and the HOGs in Sec.~\ref{app:HOGgradient}.
Here we consider the effect of the derivative kernel size in the distribution of \prs\ and $r$.
Figure~\ref{fig:corrPlaneMultiKernel} show the values of both quantities for derivative kernels with 46\arcsec, 60\arcsec, 75\arcsec, and 105\arcsec\ FWHM.
It is clear from the distribution of these two quantities that the kernel size does not significantly affect the positions of the high-$V$ velocity ranges or their corresponding range of $r$ values.
However, the level of significance of the off-diagonal high-$V$ regions around \vhi\,$\approx$\,\vco\,$\approx$\,55\,\kps\ changes depending on the size of the derivative kernel.

\section{Statistical significance of the HOG method}\label{app:HOGstats}

\subsection{Impact of noise in the HOG results}\label{app:HOGstatsMCnoise}

\juan{Throughout this paper, we have reported the results of the HOG analysis based on the selection of gradients in pixels with intensities a number of times above the noise level.
Here we detail the effect of this gradient selection threshold and present an alternative method based on Monte Carlo sampling to quantify the effect of noise and propagate the observation errors into the HOG results}.

\subsubsection{Gradient selection threshold}

\begin{figure}[ht!]
\centerline{
\includegraphics[width=0.45\textwidth,angle=0,origin=c]{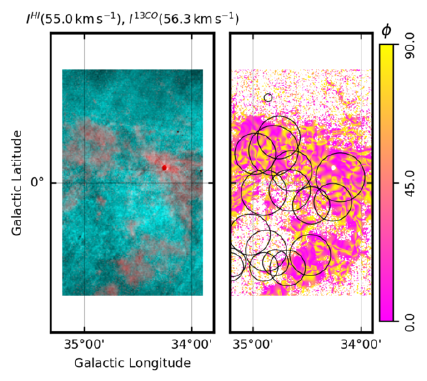}
}
\vspace{-0.1cm}
\centerline{
\includegraphics[width=0.45\textwidth,angle=0,origin=c]{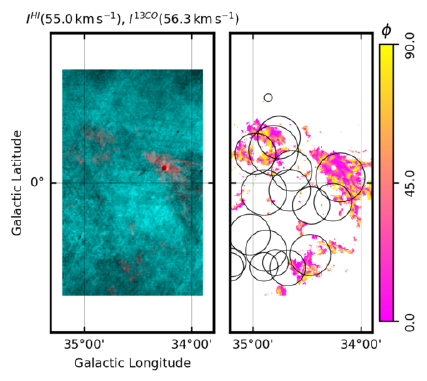}
}
\vspace{-0.2cm}
\caption{
Intensity and relative orientation angle maps from the THOR H{\sc i} and GRS $^{13}$CO observations presented in Fig,~\ref{fig:HIandCOmaps}.
\emph{Left.} H{\sc i} (teal) and $^{13}$CO emission (red) in the velocity channels with the largest spatial correlation in the velocity range 30\,$\leq$\,\vlsr\,$\leq$\,60\,\kps, as inferred from the \prs\ values shown in Fig.~\ref{fig:corrPlane}.
\emph{Right.} Relative orientation angle $\phi$, Eq.~\eqref{eq:phi}, between the gradients of the H{\sc i} and $^{13}$CO intensity maps in the indicated velocity channels.
The white \juan{portions of} the $\phi$ map correspond to areas where the gradient is not significant in either tracer, as estimated using the rejection criteria $I \geq \sigma_{I}$ (top) and $I \geq 7\sigma_{I}$ (bottom).
The black circles in the right panels correspond to the positions and effective sizes of the MC candidates from the \cite{rathborne2009} catalog in the aforementioned velocity range.
}\label{fig:hogPanelMultiSigma}
\end{figure}

\juan{One approach to the estimation of the $V$ values is the selection of the gradient vectors based on the intensity signal-to-noise ratio (S/N), as described in Sec.~\ref{sec:gradselection}.
Given that the gradients are independent of the intensity, this selection aims to guarantee that the relative orientation angles used in the calculation of $V$ come from regions that are not dominated by noise.
Although this approach is very practical, it reduces the number of samples on which to derive the projected Rayleigh statistic, as illustrated in Fig.~\ref{fig:hogPanelMultiSigma}}.

\begin{figure}[ht!]
\centerline{
\includegraphics[width=0.25\textwidth,angle=0,origin=c]{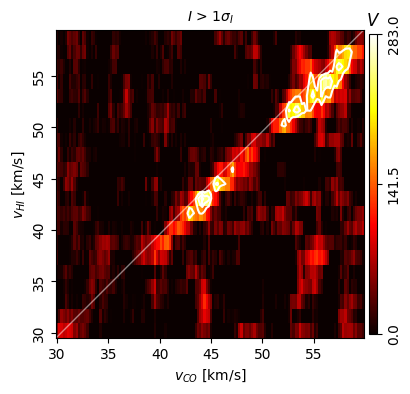}
\includegraphics[width=0.25\textwidth,angle=0,origin=c]{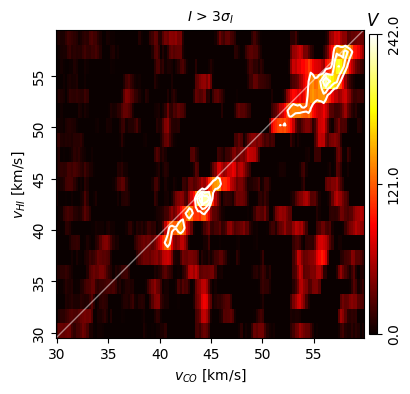}
}
\centerline{
\includegraphics[width=0.25\textwidth,angle=0,origin=c]{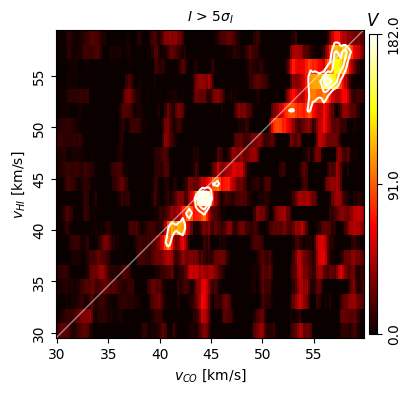}
\includegraphics[width=0.25\textwidth,angle=0,origin=c]{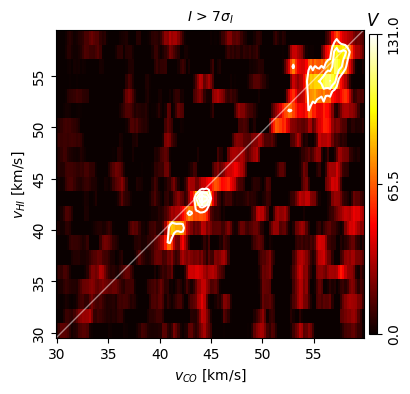}
}
\caption{
Projected Rayleigh statistic, $V$, the statistical test of spatial correlation between H{\sc i} and $^{13}$CO velocity-channel maps in the HOG method.
The panels correspond to the rejection criteria $I \geq \sigma_{I}$, $I \geq 3\sigma_{I}$, $I \geq 5\sigma_{I}$, and $I \geq 7\sigma_{I}$.
The contours correspond to 3, 4, and 5$\varsigma_{V}$, where $\varsigma_{V}$ is the population variance defined in Eq.~\eqref{eq:sigmav}.
}\label{fig:corrPlaneMultiSigma}
\end{figure}

\juan{To illustrate the effect of the gradient selection on the values of $V$, we present in Fig.~\ref{fig:corrPlaneMultiSigma} the results of the HOG analysis applied to the H{\sc i} and $^{13}$CO PPV cubes in the velocity range 30\,$\leq$\,\vlsr\,$\leq$\,60\,\kps\ after four different selections of the intensity S/N threshold.
It is clear from the general distribution of $V$ in this particular velocity range that the results of the HOG analysis do not critically depend on the selection of the threshold.
There are two main reasons that potentially explain this result.
First, the regions that dominate the signal in $V$ are those with high intensity S/N and consequently, the $V$ values are unaffected by the intensity threshold.
Second, if the noise were uniform across the velocity-channel maps, the regions with low signal to noise would have randomly-oriented gradients that are uncorrelated in both tracers and thus would not significantly affect the distribution of $V$ across velocity channels.}

\subsubsection{Error propagation in HOG}

\juan{An alternative to the selection of gradients based on the intensity signal-to-noise ratio (S/N) is the use of Monte Carlo sampling to propagate the uncertainties in the observations into the HOG results.
To do this, we generate draws $I^{\rm mc}_{ij,l}$ from a Gaussian probability distribution function described by the mean value $I_{ij,l}$ and the variance $(\sigma_{I})_{ij,l}$, where $I_{ij,l}$ are the observed intensities in the PPV cubes and $(\sigma_{I})_{ij,l}$ is estimated from the low intensity S/N channels.
As a zeroth-order approximation, we assume that $(\sigma_{I})_{ij,l}$\,$=$\,$\sigma_{I}$, that is, that the noise is constant in each velocity-channel map and across the velocity channels.
The impact of the non-uniform noise distribution is further explored in Sec.~\ref{app:HOGstatsTHORnoise}.}

\begin{figure}[ht!]
\centerline{
\includegraphics[width=0.25\textwidth,angle=0,origin=c]{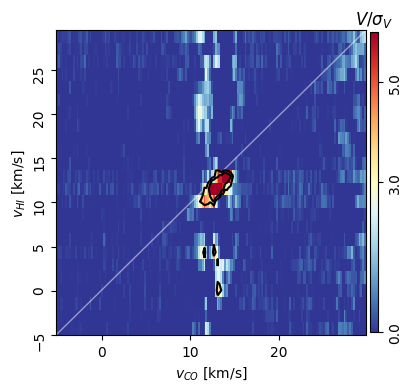}
\includegraphics[width=0.25\textwidth,angle=0,origin=c]{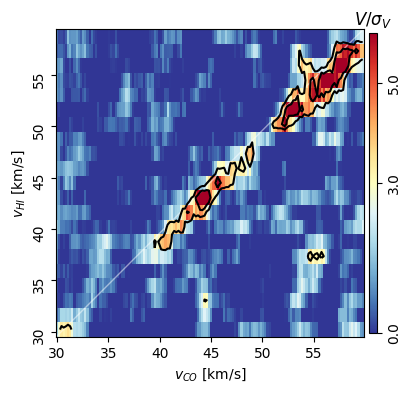}
}
\centerline{
\includegraphics[width=0.25\textwidth,angle=0,origin=c]{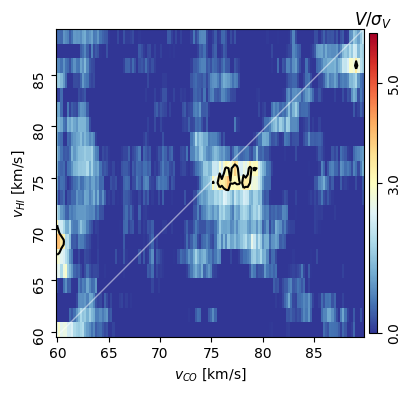}
\includegraphics[width=0.25\textwidth,angle=0,origin=c]{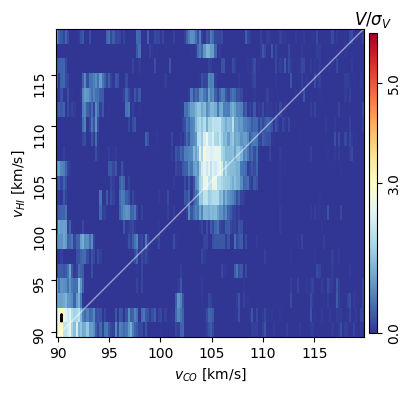}
}
\caption{
Signal-to-noise ratio, $V/\sigma_{V}$, of the projected Rayleigh statistic in four different velocity ranges.
The contours correspond to $V/\sigma_{V} \geq 3$ and $V/\sigma_{V} \geq 5$, where $\sigma_{V}$ is estimated using the Monte Carlo realizations of the data introduced in Sec.~\ref{app:HOGstatsMCnoise}.
}\label{fig:corrPlaneMCerr}
\end{figure}

\juan{We report in Fig.~\ref{fig:corrPlaneMCerr} the signal-to-noise ratios, $V/\sigma_{V}$, obtained from a Monte Carlo sampling in the velocity ranges $-$5,$\leq$\,\vlsr\,$\leq$30, 30\,$\leq$\,\vlsr\,$\leq$60, 60\,$\leq$\,\vlsr\,$\leq$90, and 90\,$\leq$\,\vlsr\,$\leq$120\,\kps.
These were obtained using 100 realizations of each $I^{\rm mc}_{ij,l}$ value in the H{\sc i} and $^{13}$CO PPV cubes.
It is clear from the distribution of $V/\sigma_{V}$ that the confidence intervals are very similar to those obtained using the population variance $\varsigma_{V}$ introduced in Eq~\eqref{eq:sigmav}.
This result reassures our assumption that velocity-channel maps in a broad range of velocities provide a good set of independent samples to determine the statistical significance of the observed $V$ values.}

\juan{In contrast to the intensity S/N selection, the Monte Carlo sampling does not reduce the number of samples from which to derive the projected Rayleigh statistic and fully propagates the measurement errors in the values of $V$.
Additionally, it does not depend on the velocity range selected for the estimation of $\varsigma_{V}$.
However, it is computationally costly as it requires the HOG calculations of multiple realizations of each pair of velocity channels maps; the estimation of the values presented in Fig.~\ref{fig:corrPlaneMCerr} requires a considerably larger amount of time and computational resources than those reported in Fig.~\ref{fig:corrPlane}.
For the sake of simplicity and given the proximity in the results of both methods, we have chosen to report the statistical significance of the HOG results in terms of $\varsigma_{V}$ in the main body of this paper}.

\subsection{Impact of non-uniform noise in the HOG results}\label{app:HOGstatsTHORnoise}

\juan{In radio interferometry, the imaging of a source requires the measurement of the two-dimensional spatial frequency spectrum.
This spectrum is not fully sampled, as there are gaps between the antennas. 
Hence, the Fourier transformation of the spatial frequency spectrum coverage, which is the spatial resolution element of the array, is not a simple two-dimensional Gaussian function but a rather complicated structure commonly called the ``dirty beam''.
During the imaging process, the ``dirty beam'' produces artifacts in the intensity distribution around a strong source that are commonly called side lobes.
There are several different methods to remove the side lobes, the most common of them is the {\tt clean} algorithm.
One of the products of the {\tt clean} algorithm is the residual image, which can be used as a good estimate for the level and spatial distribution of the noise in the observation \citep{bihr2016}.
Note that there is an additional contribution to the noise from the single-dish data used for the construction of the THOR H{\sc i} products, but that contribution is expected to be uniform across the observed region.}

\begin{figure}[ht!]
\centerline{
\includegraphics[width=0.3\textwidth,angle=0,origin=c]{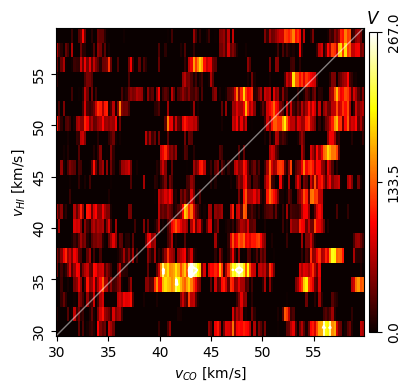}
}
\caption{
\juan{Projected Rayleigh statistic, $V$, resulting from the HOG analysis of the H{\sc i} noise cube and the $^{13}$CO PPV cube in the indicated velocity range.}
}\label{fig:HInoiseHOG}
\end{figure}

\juan{Although in principle uncorrelated, the spatial distribution of the THOR H{\sc i} noise can have some fortuitous correlation with the distribution of GRS $^{13}$CO.
In order to test if such a correlation has a critical impact in the distribution of $V$ across velocity channels, we calculated the HOG correlation between the GRS $^{13}$CO observations and the H{\sc i} noise cube constructed with the residual images of each velocity channel map.
The results of this test, presented in Fig.~\ref{fig:HInoiseHOG}, indicate that there is some chance correlation between both data sets, although the spatial window used to determine the noise maps from the {\tt clean} process introduces spatial correlations (see \cite{bihr2016} for details) that increase the values of $V$ and they cannot be directly compared with those in Fig.~\ref{fig:corrPlaneMultiSigma}.
However, the distribution of $V$ values across velocity channels is very different from the distribution of high $V$ values reported in Fig.~\ref{fig:corrPlaneMultiSigma}, which is sufficient to demonstrate that the spatial distribution of the noise has no significant impact in the results presented in the body of this paper}.

\subsection{Impact of chance correlation in the HOG results}\label{app:HOGstatsJacknives}

Given that reproducing the statistical properties of each studied PPV cube is almost impossible, we test for chance correlation by using the same PPV cubes but flipping its spatial coordinates.
Explicitly, we repeat the analysis using three test configurations:
\begin{itemize}
\item Keeping the original H{\sc i} PPV cube but flipping the $^{13}$CO PPV cube in the vertical (galactic latitude) direction.
\item Keeping the original H{\sc i} PPV cube but flipping the $^{13}$CO PPV cube in the horizontal (galactic longitude) direction.
\item Keeping the original H{\sc i} PPV cube but flipping the $^{13}$CO PPV cube in both the vertical and horizontal directions.
\end{itemize}

\begin{figure}[ht!]
\centerline{
\includegraphics[width=0.25\textwidth,angle=0,origin=c]{{HOGcorrTHOR-GRS_La34.55Lb34.55_k90_v1from30.0to60.0_v2from30.0to60.0_Vplane}.png}
\includegraphics[width=0.25\textwidth,angle=0,origin=c]{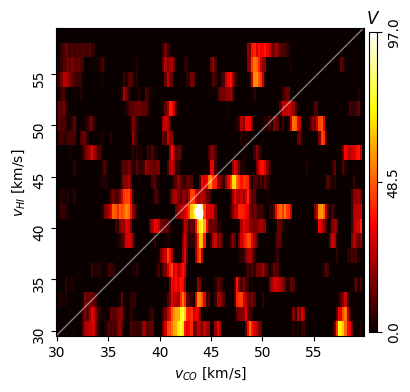}
}
\centerline{
\includegraphics[width=0.25\textwidth,angle=0,origin=c]{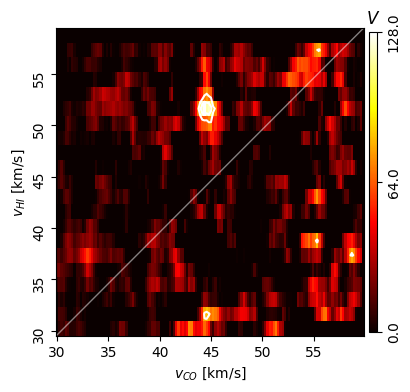}
\includegraphics[width=0.25\textwidth,angle=0,origin=c]{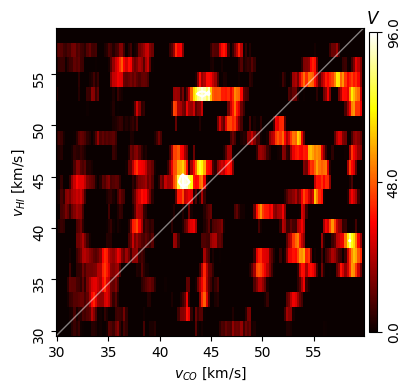}
}
\caption{ 
Projected Rayleigh statistic, $V$, the statistical test of spatial correlation between H{\sc i} and $^{13}$CO velocity-channel maps in the HOG method, as defined in Eq.~\eqref{eq:myprs}, after flipping the velocity channels maps in the following directions.
\emph{Top left}. Original THOR H{\sc i} and GRS $^{13}$CO tiles.
\emph{Top right}. Original THOR H{\sc i} tile and GRS $^{13}$CO flipped in the vertical (galactic latitude) direction 
\emph(Bottom left). Original THOR H{\sc i} tile and GRS $^{13}$CO flipped in the horizontal (galactic longitude) direction.
\emph{Bottom right}. Original THOR H{\sc i} tile and GRS $^{13}$CO flipped in both the vertical and horizontal directions.
The contours correspond to the 3$\varsigma_{V}$, 4$\varsigma_{V}$, and 5$\varsigma_{V}$ values in the indicated velocity range.
\juan{The large differences between the top-left and the other panels indicates that the effect of chance correlation in the results of the HOG method is small.}
}\label{fig:corrPlaneNullTests}
\end{figure}

The results of these ``flipping'' tests, presented in Fig.~\ref{fig:corrPlaneNullTests}, show that the high-$V$ regions around \vhi\,$\approx$\,\vco\ are not present when the spatial distribution of the $^{13}$CO emission is not the observed one.
This indicates that the high $V$ values around \vhi\,$\approx$\,\vco\ are not the product of the concentration of the emission in particular velocity channels, but rather a significant correlation between the contours of both tracers in the corresponding velocity ranges.

\begin{figure}[ht!]
\centerline{
\includegraphics[width=0.25\textwidth,angle=0,origin=c]{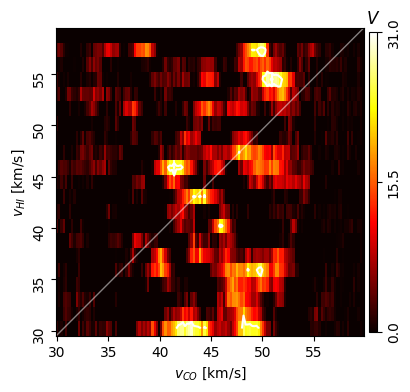}
\includegraphics[width=0.25\textwidth,angle=0,origin=c]{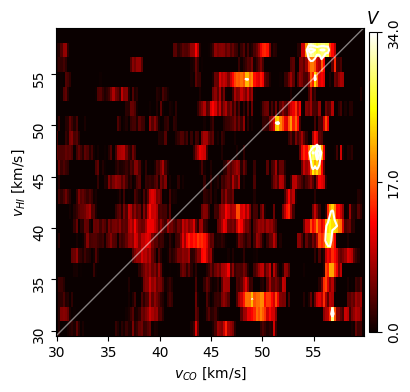}
}
\centerline{
\includegraphics[width=0.25\textwidth,angle=0,origin=c]{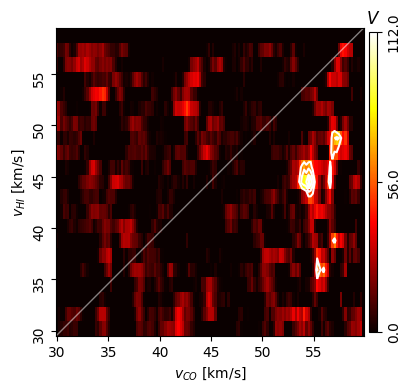}
\includegraphics[width=0.25\textwidth,angle=0,origin=c]{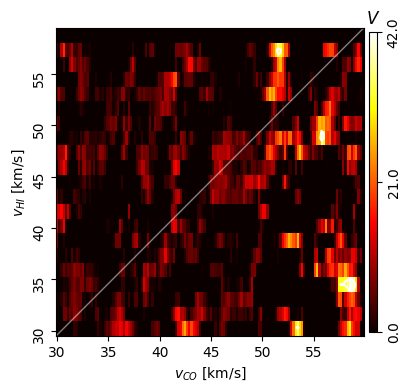}
}
\caption{Same as Fig.~\ref{fig:corrPlaneNullTests} for comparisons between the $\Delta l$\,$=$\,1\pdeg25 THOR H{\sc i} observations centered on $l$\,$=$\,34\pdeg55 and the $\Delta l$\,$=$\,1\pdeg25 GRS $^{13}$CO tiles centered on $l$\,$=$\,32\pdeg05 (top left), 33\pdeg3 (top right), 35\pdeg8 (bottom left), and 37\pdeg05 (bottom right).
\juan{The large differences between the top-left in Fig.~\ref{fig:corrPlaneNullTests} and these panels indicates that the effect of chance correlation in the results of the HOG method is small.}
}\label{fig:corrPlaneNullTests2}
\end{figure}

In addition to the ``flipping'' tests we compute HOG for the H{\sc i} PPV cube centered on $l$\,$=$\,34\pdeg55, which is the subject of the analysis presented in this paper, with $^{13}$CO PPV cubes with the same size but centered on $l$\,$=$\,32\pdeg05, 33\pdeg3, 35\pdeg8, and 37\pdeg05.
The results of these tests, presented in Fig.~\ref{fig:corrPlaneNullTests2}, indicate that the high-$V$ regions around \vhi\,$\approx$\,\vco\ are not present when comparing these PPV cubes.
These reinforces the conclusion that the high $V$ values around \vhi\,$\approx$\,\vco\ are not the product of the concentration of the emission in particular velocity channels.
The maximum values from these ``offset'' tests are significantly lower than those obtained in the original HOG analysis.
Furthermore, the distribution of the $V$ values in either the ``flipping'' or the ``offset'' tests is not related with those found in the original HOG analysis.

\section{HOG in synthetic observations of MHD sims}\label{app:MHDsims}

\begin{figure*}[ht!]
\centerline{
\includegraphics[width=1.0\textwidth,angle=0,origin=c]{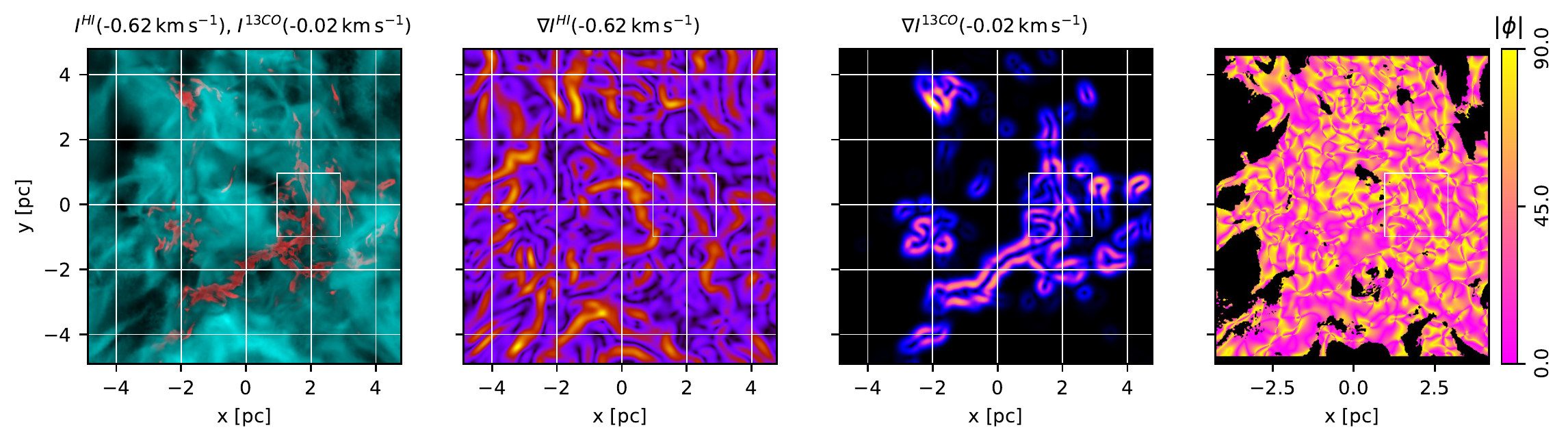}
}
\centerline{
\includegraphics[width=1.0\textwidth,angle=0,origin=c]{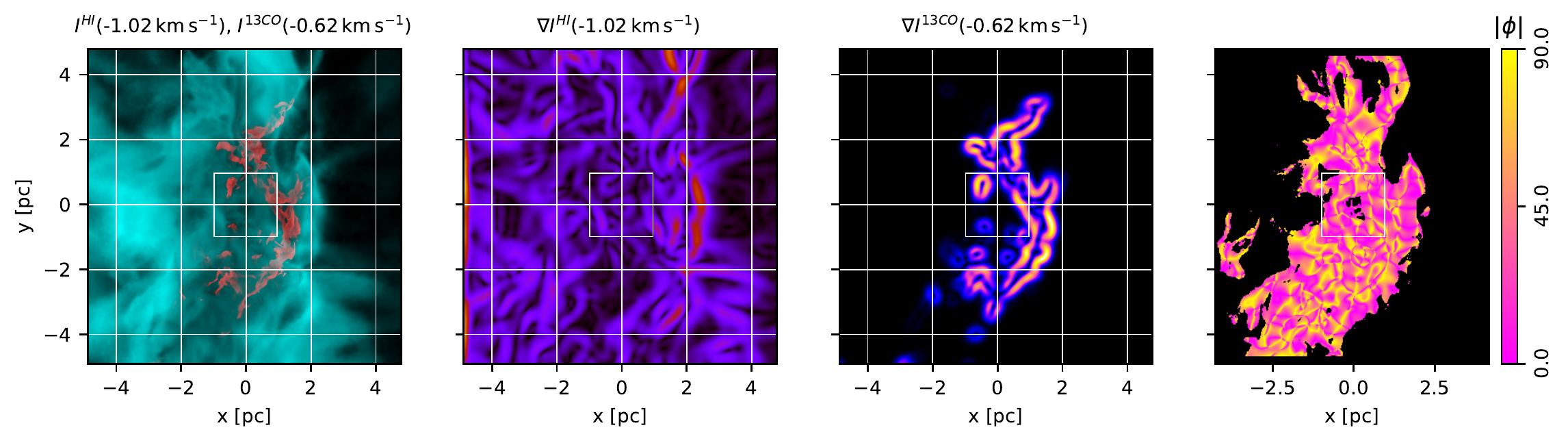}
}
\caption{
Intensity, intensity gradients, and relative orientation angle maps from the synthetic observations of the \cite{clarkInPrep} simulations.
In contrast with the simulations presented in Sec.~\ref{section:mhd}, the synthetic observation do not include the 100-K H{\sc i} background emission.
\emph{Left.} H{\sc i} (teal) and $^{13}$CO emission (red) in the velocity channels with the largest spatial correlation, as inferred from the \prs\ values shown in Fig.~\ref{fig:HOGcorrPCsims}.
\emph{Middle left.} Norm of the gradient of the H{\sc i} intensity map in the indicated velocity channel.
\emph{Middle right.} Norm of the gradient of the $^{13}$CO intensity map in the indicated velocity channel.
\emph{Right.} Relative orientation angle $\phi$, Eq.~\eqref{eq:phi}, between the gradients of the H{\sc i} and $^{13}$CO intensity maps in the indicated velocity channels.
The white color in the $\phi$ map corresponds to areas with no significant gradient in either tracer.
The square indicates the block, selected from a 7\,$\times$\,7 spatial grid, with the largest values of \prs.
The top and bottom panels correspond to the \juan{face-on and edge-on synthetic observations}, respectively.
}\label{fig:HOGpanelPCsimsHIand13CO}
\end{figure*}

\begin{figure}[ht!]
\centerline{
\includegraphics[width=0.5\textwidth,angle=0,origin=c]{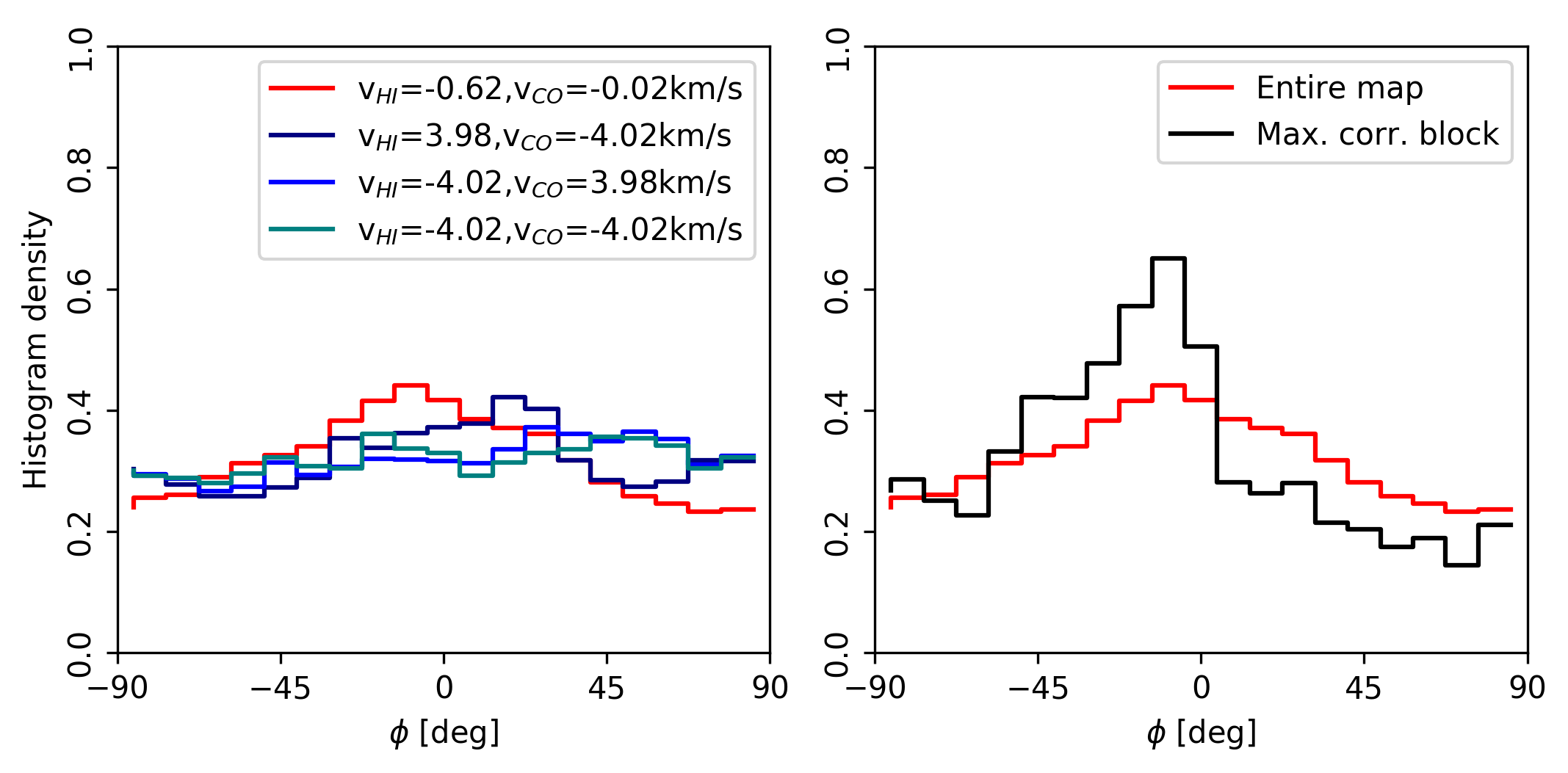}
}
\centerline{
\includegraphics[width=0.5\textwidth,angle=0,origin=c]{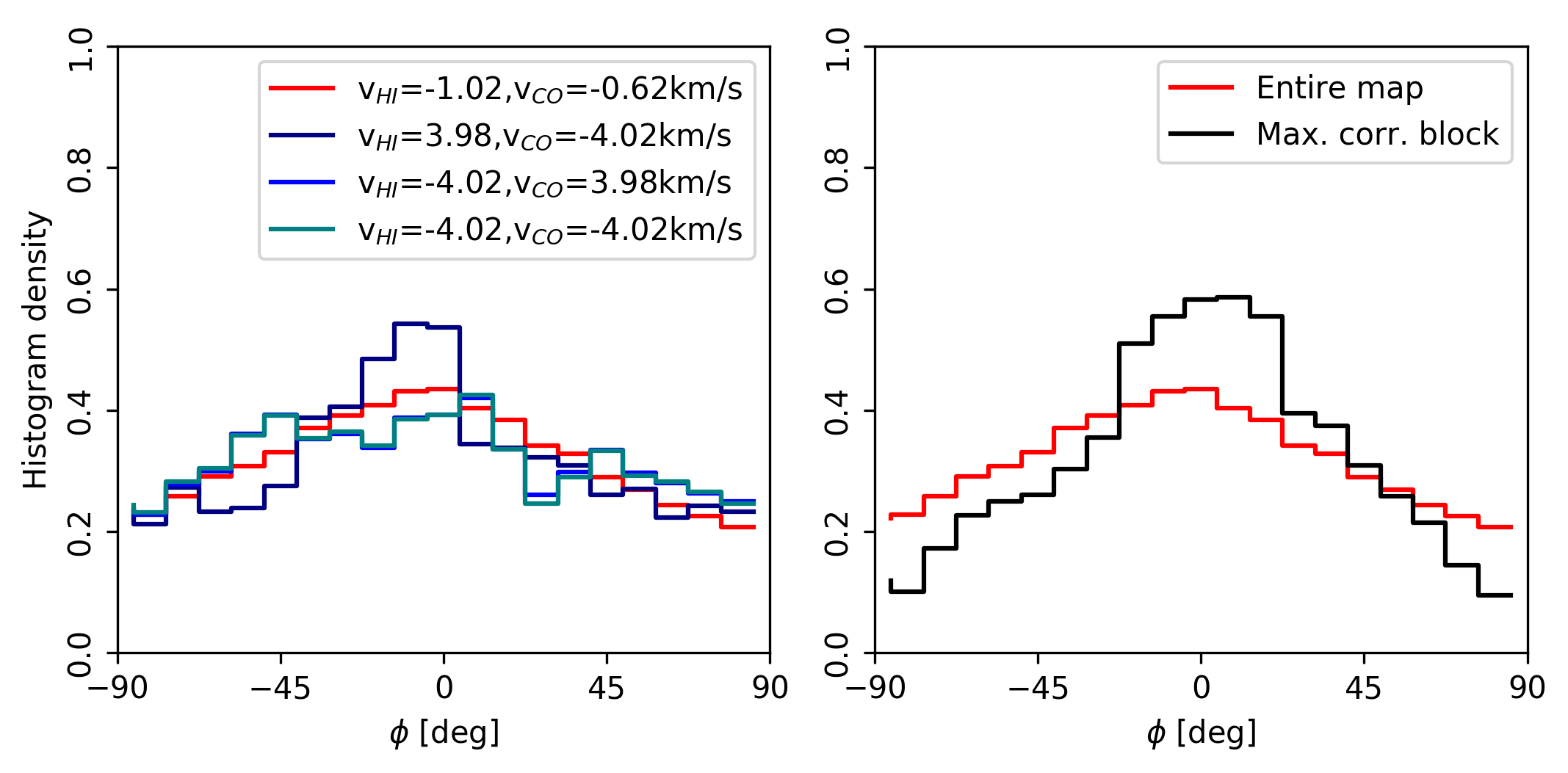}
}
\caption{
\emph{Left}. Histograms of oriented gradients (HOGs) corresponding to the pair of velocity-channel maps with the largest spatial correlation, as inferred from the \prs\ values shown in Fig.~\ref{fig:HOGcorrPCsimsHIand13CO}, and three pairs of arbitrarily selected velocity channels in the synthetic observations of the \cite{clarkInPrep} simulations without the 100-K H{\sc i} background emission considered in Sec.~\ref{section:mhd}.
\emph{Right}. For the pair of velocity-channel maps with the largest spatial correlation, HOGs corresponding to the entire map and just the block with the largest \prs\ indicated in Fig.~\ref{fig:HOGpanelPCsimsHIand13CO}.
The top and bottom panels correspond to the \juan{face-on and edge-on synthetic observations}, respectively.
}\label{fig:HOGPCsimsHIand13CO}
\end{figure}

\begin{figure}[ht!]
\centerline{
\includegraphics[width=0.25\textwidth,angle=0,origin=c]{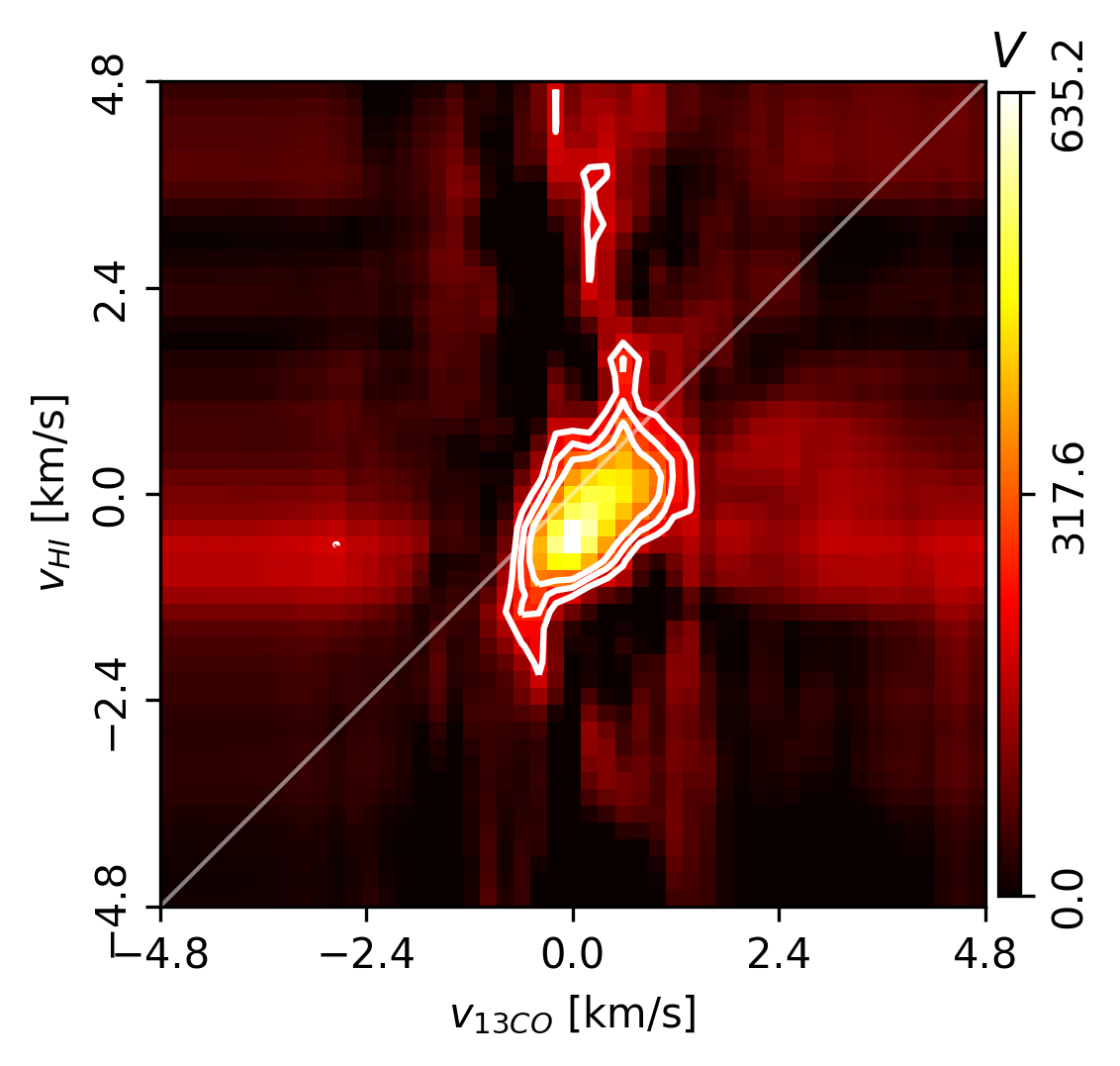}
\includegraphics[width=0.25\textwidth,angle=0,origin=c]{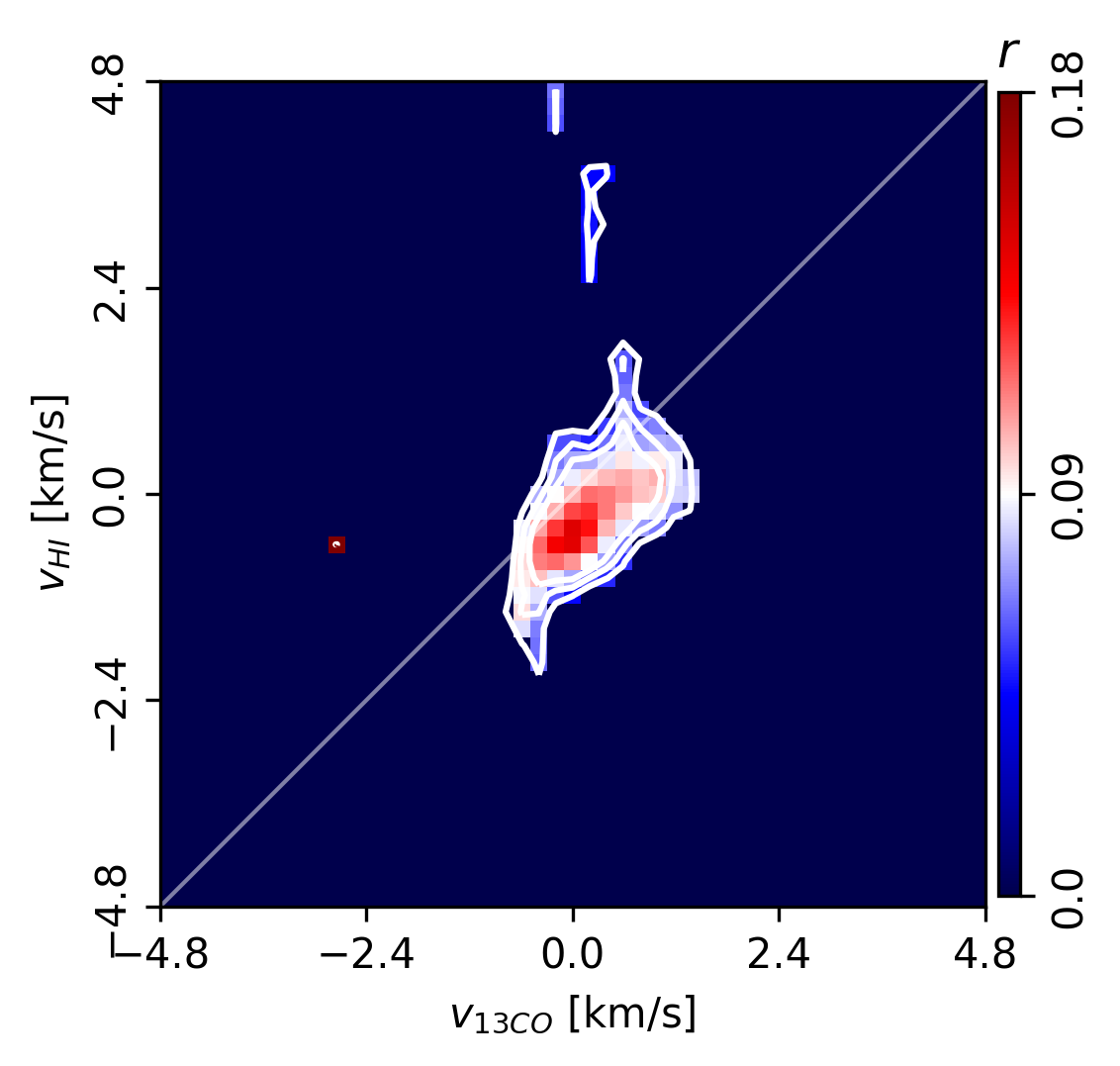}
}
\centerline{
\includegraphics[width=0.25\textwidth,angle=0,origin=c]{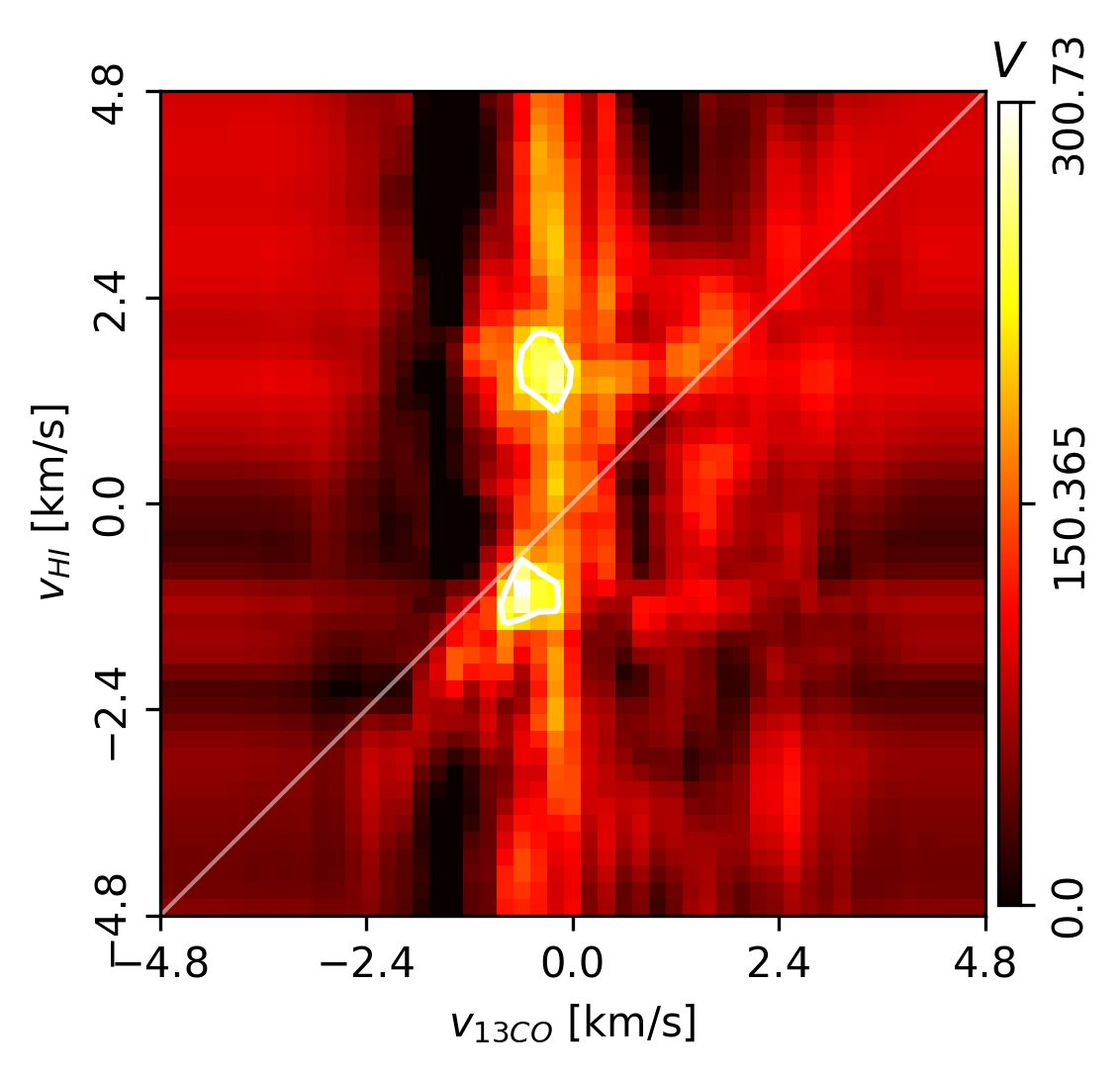}
\includegraphics[width=0.25\textwidth,angle=0,origin=c]{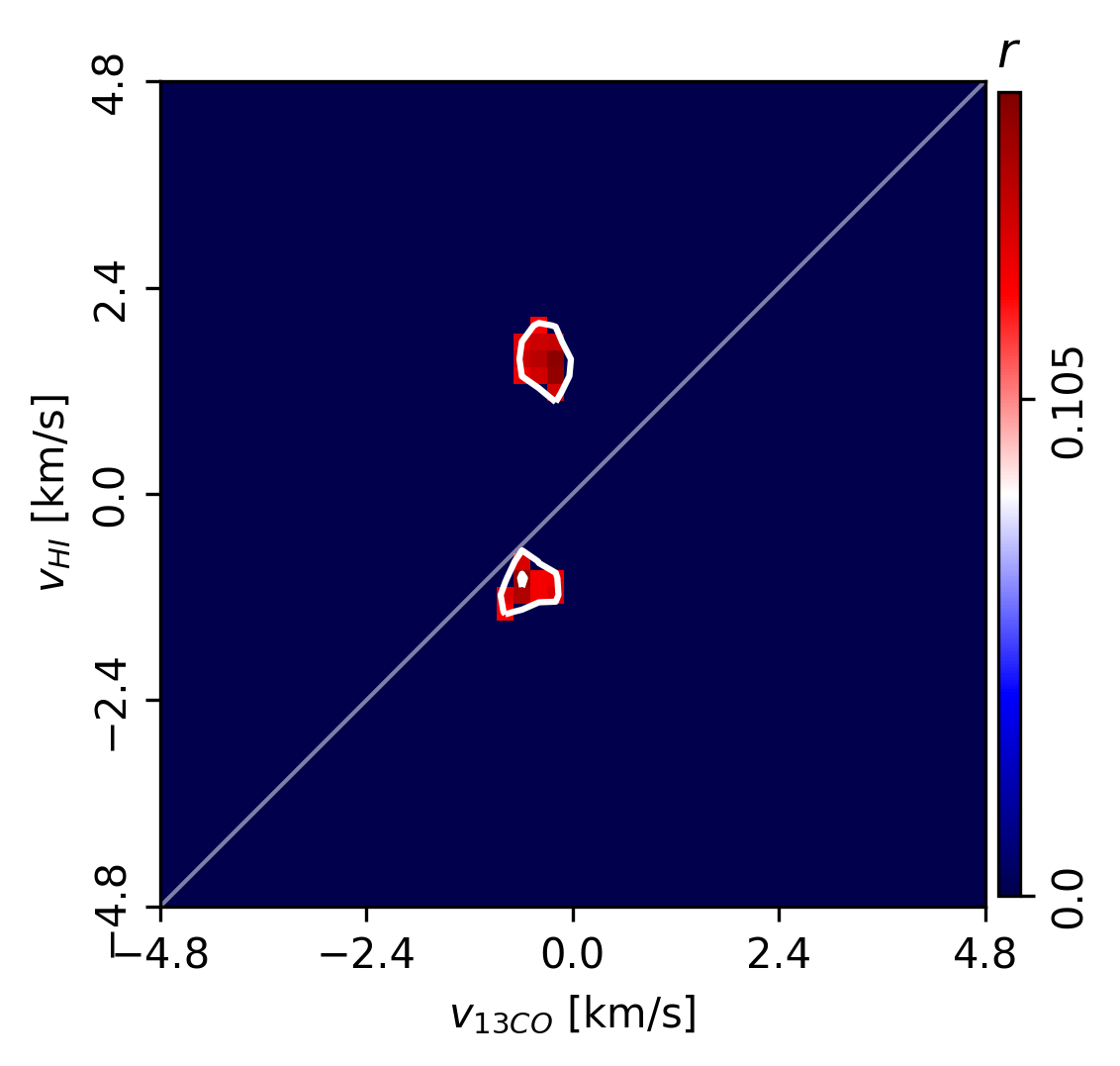}
}
\caption{
Results of the HOG analysis of the H{\sc i} (without 100-K H{\sc i} background emission) and $^{13}$CO synthetic observations of the \cite{clarkInPrep} simulations.
\emph{Left.} Projected Rayleigh statistic, $V(v_{\rm 13CO},v_{\rm HI})$, the HOG statistical test of spatial correlation between H{\sc i} and $^{13}$CO velocity-channel maps, defined in Eq.~\eqref{eq:myprs}.
\juan{The contours indicate the 3$\varsigma_{V}$, 4$\varsigma_{V}$, and 5$\varsigma_{V}$ levels in the corresponding velocity range.}
\emph{Right.} Mean resultant vector length, $r(v_{\rm 13CO},v_{\rm HI})$, within the 3$\varsigma_{V}$ confidence interval, a HOG metric that is roughly equivalent to the percentage of gradient pairs that imply the spatial correlation between the velocity-channel maps, defined in Eq.~\eqref{eq:mymrv}.
The top and bottom panels correspond to the \juan{face-on and edge-on synthetic observations}, respectively.
}\label{fig:HOGcorrPCsimsHIand13CO}
\end{figure}

\subsection{Further analysis of the cloud collision simulations}\label{app:MHDsimsClark}

We complement the analysis presented in Sec.~\ref{section:mhd} by considering the correlation between $^{13}$CO synthetic observations and the H{\sc i} emission, this time without the 100\,K H{\sc i} background emission.
The results of the calculation of the oriented gradients in a pair of velocity-channel maps with high $V$ values are presented in Fig.~\ref{fig:HOGpanelPCsimsHIand13CO}.
The spatial distribution of gradients with relative orientation angles $\phi$\,$\approx$\,0\deg\ indicates that the correlation between the H{\sc i} and $^{13}$CO emission is not exclusively found in regions with H{\sc i} self-absorption.

The HOGs, presented in Fig.~\ref{fig:HOGPCsimsHIand13CO}, also confirm that there is a significant population of parallel gradients, $\phi$\,$\approx$\,0\deg\, in some channels, despite the fact that these synthetic observations do not include the H{\sc i} self-absorption features produced with the 100\,K H{\sc i} background emission.
These results confirm that the H{\sc i} and $^{13}$CO spatial correlation is not exclusively related to the H{\sc i} self-absorption, where it is mostly expected, but that also the general distribution of the H{\sc i} is related to that of the molecular gas.
This does not contradict the conclusion that the H{\sc i} and $^{13}$CO correlation is mostly associated to the CNM, but indicates that the H{\sc i} structure is correlated with the molecular gas in emission and not only in absorption.

There is, however, one significant difference between the spatial correlation of $^{13}$CO and H{\sc i} with and without the 100\,K H{\sc i} background emission that is evident in the distribution of the \prs\ and $r$ values across multiple velocities, as shown in Fig.~\ref{fig:HOGcorrPCsimsHIand13CO}.
While the \prs\ and $r$ corresponding to the \juan{face-on synthetic observations} seem unchanged and reveal high values centered \juan{exclusively} around \vhi\,$\approx$\,\vco\,$\approx$\,0\,\kps, the \juan{edge-on synthetic observations present correlation around two pairs of velocities: at \vhi\,$\approx$\,\vco\,$\approx$\,0\,\kps\ and around \vhi\,$\approx$\,2\,\kps\ and \vco\,$\approx$\,0\,\kps}.
\juan{Most likely, this correlation is a consequence of the H{\sc i} edge of the shocked interface appearing in multiple velocity channels and being correlated with the $^{13}$CO clouds at \vco\,$\approx$\,0\,\kps, an observation that is also consistent with the less-significant vertical stripe in the $V$ values seen in Fig.~\ref{fig:HOGcorrPCsimsHIand13CO}.}
\juan{Unlike the background-emission examples discussed in Sec.~\ref{section:discussionEmissionBackground}, the edge-on synthetic observations constitute a very rare scenario where the common boundary is maintained by the motions of the colliding clouds and highlighted by the particular line of sight orientation.
It is implausible that this singular configuration provides an explanation for the spatial correlations at \vhi\,$\neq$\,\vco\ reported in Fig.~\ref{fig:corrPlane47to62HOGtowardsMCs}.}

\subsection{Analysis of FRIGG simulations}\label{app:FRIGGsims}

In addition to the numerical simulations presented in Sec.~\ref{section:mhd} and Sec.~\ref{app:MHDsimsClark}, we applied the HOG method to a set of synthetic observations from the stratified, 1-kpc scale, magneto-hydrodynamical simulations part of the FRIGG project \citep{hennebelle2018}.
The ISM in these simulations is regulated by supernovae and include self-gravity, magnetic fields, cooling and heating processes, and a gravity profile that accounts for the distribution of stars and dark matter.
The FRIGG simulations were designed to cover spatial scales between the intermediate galactic scales and the self-gravitating prestellar cores, ranging between the 1-kpc side simulation domain down to a maximum resolution of 3.8\,$\times$\,10$^{-3}$\,pc.
Further details on the initial conditions, zoom-in resolution strategies, and included physics are presented in \cite{hennebelle2018}.

We extracted segments of the simulation around a set of density structures identified using a friend-of-friends (FoF) algorithm with a threshold density $n_{0}$\,$=$\,10$^{3}$\,cm$^{-3}$ in a snapshot of the simulation taken at $t$\,$=$\,9.11\,Myr.x
The size of the extracted volumes was set to be twice the effective diameter of the identified structures.
Explicitly, the extraction yields to regular density, temperature, velocity, and magnetic field cubes with a common grid resolution, which we choose to be 0.12\,pc.
For the sake of simplicity, we focused in a structure with an effective radius of 10\,pc, which was the fourth largest linked density structure identified by the FoF algorithm. 
Using the estimated proper motion of this structure, we traced back its evolution and extracted its properties in snapshots taken at $t$\,$=$\,9.06, 9.01, 8.96 and 8.92\,Myr.

\subsubsection{Synthetic H{\sc i} observations}

We produce synthetic H{\sc i} observations by applying \juan{the procedure described} in \cite{miville-deschenes2007}, which is itself based on the general radiative transfer equations presented in \cite{spitzer1978} in the optically-thin regime.
Given a density cube $n_{ijk}$; where the indexes $i$, $j$, and $k$ run over the spatial coordinates $x$, $y$, and $z$, respectively; the brightness temperature map in the $xy$-plane at velocity $u_{l}$ for a distant cloud with respect to its size is
\begin{equation}
[T_{B}]_{ij,l} = \frac{1}{C\sqrt{2\pi}}\sum_{k}\frac{n_{ijk}}{\sigma}\times\exp\left(-\frac{(u_{l}-[v_{z}]_{ijk})^{2}}{2\sigma^{2}}\right)\Delta z,
\end{equation}
where $[v_{z}]_{ijk}$ is the velocity along the line of sight, which we assumed to be the $z$ axis.
The term $\sigma^{2}$\,$\equiv$\,$k_{\rm B}T_{ijk}/m$ represents the thermal broadening of the 21-cm line; $T_{ijk}$ is the temperature cube, $m$ is the hydrogen atom mass, and $k_{B}$ is the Boltzmann constant.
Finally, the constant $C$\,$=$\,$1.813\times10^{18}$\,cm$^{-2}$ \citep{spitzer1978}.

Besides the assumption that the emission is optically thin, this treatment assumes that \juan{the excitation temperature} (spin temperature) of the 21-cm line is the same as the kinetic temperature of the gas.
The latter assumption is not representative of the general conditions in the ISM \citep[see for example][and the references therein]{kim2014}, but it is sufficient for the our current exploration of the simulations.
We present some selected spectra in the left-hand-side panel of Fig.~\ref{fig:syntheticFRIGGspectra} and integrated emission maps in the left column of Fig.~\ref{fig:HOGcorrFRIGGsims}.

\subsubsection{Synthetic $^{13}$CO observations}

\begin{figure}[ht!]
\centerline{
\includegraphics[width=0.5\textwidth,angle=0,origin=c]{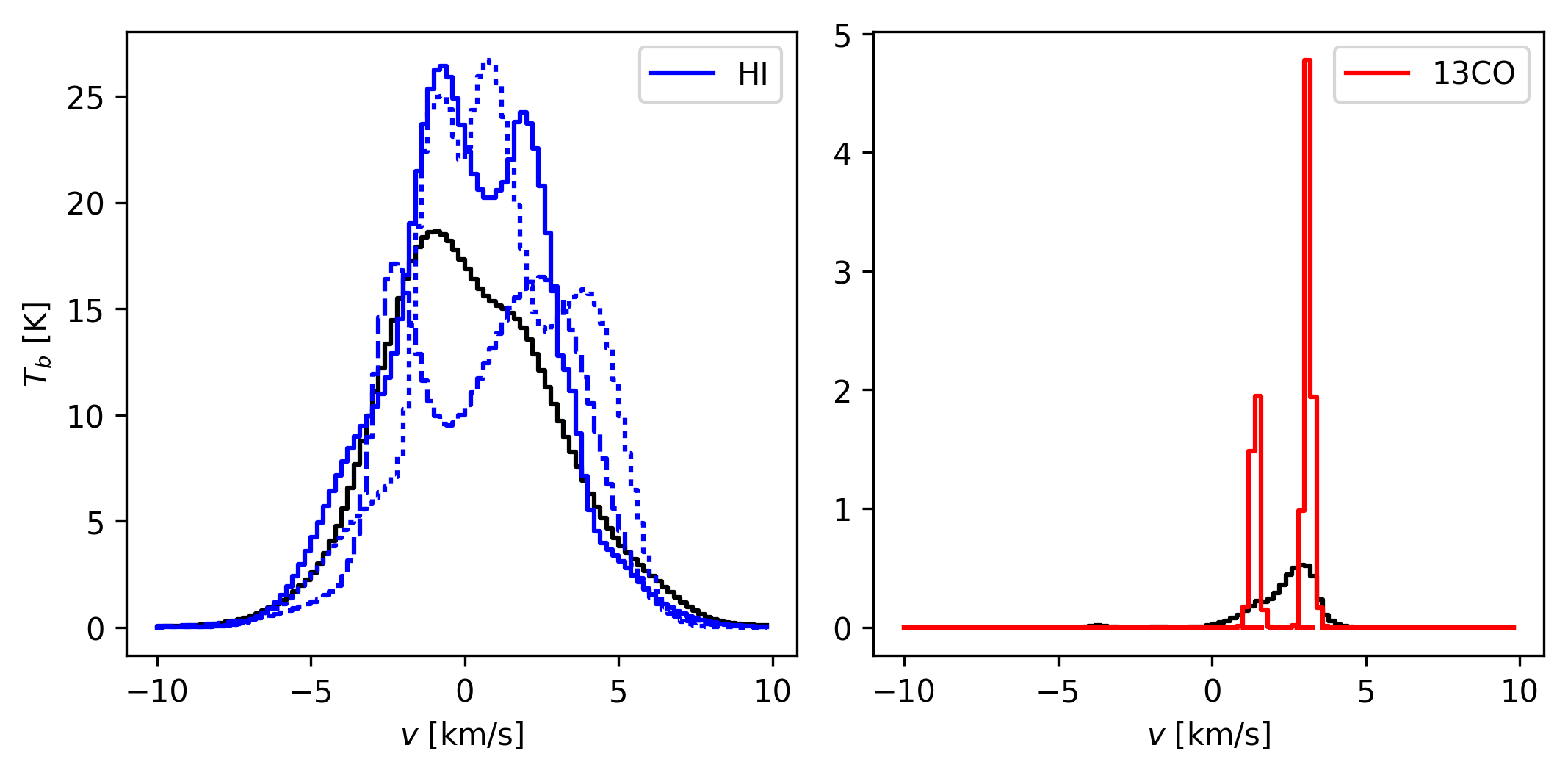}
}
\caption{
Spectra from the synthetic observations of H{\sc i} and $^{13}$CO emission from a segment of the FRIGG simulations in a snapshot taken at $t$\,$=$\,9.11\,Myr, also shown in the top panels of Fig.~\ref{fig:HOGcorrFRIGGsims}.
The black lines correspond to the average spectra over the whole map.
The solid, dashed, and segmented colored lines correspond to the spectra toward the positions $[x,y]$\,$=$\,[0,0], [5,0], and [0,5]\,pc, respectively.
}
\label{fig:syntheticFRIGGspectra}
\end{figure}

We produced synthetic $^{13}$CO($J$\,$=$\,1\,$\rightarrow$\,0) emission observations using an approach similar to the one described for H{\sc i}, changing the corresponding coefficients and masses.
To account for the $^{13}$CO distribution in the simulated volume, we considered a critical density $n_{\rm c}$\,$=$\,500\,cm$^{-3}$ and a critical temperature $T_{\rm c}$\,$=$\,100\,K.
That is, we considered that only gas at $n$\,$>$\,$n_{\rm c}$ and $T$\,$<$\,$T_{\rm c}$ produces $^{13}$CO emission.
This strategy is evidently an over-simplification of the model and does not replace the proper treatment of a chemical network in the simulation, as in the \cite{clarkInPrep} simulations.
However, this simple experiment can shed light on the spatial correlation between the dense and the diffuse gas across a range of velocities without intending to fully reproduce the complexity of the observations.
We present some selected spectra in the right-hand-side panel of Fig.~\ref{fig:syntheticFRIGGspectra} and integrated emission maps in the center column of Fig.~\ref{fig:HOGcorrFRIGGsims}.

\subsubsection{HOG analysis}

\begin{figure*}[ht!]
\centerline{
\includegraphics[width=1.0\textwidth,angle=0,origin=c]{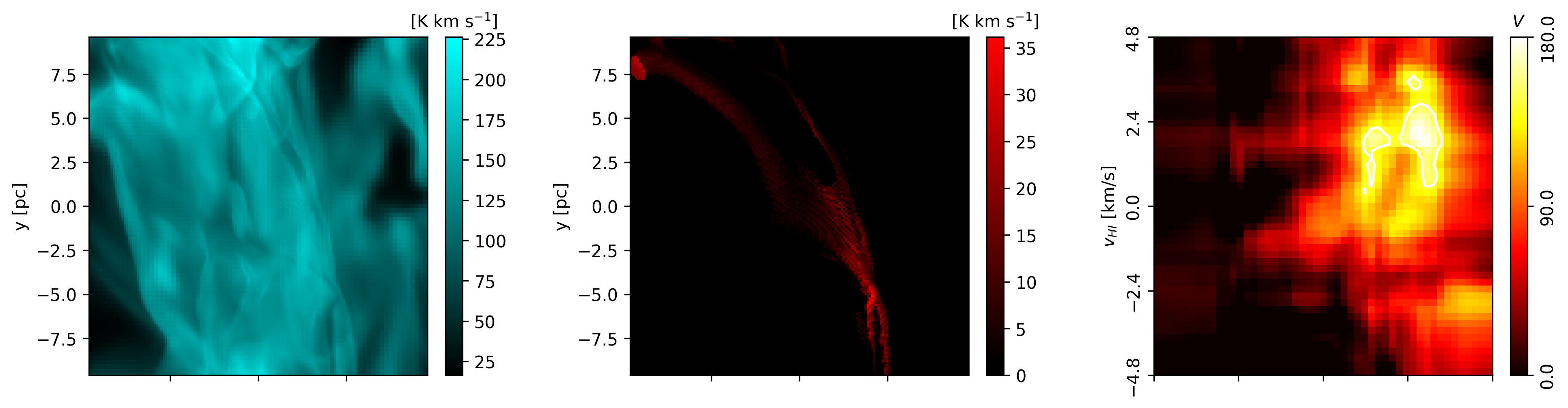}
}
\vspace{-0.2cm}
\centerline{
\includegraphics[width=1.0\textwidth,angle=0,origin=c]{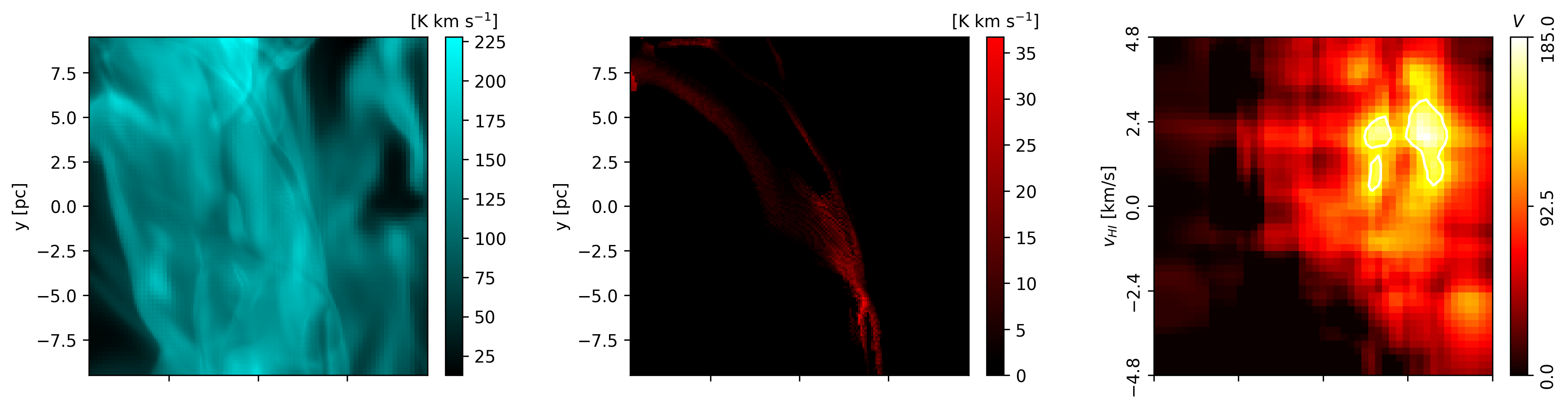}
}
\vspace{-0.2cm}
\centerline{
\includegraphics[width=1.0\textwidth,angle=0,origin=c]{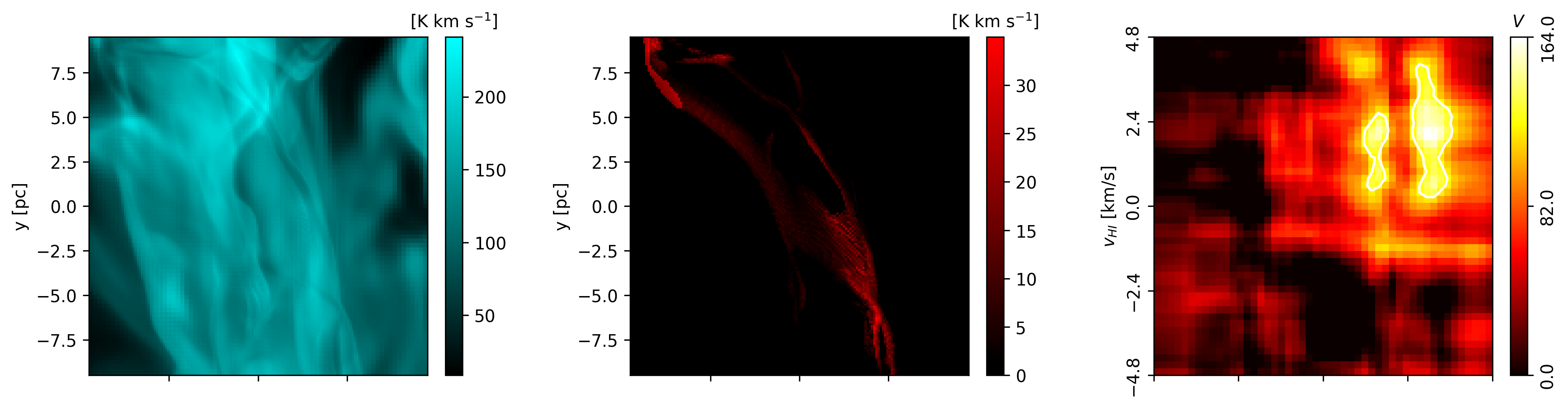}
}
\caption{
Integrated H{\sc i} (left), $^{13}$CO emission (center), and projected Rayleigh statistic ($V$, right) corresponding to the synthetic observations of a segment of the MHD simulations presented in \cite{hennebelle2018}.
From top to bottom, the panels represent snapshots taken at $t$\,$=$\,9.11, 9.06, and 9.01\,Myr, respectively.
}\label{fig:HOGcorrFRIGGsims}
\end{figure*}

\begin{figure*}[ht!]
\centerline{
\includegraphics[width=1.0\textwidth,angle=0,origin=c]{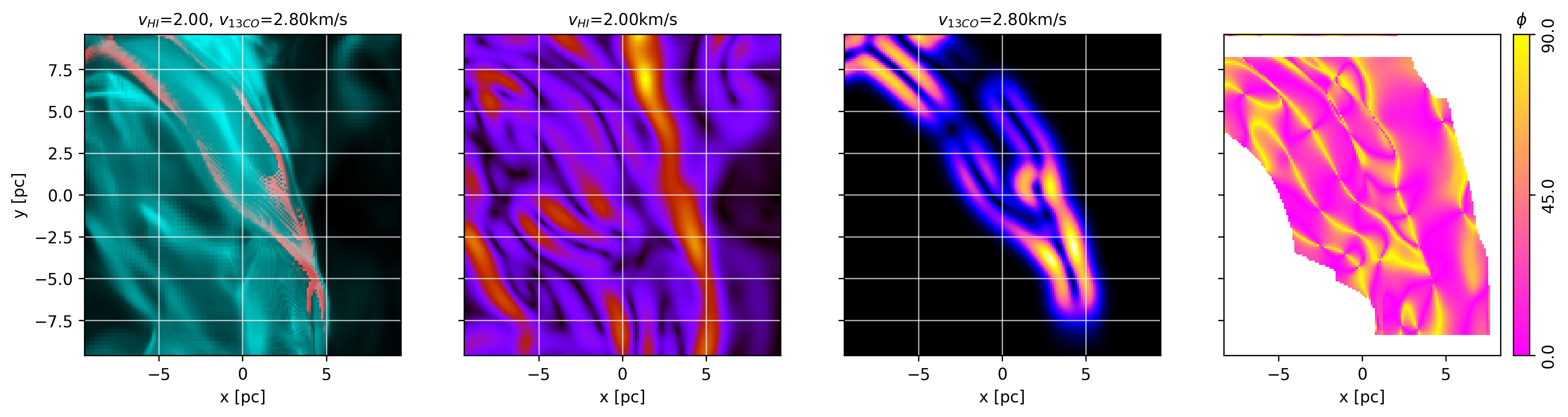}
}
\caption{
Intensity, intensity gradients, and relative orientation angle maps from the synthetic observations of a segment of the FRIGG simulation in the snapshot taken at $t$\,$=$\,9.11\,Myr.
\emph{Left.} Synthetic H{\sc i} (teal) and $^{13}$CO emission (red) in the velocity channels with the largest HOG correlation, as inferred from the values shown in Fig.~\ref{fig:HOGcorrFRIGGsims}.
\emph{Middle left.} Norm of the gradient of the H{\sc i} intensity map in the indicated velocity channel.
\emph{Middle right.} Norm of the gradient of the $^{13}$CO intensity map in the indicated velocity channel.
\emph{Right.} Relative orientation angle $\phi$, Eq.~\eqref{eq:phi}, between the gradients of the intensity maps shown in the left-hand-side panel.
}
\label{fig:syntheticFRIGGhogpanel}
\end{figure*}

The results of the HOG analysis of the synthetic observations of the selected portion of the FRIGG simulations in different time steps is presented on the right-hand-side column of Fig.~\ref{fig:HOGcorrFRIGGsims}.
It is clear from the distribution of values of the projected Rayleigh statistic, $V$, that the velocity channels with the highest correlation are not concentrated around \vhi\,$\approx$\vco, but there is a significant level of correlation between velocity channel maps separated by approximately 1\,\kps.
This is evident in Fig.~\ref{fig:syntheticFRIGGhogpanel}, where we present the comparison between the velocity channels centered on \vhi\,$\approx$\,2.0 and \vco\,$\approx$\,2.8\,\kps, which show the largest $V$ values.
Given the assumptions that we made to produce the synthetic H{\sc i} and $^{13}$CO synthetic observations, it is apparent that the spatial correlations in the HOG analysis of observations can be reproduced with the physical conditions included in the FRIGG simulations.
However, the particular origin of the spatial correlation in H{\sc i} and $^{13}$CO velocity channels separated by a few kilometers per second is yet to be determined.

\end{document}